%% file: main.tex
\documentclass{article}

\usepackage{PRIMEarxiv}

\usepackage[utf8]{inputenc} 
\usepackage[T1]{fontenc}    
\usepackage{hyperref}       
\usepackage{url}            
\usepackage{booktabs}       
\usepackage{amsmath}
\usepackage{amsfonts}       
\usepackage{nicefrac}       
\usepackage{microtype}      
\usepackage{fancyhdr}       
\usepackage{graphicx}       
\usepackage{multirow}
\usepackage{enumitem}
\usepackage[ruled,vlined]{algorithm2e}
\usepackage{natbib}
\usepackage{acronym}
\usepackage{float}
\graphicspath{{media/}} 

\title{Integrating Prognostics, Maintenance, and Tail Assignment under Remaining Useful Life Uncertainty: A Stochastic Optimisation Approach for Airline Reliability}

\date{\today}

\makeatletter
\renewcommand{\@maketitle}{%
  \vbox{%
    \hsize\textwidth
    \linewidth\hsize
    \vskip 0.1in
    \@toptitlebar
    \centering
    {\LARGE\scshape \@title\par}
    \@bottomtitlebar
    {\large\scshape A Preprint\par}
    \vskip 0.1in
    \def\And{%
      \end{tabular}\hfil\linebreak[0]\hfil%
      \begin{tabular}[t]{c}\bf\rule{\z@}{24\p@}\ignorespaces%
    }
    \def\AND{%
      \end{tabular}\hfil\linebreak[4]\hfil%
      \begin{tabular}[t]{c}\bf\rule{\z@}{24\p@}\ignorespaces%
    }
    \begin{tabular}[t]{c}\bf\rule{\z@}{24\p@}\@author\end{tabular}%
    \vskip 0.4in \@minus 0.1in \center{ } \vskip 0.2in
  }
}
\makeatother

\author{
  Benno Käslin \\
  Delft University of Technology \\
  Zurich University of Applied Sciences \\
  Swiss International Air Lines Ltd. \\
  \texttt{b.kaeslin@tudelft.nl} \\
  \And
  Marta Ribeiro \\
  Delft University of Technology \\
  292F+VC Delft \\
  Netherlands \\
  \texttt{m.j.ribeiro@tudelft.nl} \\
  \And
  Dimitrios Zarouchas \\
  Delft University of Technology \\
  292F+VC Delft \\
  Netherlands \\
  \texttt{d.zarouchas@tudelft.nl} \\
  \And
  Manuel Arias Chao \\
  Delft University of Technology \\
  Zurich University of Applied Sciences \\
  \texttt{m.a.c.ariaschao@tudelft.nl} \\
}

\begin{document}
\maketitle
\begin{abstract}
Ensuring reliability, safety, and economic efficiency in airline operations requires maintenance and fleet scheduling strategies that explicitly account for uncertainty in Remaining Useful Life (RUL) predictions. However, the integration of prognostic uncertainty into operational decision-making remains a major challenge. In practice, tail assignment (TA) and maintenance scheduling (MS) are typically optimized separately or sequentially, thereby limiting the effective use of predictive health information despite their strong interdependencies.
This paper proposes a unified optimisation framework that jointly integrates TA, MS, and predictive maintenance (PdM) under RUL with confidence intervals. The problem is formulated as a stochastic mixed-integer linear program, and a scalable solution approach is developed by embedding a neural network surrogate to approximate expected disruption costs resulting from RUL uncertainty.
The proposed framework is evaluated using operational scenarios derived from real-world airline data. Results show that explicitly incorporating prognostic uncertainty in a joint planning model reduces operational risk, i.e., downstream disruption costs and flight cancellations, compared to deterministic and sequential approaches, at the expense of moderate increases in planning cost.
These findings highlight the value of tightly coupling predictive maintenance with operational planning and demonstrate the potential of surrogate-assisted stochastic optimisation for scalable, uncertainty-aware airline decision-making.
\end{abstract}

\newpage
\input{01_introduction}

\input{02_problem_description}

\input{03_related_work}

\input{04_methodology}

\input{05_experimental_setup}

\input{06_results}

\input{07_discussion}

\input{08_conclusion}

\input{09_acknowledgement_coi_data_and_code}
\newpage
\input{10_appendix}
\newpage
\input{11_abbreviations}
\newpage
\bibliographystyle{unsrt}
\bibliography{references}

\end{document}

%% file: 01_introduction.tex
\section{Introduction}
\label{sec:introduction}

In the highly competitive aviation sector, even incremental operational improvements translate into substantial strategic gains. Maintenance activities alone account for approximately 11.5\% of total airline operational cost~\cite{InternationalAirTransportAssociationIATA2024MaintenanceData}, making them a critical lever for cost control and operational reliability. Effective maintenance management directly impacts three operational dimensions: (i) costs efficiency through efficient resource allocation, (ii) fleet availability to meet dynamic passenger demand, and (iii) safety and reliability standards, which are essential for public trust and regulatory compliance.

Airlines face a fundamental coordination problem in jointly managing flight operations and aircraft maintenance. Traditionally, airline planning addresses these decisions sequentially: aircraft are first assigned to flights based on operational objectives, then maintenance activities are subsequently scheduled around the available ground windows~\cite{Feo1989FlightPlanning}. This sequential approach fails to exploit natural interdependencies between the two decision layers. On the one hand, maintenance activities directly constrain aircraft availability for flight assignments and on the other hand, \ac{TA} decisions determine aircraft routing and positioning~\cite{Eltoukhy2017AirlineDirections}, thereby influencing feasible maintenance opportunities in both time and location~\cite{SriramAnRe-assignment}. Consequently, globally optimal solutions may require accepting locally suboptimal assignment decisions (e.g., repositioning aircraft) to enable more efficient \ac{MS} and reduce overall system costs~\cite{Barnhart1998Flight1}.

Additionally, sequential approaches ignore how operational uncertainty about component health influences the value of different \ac{TA} and \ac{MS} decisions ~\cite{Lagos2019DynamicOperations}. Component degradation is inherently uncertain, arising from multiple sources: complex and nonlinear degradation mechanisms~\cite{SankararamanWhyUncertain}, operational variability (usage patterns, environmental conditions), and measurement noise in sensor systems~\cite{Bieber2021Data-DrivenMaintenance}. Deterministic planning based solely on expected \ac{RUL} values ignores this uncertainty, producing brittle schedules: when actual component degradation exceeds predictions, a frequent occurrence due to the non-Gaussian nature of failure distributions, unplanned maintenance becomes necessary, forcing costly flight cancellations and reactive decisions~\cite{dePater2022Alarm-basedPrognostics}. For instance, unexpected component failures force reactive maintenance, triggering flight cancellations and cascading delays that disrupt operational plans and passenger itineraries~\cite{Vink2020DynamicFramework}.

Operational uncertainty can be reduced through a transition from scheduled preventive maintenance toward \ac{CBM} strategies that leverage real-time sensor data, condition and health monitoring~\cite{Si2011RemainingApproaches,Khan2018AManagement}. \ac{CBM} encompasses a spectrum from simple threshold-based condition monitoring to prognostic approaches that forecast component degradation and predict \ac{RUL}~\cite{Pecht2010ASystems,AriasChao2022FusingPrognostics}. This latter variant, known as \ac{PdM}, uses probabilistic \ac{RUL} predictions to schedule maintenance interventions proactively, before failure occurs, rather than reacting to threshold exceedances or adhering to fixed intervals. This paper focuses specifically on \ac{PdM}: all maintenance decisions driven by \ac{RUL} predictions are referred to as \ac{PdM} throughout.
However, despite maturity in individual optimisation techniques and growing \ac{PdM} adoption, a critical integration gap still persists in the literature. Existing literature predominantly addresses two disjoint problem classes. First, integrated \ac{TA} and \ac{MS} models typically assume deterministic maintenance requirements based on predefined intervals~\cite{Lagos2019DynamicOperations, VarennaAOperations, Deng2020AOptimization}. While some studies consider stochastic task arrivals~\cite{Lagos2019DynamicOperations}, they do not incorporate prognostics. Second, \ac{PdM} approaches that account for prognostic uncertainty focus on maintenance decision-making in isolation, without joint optimisation with flight assignment decisions~\cite{dePater2021PredictiveComponents,Lee2023DeepPrognostics,Tseremoglou2024Condition-BasedApproach}. Consequently, prognostic information is either ignored in operational planning or incorporated heuristically, rather than through a unified optimisation framework. 

To the best of our knowledge, there exists no formulation that jointly optimises \ac{TA}, \ac{MS}, and \ac{PdM} decisions while explicitly modelling \ac{RUL} uncertainty and its impact on fleet-level operations.
This paper addresses this gap by proposing a unified stochastic optimisation framework that integrates \ac{TA}, \ac{MS}, and \ac{PdM} decisions under explicit \ac{RUL} uncertainty. We hypothesize that such integration enables more robust and cost-effective operational plans by explicitly accounting for the risk of premature component failures.
Beyond the technical challenge of integrating prognostics into operational planning, a fundamental question arises: what is the operational and economic value of the underlying prognostic model? Uncertainty-aware maintenance decisions depend not only on the expected \ac{RUL} estimate but also on the calibration of the associated uncertainty. A predictor with poorly calibrated confidence intervals may lead to overly optimistic or overly conservative maintenance actions, both of which incur significant operational costs. The integrated framework proposed in this paper creates a principled basis for addressing this question, since it explicitly links prognostic model outputs to downstream planning costs and operational outcomes, making such a systematic assessment tractable as a natural extension.

The main contributions of this paper are then as follows:

\begin{enumerate}
    \item \textbf{Integrated \ac{MS}--\ac{TA}--\ac{PdM} optimisation framework}: We formulate the joint optimisation of all three decision layers as a single programme, capturing the bidirectional interdependencies between flight assignment, \ac{MS}, and \ac{CBM} decisions.

    \item \textbf{Explicit modelling of \ac{RUL} uncertainty in integrated planning}: Unlike prior \ac{TA}--\ac{MS} work that assumes deterministic maintenance requirements and prior \ac{PdM} approaches that optimise maintenance in isolation, we embed stochastic \ac{RUL} constraints directly into the aircraft-to-flight assignment feasible region, capturing how component health uncertainty explicitly influences aircraft availability and fleet-level planning decisions.
    
    \item \textbf{Scalable neural network-enhanced stochastic optimisation}: We employ offline-trained, ReLU-linearizable \acp{NN} to approximate evaluation-stage costs, enabling tractable large-scale \ac{SP} without explicit scenario enumeration at solve time.

    \item \textbf{Validation and quantification on real airline data}: We implement and validate the framework using actual operational data from Swiss International Air Lines Ltd., demonstrating practical applicability and quantifying the marginal value of prognostic information in integrated planning.
\end{enumerate}

The remainder of this paper is organized as follows. Section~\ref{sec:problem_description} formally introduces the three core optimisation problems (\ac{TA}, \ac{MS}, and \ac{PdM}) and motivates their joint optimisation by illustrating key interdependencies. Section~\ref{sec:related_work} then positions this work within the broader literature, identifying four critical research gaps in the integration of MS, TA, and PdM under prognostic uncertainty. Sections~\ref{sec:methodology} and~\ref{sec:experimental_setup} together present our integrated stochastic optimisation framework and solution approach, including the neural network cost approximation strategy and the experimental design on real airline data, respectively. Section~\ref{sec:results} reports the computational results, demonstrating measurable improvements from joint optimisation and the value of prognostic information. Section~\ref{sec:discussion} interprets these findings in the context of practical airline operations. Finally, Section~\ref{sec:conclusion} synthesizes the contributions and outlines future research directions.

%% file: 02_problem_description.tex
\section{Problem Description}
\label{sec:problem_description}

\begin{figure}[h]
    \centering
    \includegraphics[width=0.85\linewidth]{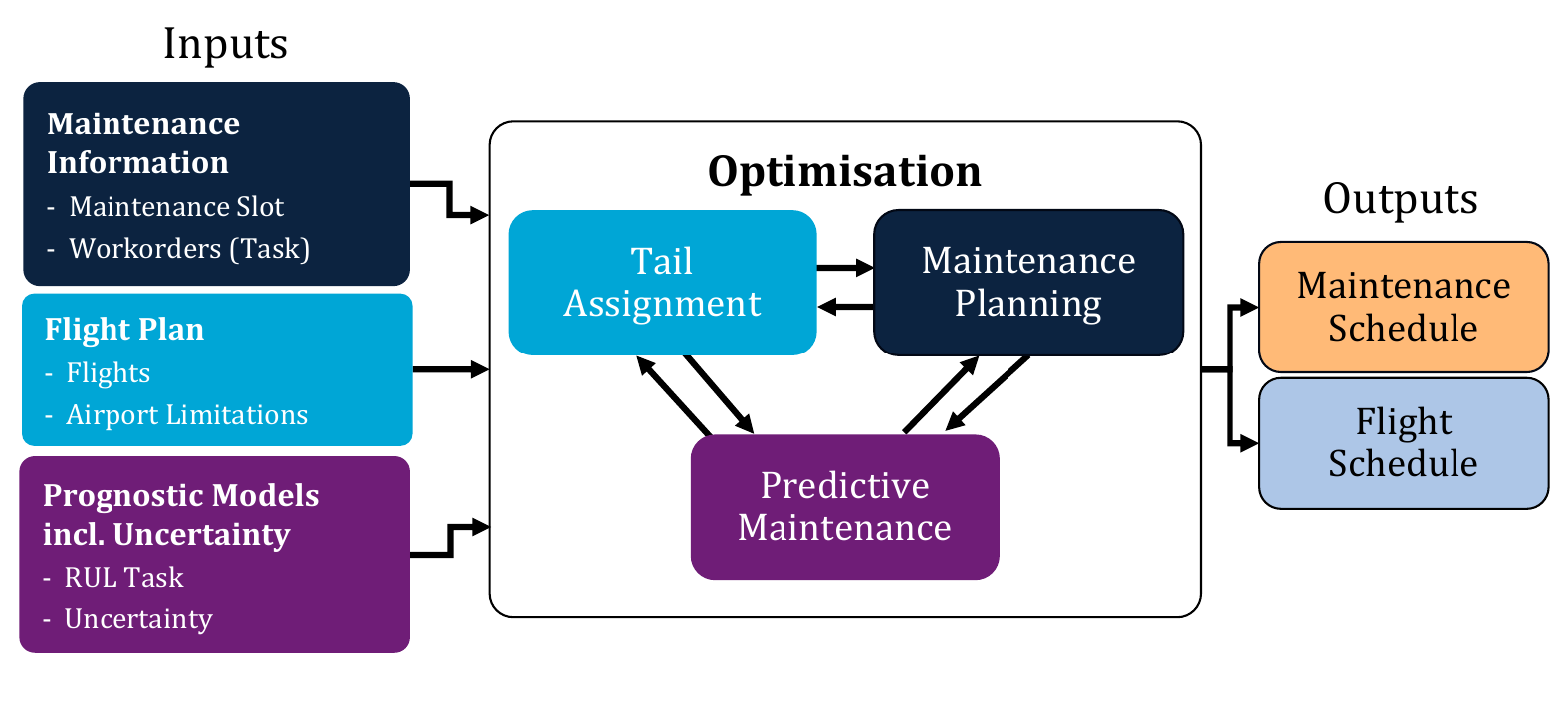}
    \caption{Problem structure: The three decision problems (\ac{TA}, \ac{MS}, \ac{PdM}) are 
             coupled through aircraft positioning, maintenance availability, and 
             component health. Traditional approaches optimize these sequentially. 
             This work optimises them jointly under explicit \ac{RUL} uncertainty.}
    \label{fig:problem_structure}
\end{figure}

Airlines face the complex challenge of simultaneously managing flight operations and aircraft maintenance under operational uncertainty. This section introduces the three interconnected optimisation problems; \ac{TA}, \ac{MS}, and \ac{PdM}. These elements form the foundation of our integrated framework, in conjunction with the uncertainty inherent in prognostic models. As shown in Figure \ref{fig:problem_structure}, these three decision problems are inherently coupled. \ac{TA} determines aircraft positioning, which constrains where and when maintenance can occur. \ac{MS}, in turn, determines aircraft unavailability, which constrains which flights an aircraft can operate. Component health (implicitly captured by \ac{RUL} predictions) determines urgency of maintenance, which feeds back to both \ac{TA} and \ac{MS} decisions. Traditional airline operations treat these problems separately, leading to suboptimal and fragile plans. 

We organize this section into five core themes: (i) \ac{TA} and aircraft-to-flight assignment decisions (Section \ref{subsec:ta}), (ii) \ac{MS} and slot allocation (Section \ref{subsec:ms}), (iii) \ac{PdM} and prognostic uncertainty (Section \ref{subsec:pdm}), (iv) the integrated optimisation problem that jointly optimises all three decision layers (Section \ref{subsec:integrated}), and (v) current industry practice, which highlights the suboptimality of sequential decision-making (Section \ref{subsec:current_practice}). Together, these subsections formalize the three problems and their interdependencies.

\subsection{Tail Assignment}
\label{subsec:ta}
The \ac{TA} problem involves determining which specific aircraft (tail) executes each flight in a given schedule. While flight schedules are typically determined approximately one year in advance, tail assignments are made within a planning horizon of several days to weeks, depending on airline operations and fleet composition. Each flight must be assigned to exactly one aircraft from the available fleet, while respecting operational constraints such as number of booked passenger, aircraft location, turnaround times, and maintenance requirements. The assignment decision directly impacts maintenance feasibility, as aircraft must be positioned at appropriate locations with sufficient ground time to enable maintenance activities.

Formally, \ac{TA} decisions are captured through binary variables $F_{p,f} \in \{0,1\}$ indicating whether aircraft $p$ is assigned to flight $f$ when equal to 1, and $C_f \in \{0,1\}$ indicating flight cancellation when equal to 1. The basic constraint ensures each flight is either assigned to one aircraft or cancelled:
\begin{align}
\sum_{p} F_{p,f} + C_f = 1 \quad \forall f \in \mathcal{F}
\label{eq:ta_basic}
\end{align}

\subsection{Maintenance Scheduling}
\label{subsec:ms}
The \ac{MS} problem determines when and where to perform scheduled maintenance tasks on each aircraft. Two different types of maintenance are considered: line maintenance, and hangar maintenance tasks. Line maintenance, which occurs on the tarmac between flights, is excluded from the scope of this paper. Hangar maintenance tasks have specific duration requirements and can only be performed at designated stations during available maintenance slots. Ground time between flights provides opportunities for maintenance, but task scheduling must account for hangar slot availability, technician capacity and specialization, and parts availability. The timing of maintenance activities constrains which flights an aircraft can operate, creating strong interdependencies with \ac{TA} decisions. Two types of scheduled hangar tasks are considered: preventive maintenance, prescribed by the \ac{AMP} at fixed flight-hour, flight-cycle, or calendar-based intervals; and corrective maintenance, arising from operational defect reports such as \acp{PIREP} and \acp{MAREP}.

Formally, \ac{MS} is governed by binary variables $M_{s,p} \in \{0,1\}$, indicating whether maintenance slot $s$ is assigned to aircraft $p$, and $T_{g,s} \in \{0,1\}$, indicating whether maintenance task $g$ is scheduled in slot $s$. A core constraint ensures that each maintenance task must be assigned to exactly one slot:
\begin{align}
\sum_{s \in \mathcal{S}} T_{g,s} = 1 \quad \forall g \in \mathcal{G}
\label{eq:ms_task_assignment}
\end{align}
Additionally, a task can only be scheduled in a slot if that slot is assigned to the aircraft requiring the task:
\begin{align}
\sum_{g \in \mathcal{G}_p} T_{g,s} \leq \mathrm{MaxTasks}_s \cdot M_{s,p} \quad \forall s \in \mathcal{S}, p,
\label{eq:ms_slot_coupling}
\end{align}
where $\mathcal{G}_p$ denotes the set of maintenance tasks applicable to aircraft $p$. This constraint directly couples \ac{MS} with \ac{TA} through the aircraft location and availability variables, embodying the interdependency between the two decision layers.

\subsection{Predictive Maintenance}
\label{subsec:pdm}

\ac{PdM} leverages prognostic models to assess component health and predict \ac{RUL}. Multiple critical components per aircraft require monitoring, each with its own degradation process. However, \ac{RUL} predictions inherently contain uncertainty due to complex degradation mechanisms, operational variability, and measurement noise. Unlike time-based maintenance, \ac{PdM} timing adapts to actual component condition, but deterministic planning based on expected \ac{RUL} values alone produces brittle schedules that incur high costs when components degrade faster than anticipated. \ac{PdM} decisions must ensure that no component exceeds its safe operating threshold during flight operations, triggering maintenance interventions before failure occurs. This prognostic uncertainty must be explicitly modeled to generate robust operational plans that hedge against early component failures, with intervention timing and necessity driven by \ac{RUL} predictions with confidence intervals while constrained by aircraft positioning from \ac{TA} and maintenance slot availability from \ac{MS}.

Formally, the \ac{PdM} constraint enforces that each component's \ac{RUL} does not fall below its safe operating threshold for each component $e$ on each aircraft $p$ throughout the planning horizon:
\begin{align}
\mathrm{RUL}_{e,p,t} \geq 0 \quad \forall e \in \mathcal{E}, p \in \mathcal{P}, t \in \mathcal{T}
\label{eq:cbm_nonneg}
\end{align}
where $\mathrm{RUL}_{e,p,t}$ decreases with flight assignments $F_{p,f}$ and increases when prognostic maintenance tasks are executed (scheduled through $M_{s,p}$ and $T_{g,s}$). This coupling embeds prognostic uncertainty directly into the joint optimisation, ensuring that maintenance decisions are informed by component health predictions.

\subsection{Integrated Optimization Problem}
\label{subsec:integrated}
The integrated problem jointly optimises \ac{TA}, \ac{MS}, and \ac{PdM} decisions under prognostic uncertainty. The objective balances multiple cost components: flight assignment costs, maintenance execution costs, and the downstream costs of reactive maintenance and flight cancellations when uncertainty is realised in operations. The integration exploits the natural interdependencies between these decisions: \ac{TA} determines aircraft positioning and availability, \ac{MS} determines when aircraft are unavailable for flights, and \ac{PdM} constraints based on \ac{RUL} predictions with confidence intervals drive the timing and necessity of maintenance interventions. By optimizing these decisions jointly rather than sequentially, the framework captures operational trade-offs and generates coordinated plans that improve reliability and reduce total system costs under uncertainty. 

\subsection{Current Industry Practice}
\label{subsec:current_practice}
At Swiss International Air Lines, as at most major carriers, \ac{TA} and line maintenance planning are managed by separate functions on distinct time horizons. \ac{TA} is initially planned for the full month ahead and continuously refined as operational needs evolve. Line maintenance workpackages are prepared over a horizon of several days and finalised on the day of operation, a few hours before the aircraft arrives for its overnight ground time. This structure reflects deliberate organisational design: regulatory requirements under the \ac{AMP} prescribe fixed flight-hour, flight-cycle, and calendar-based intervals, and separating planning responsibilities ensures compliance, resource predictability, and clear accountability.

The inherent separation between planning layers means that component health information does not systematically feed into \ac{TA} decisions. Where \ac{RUL} predictions with confidence intervals are considered, a common heuristic is to plan against a conservative percentile of the distribution. While this provides a safety buffer, it does so at significant cost: interventions are scheduled far earlier than necessary, leading to excessive preventive maintenance and inefficient use of maintenance slots and technician capacity. As a result, the natural interdependencies between \ac{TA}, \ac{MS}, and component health are not jointly exploited, which motivates the integrated framework proposed in this paper.

%% file: 03_related_work.tex
\section{Related Work}
\label{sec:related_work}

This section reviews the state of the art relevant to the joint stochastic optimisation of \ac{MS}, \ac{TA}, and \ac{CBM} under prognostic uncertainty. We organize the literature into four core themes that progressively narrow toward the research gaps this work addresses.

\subsection{Integrated Maintenance Scheduling and Tail Assignment}
\label{subsec:rw_ms_ta}

Efficient airline operations require balancing downtime through \ac{MS} and uptime through \ac{TA}. Early foundational work on fleet assignment, such as Hane et al.~\cite{Hane1995TheProgram}, established the network-flow
formulations that underpin modern aircraft routing and assignment. Barnhart et al.~\cite{Barnhart1998Flight1} extended these formulations to flight string models that explicitly capture maintenance routing constraints. Sriram and
Haghani~\cite{SriramAnRe-assignment} provided one of the first integrated \ac{MS}--\ac{TA} formulations using \ac{IP}, demonstrating cost savings over sequential optimisation by enabling flexible reassignment of flight segments around maintenance events.

Varenna et al.~\cite{Varenna2025StochasticOperations} introduced the ANEMOS model, integrating network planning and \ac{MS} with modular \ac{DES}. The model evaluates the impact of maintenance resource allocation, such as reserve aircraft, on flight cancellations and disruption costs in long-haul operations. Similarly, Iwata and Mavris~\cite{Iwata2013Object-orientedProcess} applied \ac{DES} to joint \ac{MS}--\ac{TA} for military fighter fleets, highlighting the role of maintenance capacity in fleet-wide readiness. In high-disruption commercial contexts, van Kessel et al.~\cite{vanKessel2023AirlineEnvironment} combined \ac{MILP} with dynamic rescheduling, minimizing ground time and schedule instability through continuous real-time updates.

For long-term \ac{HMC} scheduling, Deng et al.~\cite{Deng2020AOptimization} introduced a lookahead \ac{ADP} approach achieving a reduction in check frequency. Their stochastic extension~\cite{Deng2022LookaheadOptimization} incorporated uncertain daily utilization and check elapsed times, further reducing check counts via stochastic forecasting. Weide et al.~\cite{Weide2022RobustUncertainty} addressed \ac{HMC} robustness via a genetic algorithm generating multi-year schedules resilient to duration and utilization variability, reporting utilization gains and check reduction. Witteman et al.~\cite{Witteman2021AProblem} studied task allocation within \ac{HMC} windows as a time-constrained variable-sized bin packing problem, achieving near-optimal solutions.

More recently, fleet-level decision-making has begun to incorporate predictive signals alongside conventional scheduling. Crespo del Castillo and Parlikad~\cite{CrespodelCastillo2024DynamicCost} developed an optimisation model that integrates predictive and preventive maintenance planning with operational workload balance, showing that \ac{RUL} distribution precision significantly affects total fleet cost. Zhang et al.~\cite{Zhang2026DistributedFleets} proposed a hierarchical multi-agent \ac{DRL} framework for dynamic \ac{MS} of airline fleets, tested on up to 200 aircraft across multiple airports. Yu et al.~\cite{Yu2026AircraftErrors} modeled execution uncertainty via human-error effects in aircraft \ac{MS} on real flight schedules, finding a 16\% cost impact. Despite these advances, maintenance timing across all these frameworks remains governed by fixed calendar, flight-hour, or flight-cycle triggers rather than prognostic component health, leaving the interdependency between \ac{RUL} predictions with confidence intervals and fleet-level assignment decisions unexploited.


\subsection{Condition-Based and Predictive Maintenance in Aviation}
\label{subsec:rw_cbm}

Aviation is transitioning from time-based preventive maintenance toward \ac{CBM} strategies that exploit real-time health monitoring and prognostic models. \ac{CBM} encompasses diagnostics (fault detection) and prognostics (predicting \ac{RUL}); its prognostic variant, \ac{PdM}, specifically uses \ac{RUL} forecasts to schedule interventions proactively before failure occurs. Reviews by Si et al.~\cite{Si2011RemainingApproaches} on \ac{RUL} estimation and by Khan and Yairi~\cite{Khan2018AManagement} provide a comprehensive overview of methodologies for \ac{RUL} prediction and data-driven system health management, establishing a broad methodological landscape.

However, despite these advances, the adoption of \ac{CBM} in aviation remains limited by practical challenges, including data availability, integration with existing \ac{MRO} processes, and certification constraints, as highlighted by Verhagen et al.~\cite{Verhagen2023Condition-BasedOpportunities}. These barriers motivate the need for frameworks that can integrate prognostic information more seamlessly into operational decision-making.

The effectiveness of such integration critically depends on the accurate quantification and propagation of prognostic uncertainty into maintenance and operational planning. Pecht and Jaai~\cite{Pecht2010ASystems} and Sankararaman and Goebel~\cite{SankararamanWhyUncertain} emphasize that propagating prognostic uncertainty is essential to avoid unsafe or overly conservative maintenance actions. Arias Chao et al.~\cite{AriasChao2022FusingPrognostics} demonstrated fusion of physics-based and \ac{ML} models for turbofan \ac{RUL} estimation, showing how hybrid approaches can improve prediction fidelity. Zhuang et al.~\cite{Zhuang2023ALearning} further showed that Bayesian deep learning can generate calibrated \ac{RUL} distributions that propagate uncertainty directly into spare-parts and maintenance decisions under operational constraints. However, accurate \ac{RUL} estimates alone are insufficient; uncertainty must be explicitly integrated into the scheduling of both maintenance tasks and aircraft assignments to generate robust operational plans.

Recent research has begun to explore this integration. De Pater and Mitici~\cite{dePater2021PredictiveComponents} proposed an \ac{ILP} formulation that integrates \ac{PdM} planning for multi-component aircraft systems, demonstrating substantial cost reductions compared to corrective and preventive maintenance baselines. Their follow-up work~\cite{dePater2022Alarm-basedPrognostics} extended this to address alarm-based schedules where \ac{RUL} predictions are subject to false and missed alarms. Mitici et al.~\cite{Mitici2023DynamicEngines} generalized this line of work to multiple components simultaneously, integrating probabilistic \ac{CNN}-based \ac{RUL} estimates with dynamic planning and reporting a 53\% cost reduction over time-based strategies. Lee and Mitici~\cite{Lee2023DeepPrognostics} combined probabilistic \ac{RUL} estimation via \ac{CNN} with \ac{DRL} for adaptive maintenance planning. Complementary approaches employ stochastic and learning-based scheduling frameworks: Tseremoglou and Santos~\cite{Tseremoglou2024Condition-BasedApproach} introduced a two-stage dynamic scheduling framework where the first stage optimises maintenance dates via \ac{POMCP}, and the second uses \ac{DQN} for real-time task scheduling across a fleet. Their comparative study~\cite{Tseremoglou2023AFleet} evaluated \ac{MILP} versus \ac{DRL} for \ac{CBM} scheduling. Zeng and Liang~\cite{Zeng2023ASystems} similarly combined probabilistic \ac{GRU}-based \ac{RUL} forecasting with a rolling-horizon integer program for fleet maintenance, validated on aircraft data. Li et al.~\cite{Li2024PredictionLearning} applied \ac{ML} predictors to forecast non-routine maintenance workloads, while Bieber et al.~\cite{Bieber2021Data-DrivenMaintenance} demonstrated that incorporating environmental factors improves data-driven prognostics. Duan et al.~\cite{Duan2025UncertainSystem} integrated uncertain health monitoring with cost-threshold optimisation for aircraft air conditioning systems, and Guo et al.~\cite{Guo2025ProbabilisticMaintenance} assessed probabilistic failure risk of civil aircraft under \ac{CBM}, linking inspection interval design directly to on-time flight performance. Zhu et al.~\cite{Zhu2026PredictiveCost} demonstrated that probabilistic \ac{RUL} predictions can be combined with cost-risk optimisation to identify optimal maintenance timing under uncertainty. Collectively, these works establish that embedding probabilistic \ac{RUL} predictions into \ac{MS} yields substantial cost reductions over time-based and corrective strategies. However, in all cases, maintenance decisions are optimised in isolation from flight assignment, and the operational coupling between component health and fleet-level routing remains unaddressed.

\subsection{Stochastic Optimization and Uncertainty Modeling}
\label{subsec:rw_stochastic}

\ac{SP} provides a principled framework for optimisation under uncertainty, enabling decisions that hedge against variability by explicitly modelling probabilistic scenarios. Birge and Louveaux~\cite{Birge2011IntroductionProgramming} provide a comprehensive treatment of decomposition and scenario-based methods, while Shapiro et al.~\cite{Shapiro2021LecturesTheory} develop the theory of \ac{SAA}, including statistical guarantees. \ac{2SP} has been successfully applied to airline operations: Lan et al.~\cite{Lan2006PlanningDisruptions} considered aircraft routing under disruption uncertainty, and Lagos et al.~\cite{Lagos2019DynamicOperations} and Deng and Santos~\cite{Deng2022LookaheadOptimization} demonstrated robustness improvements in \ac{MS} through \ac{ADP} under stochastic utilization and degradation parameters.

However, computational tractability becomes challenging as the number of scenarios grows, particularly when considering variable-sized problem instances with complex interdependencies between decisions. This motivates the use of approximation methods and learning-based surrogates. Dumouchelle et al.~\cite{Dumouchelle2022Neur2SP:Programming} introduced Neur2SP, a \ac{NN}-enhanced \ac{SP} approach where a \ac{ReLU} \ac{NN} approximates the expected second-stage value function. After offline training, this linear network can be embedded directly into a \ac{MILP} for efficient online optimisation. This approach is particularly relevant for our problem, as disruption costs depend on complex interactions between \ac{TA}, \ac{MS}, and probabilistic \ac{RUL} trajectories-interactions that are difficult to capture through explicit scenario enumeration.

In contrast to \ac{SP}'s reliance on probabilistic information, alternative paradigms address uncertainty through fundamentally different formulations. Robust optimisation~\cite{Ben-TalLaurentElGhaouiArkadiNemirovski2021RobustOptimization, Bertsimas2010TheoryOptimization} optimises against worst-case deviations within pre-specified uncertainty sets, guaranteeing feasibility regardless of realized outcomes but potentially at significant cost. \ac{DRO} ~\cite{DelageDistributionallyProblems, Rahimian2019DistributionallyReview} extends this by hedging against uncertainty in the probability distribution itself, optimizing over an ambiguity set of plausible distributions. While these approaches provide strong worst-case guarantees, they typically yield conservative solutions that do not directly leverage the probabilistic information available from prognostic models. In contrast, \ac{SP} exploits the explicit probability distribution of \ac{RUL} predictions to generate less conservative, cost-effective plans.

Building on these foundations, the present work adopts a two-stage \ac{SP} perspective combined with \ac{NN} approximation of the expected second-stage cost. Unlike typical \ac{SP} applications that assume fixed problem sizes, our formulation must accommodate variable-size \ac{MILP} instances across different fleet compositions and planning horizons. This necessitates tailored feature engineering to bridge the variable-dimensional optimisation space with the fixed-size \ac{NN} representations required for efficient online optimisation.




\subsection{Research Gaps and Positioning}
\label{subsec:rw_gaps}

Table~\ref{tab:literature_positioning} summarizes key related works across the four thematic dimensions, highlighting scope, uncertainty treatment, and methodological approach.

\begin{table}[h]
\centering
\caption{Positioning of related work on integrated maintenance and operations planning.}
\label{tab:literature_positioning}
\resizebox{\textwidth}{!}{%
\begin{tabular}{lccccl}
\toprule
\textbf{Reference} & \textbf{\ac{MS}} & \textbf{\ac{TA}} & \textbf{\ac{CBM}/\ac{RUL}} & \textbf{Uncertainty} & \textbf{Method} \\ \midrule
Sriram \& Haghani~\cite{SriramAnRe-assignment}      
    & \checkmark &            &            & None        & Heuristic \\ 
Lagos et al.~\cite{Lagos2019DynamicOperations}      
    & \checkmark & \checkmark &            & Task        & \ac{ADP} \\ 
Varenna et al.~\cite{Varenna2025StochasticOperations}            
    & \checkmark & \checkmark &            & None        & \ac{DES} \\ 
Vink et al.~\cite{Vink2020DynamicFramework}         
    & (\checkmark) & \checkmark &            & None        & \ac{MILP}+Heuristic \\ 
Deng \& Santos~\cite{Deng2022LookaheadOptimization}
    & \checkmark &            &            & Utilization \& Checks  & \ac{ADP} \\
Zhang et al.~\cite{Zhang2026DistributedFleets}
    & \checkmark & \checkmark &            & None                   & \ac{DRL} \\
De Pater \& Mitici~\cite{dePater2021PredictiveComponents}   
    & \checkmark &            & \checkmark & \ac{RUL}  & \ac{ILP} \\ 
Lee \& Mitici~\cite{Lee2023DeepPrognostics}
    & \checkmark &            & \checkmark & \ac{RUL}  & \ac{CNN}+\ac{DRL} \\
Mitici et al.~\cite{Mitici2023DynamicEngines}
    & \checkmark &            & \checkmark & \ac{RUL}  & \ac{CNN}+Dynamic \\
Tseremoglou \& Santos~\cite{Tseremoglou2024Condition-BasedApproach}
    & \checkmark &            & \checkmark & \ac{RUL} & \ac{POMCP}+\ac{DQN} \\
Crespo del Castillo \& Parlikad~\cite{CrespodelCastillo2024DynamicCost}
    & \checkmark & (\checkmark) & (\checkmark) & \ac{RUL} dist. & \ac{MILP} \\
\midrule
\textbf{This work}                                   & \checkmark & \checkmark & \checkmark & \ac{RUL}  & \ac{MILP} + \ac{NN} \\ \bottomrule
\end{tabular}%
}
\end{table}

The survey reveals four critical gaps addressed by this work:

\paragraph{Gap 1: No integrated \ac{MS}--\ac{TA}--\ac{PdM} framework under prognostic uncertainty}
Existing \ac{MS}--\ac{TA} approaches typically treat maintenance tasks as deterministic and do not incorporate prognostic \ac{RUL}. Conversely, \ac{PdM} scheduling works do not consider \ac{TA}. There is no framework that jointly optimises \ac{MS}, \ac{TA}, and \ac{PdM} task scheduling while explicitly modelling \ac{RUL} with confidence intervals.

\paragraph{Gap 2: Limited coupling of \ac{RUL} uncertainty with flight network feasibility}
Works that model \ac{RUL} uncertainty, such as De Pater and Mitici~\cite{dePater2021PredictiveComponents,dePater2022Alarm-basedPrognostics} and Tseremoglou and Santos~\cite{Tseremoglou2024Condition-BasedApproach}, focus on \ac{CBM} planning and control without embedding those decisions into an aircraft-to-flight assignment network.

\paragraph{Gap 3: Limited validation on real-world airline planning data at the integrated level}
Many \ac{CBM} and stochastic scheduling studies rely on simulated datasets. Validation against real operational data with flight plans, maintenance programs, and airline constraints remains less common, especially for integrated \ac{MS}--\ac{TA}. This work provides validation using operational data from a European short-haul airline.

\paragraph{Gap 4: Limited quantification of the value of prognostic information in integrated planning}

The marginal benefit of incorporating \ac{PdM} health information, compared to purely deterministic maintenance, is rarely quantified in an integrated \ac{MS}--\ac{TA} framework. This paper addresses this gap through comparison of deterministic and stochastic planning variants, alongside evaluation against a real-world baseline.\\

This work addresses these gaps by developing an integrated \ac{MILP} formulation coupling \ac{MS}, \ac{TA}, and \ac{PdM} decisions under \ac{RUL} with confidence intervals, representing uncertainty via scenario sampling and neural approximation of expected costs, validated on real airline data.

%% file: 04_methodology.tex
\section{Methodology}
\label{sec:methodology}

A \ac{S-EX} formulation is proposed that optimises all planning and maintenance decisions simultaneously while explicitly accounting for prognostic uncertainty through an embedded \ac{NN} cost approximator. The key innovation is a stochastic approach that directly embeds learned \ac{NN} approximations of evaluation-stage costs into the planning-stage optimisation, enabling tractable solution of large-scale stochastic programs without explicit scenario enumeration.

\subsection{Why Neural Network Approximation?} \label{sec:why_NN}

The fundamental challenge is that uncertainty in \ac{RUL} predictions
couples planning and evaluation stages: decisions made today affect both immediate planning
costs and future robustness under \ac{RUL} with confidence intervals realisations. A naive stochastic approach
would explicitly evaluate all sampled scenarios at solve time, prohibitively expensive for
practical problem sizes. Our solution: instead of enumerating scenarios, we pre-train a \ac{NN}
to learn a compact approximation of expected evaluation-stage costs
(e.g. cancellations, reactive maintenance, ferry flights) as a function of the initial planning decisions
 and scenario characteristics (i.e. sampled \ac{RUL} realisations for each monitored component). 
This learned approximation is then embedded directly into the planning-stage \ac{MILP} via \ac{ReLU} linearisation, transforming the intractable
stochastic program into a deterministic \ac{MILP} that a standard solver can tackle efficiently.
The result is a plan that optimizes both immediate operational costs and expected future robustness
across sampled uncertainty realisations, without requiring expensive second-stage problem solves
during optimisation. 

\begin{figure}[h]
    \centering
    \includegraphics[width=\linewidth]{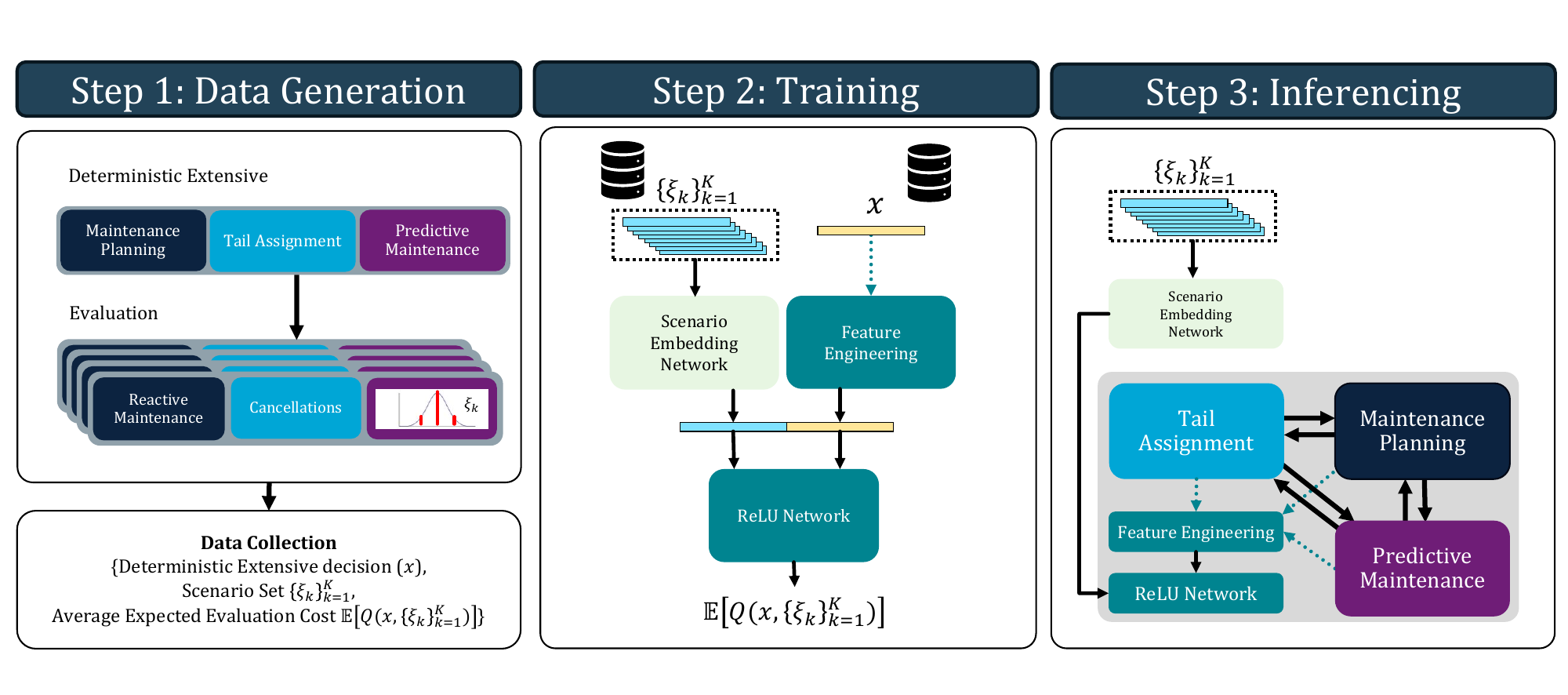}
    \caption{Three-phase stochastic optimisation framework: (1) data generation via the deterministic \ac{EX} model, (2) offline \ac{NN} training, and (3) deployment of the linearized \ac{NN} embedded in the \ac{S-EX} \ac{MILP}.}
    \label{fig:ex_nn_architecture}
\end{figure}

Figure~\ref{fig:ex_nn_architecture} illustrates the full three-phase framework.
In Step 1 (Data Generation), a deterministic \ac{EX} model is solved to generate
plans $\mathbf{x}$, which are then evaluated across sampled \ac{RUL} scenario sets
$\{\xi_k\}_{k=1}^{K}$ using a heuristic simulator. The resulting
$(\mathbf{x},\,\{\xi_k\},\,\bar{Q})$ triplets form the training dataset.
In Step 2 (Training), a \ac{NN} is trained offline on this dataset:
a Scenario Embedding Network encodes the scenario set, Feature Engineering compresses
the plan decisions, and a \ac{ReLU} predictor learns to approximate the expected
evaluation-stage cost $\mathbb{E}[Q(x,\{\xi_k\}_{k=1}^{K})]$.
In Step 3 (Inferencing), the trained \ac{ReLU} network and Feature Engineering
are linearized and embedded directly into the planning-stage \ac{MILP}, while the
Scenario Embedding Network is evaluated outside the solver on the current scenario set.
\ac{TA}, \ac{MS}, and \ac{PdM} are then optimized
jointly with the embedded cost approximator in a single \ac{MILP} solve,
yielding a plan that hedges against prognostic uncertainty.

\subsection{Problem Formulation}
\label{sec:problem_formulation}

This subsection introduces the mathematical formulation of the integrated optimisation problem.
We begin by defining the index sets, parameters, and decision variables, then present the
objective function and the three constraint groups that define the feasible region.

\subsubsection{Sets, Parameters, and Decision Variables}

Table~\ref{tab:set_form} provides an overview of the fundamental sets employed throughout the
model, while Table~\ref{tab:dec_var} summarizes the binary decision variables that capture
assignment, scheduling, and location decisions.

\begin{table}[h]
\centering
\small
\caption{Index sets used in the optimisation framework.}
\label{tab:set_form}
\begin{tabular}{cl}
\toprule
\textbf{Set} & \textbf{Definition} \\
\midrule
$\mathcal{P}$ & Fleet of aircraft \\
$\mathcal{F}$ & Scheduled flights \\
$\mathcal{G}$ & Maintenance tasks; $\mathcal{G}_{\text{prog}}$ (prognostic), $\mathcal{G}_{\text{prev}}$ (preventive) \\
$\mathcal{S}$ & Maintenance slots ($\mathcal{S}_{\text{real}}$: real slots; $\mathcal{S}_{\text{fict}}$: fictional/deferral slot) \\
$\mathcal{N}$ & Airports or ground locations \\
$\mathcal{T}$ & Discretized time indices ($\{0,1,\ldots,T_{\max}\}$) \\
$\mathcal{E}$ & Monitored components (predictors) \\
\bottomrule
\end{tabular}
\end{table}

\begin{table}[h]
\centering
\small
\caption{Binary decision variables in the optimisation model.}
\label{tab:dec_var}
\begin{tabular}{cll}
\toprule
\textbf{Variable} & \textbf{Domain} & \textbf{Description} \\
\midrule
$F_{p,f}$     & $\{0,1\}$ & Assignment: Aircraft $p$ is assigned to flight $f$ \\
$C_f$         & $\{0,1\}$ & Cancellation: Flight $f$ is cancelled \\
$G_{p,n,t}$   & $\{0,1\}$ & Location: Aircraft $p$ is present at location $n$ at time $t$ \\
$T_{g,s}$     & $\{0,1\}$ & \ac{MS}: Maintenance task $g$ is assigned to slot $s$ \\
$M_{s,p}$     & $\{0,1\}$ & Slot Assignment: Maintenance slot $s$ is assigned to aircraft $p$ \\
\bottomrule
\end{tabular}
\end{table}

\subsubsection{Objective Function}

The central objective of the integrated optimisation model is to minimise the expected total
operational cost, incorporating both flight-related and maintenance-related expenditures. This
cost function comprehensively accounts for direct operating expenses associated with fleet
assignment, penalties for flight cancellations, the allocation and execution of maintenance
tasks, assignment of maintenance slots, and penalties incurred due to unused maintenance
intervals.

Mathematically, the objective function of the extensive form is expressed as:

\begin{align}
\min \; Z = \;
  & \underbrace{\sum_{p,f} w_{\text{flight},p,f} F_{p,f}}_{\text{Flight operating costs}}
  + \underbrace{\sum_{f} w_{\text{cancel},f} C_{f}}_{\text{Cancellation costs}}
  + \underbrace{\sum_{g \in \mathcal{G}_{\text{prog}},\, s \in \mathcal{S}_{\text{real}}} w_{\text{prog}}\, T_{g,s}}_{\text{Prognostic maintenance task costs}} \nonumber\\
  & + \underbrace{\sum_{s \in \mathcal{S}_{\text{real}},\, p} w_{\text{slot}}\, M_{s,p}}_{\text{Slot assignment costs}}
  + \underbrace{\sum_{g \in \mathcal{G}_{\text{prev}},\, s \in \mathcal{S}_{\text{real}}} w_{\text{unused}}\,\Delta t_{g,s}\, T_{g,s}}_{\text{Penalty for unused maintenance interval}}
  \label{eq:obj_extensive}
\end{align}

where $w_{\text{flight},p,f}$ denotes the cost incurred for assigning aircraft $p$ to flight
$f$, $w_{\text{cancel},f}$ represents the penalty associated with the cancellation of flight
$f$, $w_{\text{prog}}$ and $w_{\text{slot}}$ quantify the costs of executing prognostic
maintenance tasks and the allocation of maintenance slots, respectively, and $w_{\text{unused}}$
reflects the penalty weight for unused maintenance intervals. The term
$\Delta t_{g,s} = \max(0,\,\mathrm{due}_g - \mathrm{start}_s - h_{\text{free}})$ calculates
the temporal gap between the due date of maintenance task $g$ and the scheduled start of slot
$s$, adjusted by the free horizon $h_{\text{free}}$. This formulation strategically balances
the competing goals of operational efficiency, schedule robustness, and optimal maintenance
resource allocation.

\subsubsection{Constraints}

The feasible region, denoted as $\chi$, is defined by a comprehensive set of constraints that
ensure the operational and maintenance integrity of the combined scheduling framework. These
constraints are categorized into three principal groups: \ac{TA}, \ac{MS}, and \ac{PdM}.

Formally, the feasible set $\chi$ consists of all assignments of the decision variables that
satisfy the following constraints:

\textbf{Tail Assignment Constraints}\\
This group governs the assignment of aircraft to scheduled flights and tracks their locations
throughout the planning horizon. The constraints guarantee that each flight is either assigned
to a single aircraft or cancelled, enforce flow conservation for aircraft among airports and
discrete time periods, mandate minimum turnaround times, and exclude restricted airports from
the scheduling solution.

\begin{align}
\sum_{p} F_{p,f} + C_f &= 1
  & \forall f \in \mathcal{F}
  \tag{TA.1}\\[4pt]
\sum_{f \in \mathcal{F}_{\text{arr}}(p,n,t)} F_{p,f}
- \sum_{f \in \mathcal{F}_{\text{dep}}(p,n,t)} F_{p,f}
+ G_{p,n,t} - G_{p,n,t-1} &= 0
  & \forall p,n,\, t \in \mathcal{T} \setminus \{0\}
  \tag{TA.2}\\[4pt]
F_{p,f}=1 \;\Rightarrow\; \sum_{t=t_{\text{arr}}}^{t_{\text{arr}}+\mathrm{TAT}_p-1} G_{p,n_{\text{arr}},t} &= \mathrm{TAT}_p
  & \forall p,f
  \tag{TA.3}\\[4pt]
\sum_{t} G_{p,n,t} &= 0
  & \forall p,\, n \in \mathcal{N}_{\text{forbidden}}(p)
  \tag{TA.4}
\end{align}

Constraint TA.1 enforces exclusive assignment or cancellation for each flight. TA.2 ensures
aircraft flow conservation at airports over time. TA.3 imposes minimum turnaround requirements.
TA.4 restricts aircraft from operating at forbidden airports.

\textbf{Maintenance Scheduling Constraints}\\
This group ensures proper scheduling and capacity management for maintenance events. Preventive
maintenance tasks must be scheduled exactly once, slot-level workforce and task limits are
respected, slot assignments remain unique and non-overlapping, and maintenance is only performed
when the aircraft is physically present.

\begin{align}
\sum_{s \in \mathcal{S}} T_{g,s} &= 1
  & \forall g \in \mathcal{G}_{\text{prev}}
  \tag{MS.1}\\[4pt]
\sum_{g \in \mathcal{G}_p} T_{g,s} &\leq \mathrm{MaxTasks}_s \cdot M_{s,p}
  & \forall s \in \mathcal{S}_{\text{real}},\, p
  \tag{MS.2}\\[4pt]
\sum_{g \in \mathcal{G}_p} \mathrm{Duration}_g\, T_{g,s} &\leq \mathrm{MaxTime}_s \cdot M_{s,p}
  & \forall s \in \mathcal{S}_{\text{real}},\, p
  \tag{MS.3}\\[4pt]
\sum_{p} M_{s,p} &\leq 1
  & \forall s \in \mathcal{S}_{\text{real}}
  \tag{MS.4}
\end{align}
\begin{align}
M_{s,p}=1 \;\Rightarrow\; \sum_{s' \in \mathcal{S}_{\text{overlap}}(s)} M_{s',p} &= 0
  & \forall s \in \mathcal{S}_{\text{real}},\, p
  \tag{MS.5}\\[4pt]
M_{s,p} &\leq G_{p,n_s,t_s}
  & \forall s \in \mathcal{S}_{\text{real}},\, p
  \tag{MS.6}
\end{align}

Constraint MS.1 ensures each preventive maintenance task is scheduled once. MS.2 and MS.3
enforce slot-level limits on the number of tasks and time allocation. MS.4 upholds uniqueness
of slot assignment. MS.5 prevents overlapping maintenance slots per aircraft. MS.6 requires
aircraft presence for maintenance. Because all five decision variable types ($F$, $C$, $G$, $T$,
$M$) appear simultaneously in this constraint set, the extensive form captures the full
coupling between fleet positioning and maintenance assignment---a coupling that sequential
decompositions relax.

\textbf{Predictive Maintenance Constraints}\\
The third group preserves the non-negativity of \ac{RUL} for all
monitored components, reflecting degradation with flight operations and recovery upon execution
of relevant prognostic maintenance tasks.

\begin{align}
\mathrm{RUL}_{e,p,0}
- \!\!\sum_{\substack{f \in \mathcal{F}:\\ \mathrm{onblock}(f)\leq t}}\!\! F_{p,f}
+ \!\!\sum_{\substack{g \in \mathcal{G}_{\text{prog}}:\\ \mathrm{pred}(g)=e}}\!\! \mathrm{RUL\_recover}_g \sum_{s \in \mathcal{S}} T_{g,s}
&\geq 0
  \quad \forall e,p,t
  \tag{PdM.1}
\end{align}

Constraint PdM.1 guarantees that for every predictor $e$ and aircraft $p$, the
\ac{RUL} is never negative, decreasing through flight activity and increasing through
prognostic maintenance interventions. Collectively, these constraints define the feasible
region $\chi$ that encapsulates all valid operational and maintenance schedules considered by
the model.\\ 

\begin{figure}[h]
    \centering
    \includegraphics[width=0.6\linewidth]{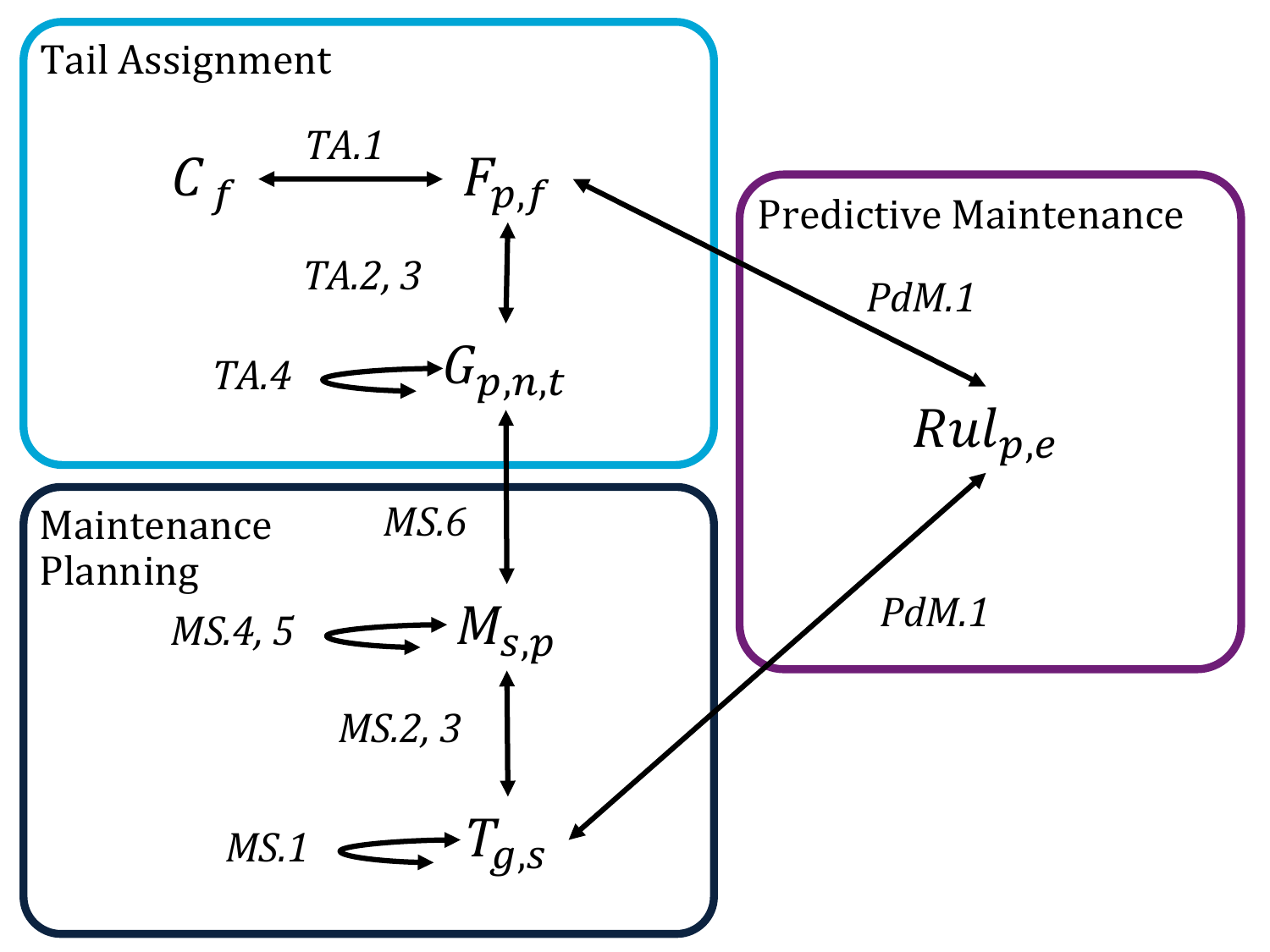}
    \caption{Variable and constraint structure of the \ac{EX} model: all decision
    variables ($F$, $C$, $G$, $T$, $M$) are optimized jointly subject to \ac{TA}.1--\ac{TA}.4,
    \ac{MS}.1--\ac{MS}.6, and \ac{PdM}.1.}
    \label{fig:ex_var_const}
\end{figure}

Figure~\ref{fig:ex_var_const} depicts the resulting variable--constraint structure of the extensive model.

\subsection{Stochastic Approach with Neural Network Approximation}
\label{subsec:stochastic_nne}

The key innovation of our approach is the integration of \ac{RUL} uncertainty directly into
the \textbf{joint} extensive optimisation through \ac{NN} approximation of evaluation-stage
costs. Rather than explicitly enumerating all scenarios during planning, a pre-trained \ac{NN}
learns to predict expected realisation costs (cancellations, reactive maintenance,
disruption) as a function of planning decisions and scenario information. This learned cost
approximation is then embedded directly into the extensive planning \ac{MILP} via \ac{ReLU}
linearisation, enabling efficient joint optimisation of all five decision variable types without
expensive second-stage problem solves.

\textbf{Mathematical Formulation}\\
Because \ac{RUL} predictions carry uncertainty, the stochastic variant augments the planning-stage
cost with a \ac{NN} estimate of expected realisation costs over a scenario set
$\Xi = \{\xi_1,\ldots,\xi_K\}$:
\begin{align}
\min_{\mathbf{x} \in \chi} \; Z_{\text{stoch}}
  = c^\top \mathbf{x} + \lambda \cdot \mathrm{NN}\!\left(\mathcal{F}(\mathbf{x}),\,\Xi\right),
\label{eq:stochastic_nne}
\end{align}

Here $\mathcal{F}(\mathbf{x})$ is a fixed-size feature vector derived from the plan (see
Section~\ref{sec:feature_eng}), $\mathrm{NN}(\cdot)$ approximates the expected realisation
cost $\bar{Q}(\mathbf{x}) = \frac{1}{K}\sum_k Q_k(\mathbf{x})$, and $\lambda$ balances
planning-stage costs with evaluation-stage robustness. The \ac{NN} uses only \ac{ReLU}
activations, enabling exact linearisation and direct embedding into the \ac{MILP}
solver for joint optimisation. Crucially, since $\mathbf{x}$ here collects all five
variable types simultaneously, the \ac{NN} approximation guides both \ac{TA} and maintenance
decisions jointly in a single solve, unlike sequential approaches that can only guide the
respective stage-level subproblem.

\begin{figure}[h]
    \centering
    \includegraphics[width=0.6\linewidth]{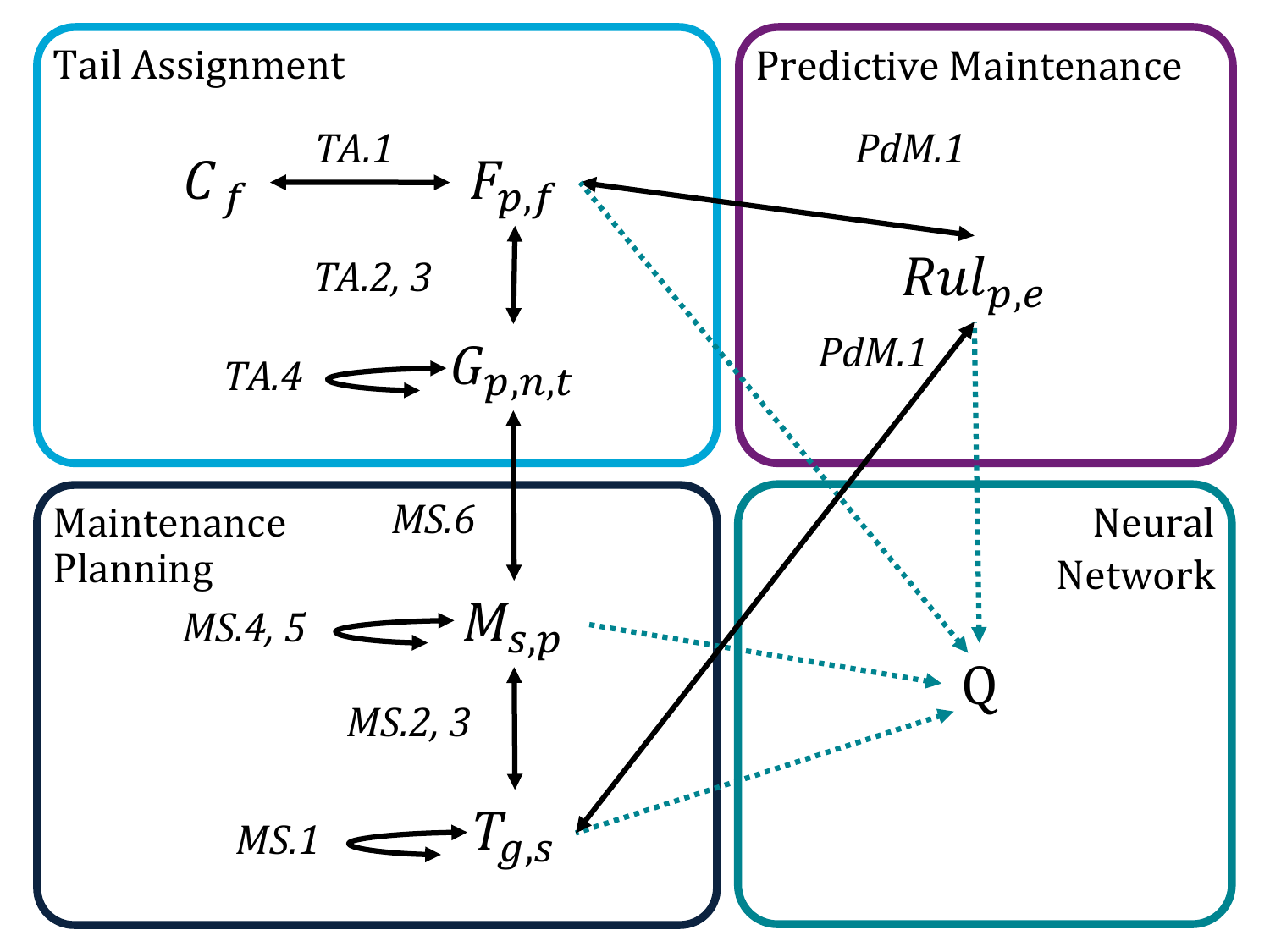}
    \caption{Variable and constraint structure of the \ac{S-EX} model, showing how decision variables from all planning blocks are extracted into a feature vector and passed to the embedded \ac{NN} approximating the expected evaluation-stage cost $Q$.}
    \label{fig:ex_var_const_stoch}
\end{figure}

Figure~\ref{fig:ex_var_const_stoch} shows the resulting augmented variable--constraint structure.

\begin{figure}[h]
    \centering
    \includegraphics[width=0.8\linewidth]{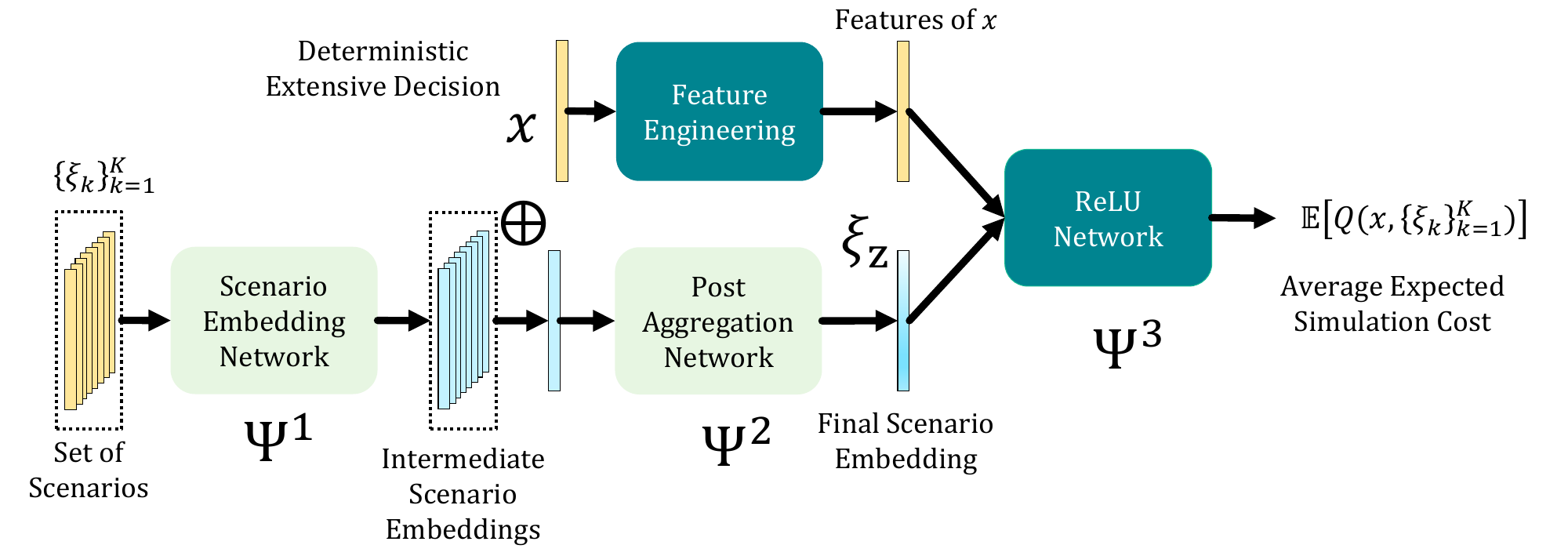}
    \caption{\ac{NN} architecture of the trained network. Only the Feature Engineering and the
    \ac{ReLU} Network are embedded into the \ac{MILP}; the Scenario Encoder does not need to be linearized and will be deployed outside of the \ac{MILP}.}
    \label{fig:nn_architecture}
\end{figure}

\subsection{Neural Network: Training and Features}
\label{sec:feature_eng}

The \ac{NN} is trained offline to approximate expected realisation costs
$\bar{Q}(\mathbf{x})$ under scenario uncertainty. The training procedure uses a two-phase
approach. In the first phase, for each training instance the deterministic plan
$\mathbf{x}_{\text{plan}}$ is obtained by solving:
\begin{align}
\min_{\mathbf{x} \in \chi} \; Z_{\text{det}} = c^\top \mathbf{x}
\label{eq:deterministic}
\end{align}
Then $K$ \ac{RUL} scenarios are sampled, and a heuristic simulator evaluates realisation
costs $Q_k = \mathcal{H}(\mathbf{x}_{\text{plan}},\xi_k)$ for each scenario
$\xi_k \in \Xi$. Two training targets are considered, corresponding to the two
uncertainty-handling strategies evaluated in Section~\ref{subsec:experimental_design}:

\begin{align}
  \bar{Q}^{\text{mean}}
    &= \frac{1}{K}\sum_{k=1}^{K} Q_k(\mathbf{x}_{\text{plan}},\xi_k)
    \label{eq:target_mean}\\[6pt]
  \bar{Q}^{95}
    &= \mathrm{Percentile}_{95}\!\left(\{Q_k(\mathbf{x}_{\text{plan}},\xi_k)\}_{k=1}^{K}\right)
    \label{eq:target_95p}
\end{align}

The mean target $\bar{Q}^{\text{mean}}$ trains the network to predict the expected
realisation cost under the sampled scenario distribution. The 95th-percentile target
$\bar{Q}^{95}$ trains the network to predict a conservative upper bound on realisation cost,
steering the optimiser away from plans that are vulnerable to high-cost tail scenarios. When
embedded in the planning \ac{MILP}, a network trained on $\bar{Q}^{95}$ penalises schedules
that carry high worst-case risk rather than only high expected cost.

In the second phase, the \ac{NN} is trained by minimizing mean squared error
$\mathcal{L} = \mathbb{E}[(\mathrm{NN}(\mathbf{z}) - \bar{Q}^{\bullet})^2]$ over
$K_{\text{train}}$ plan--cost pairs, where $\bar{Q}^{\bullet}$ denotes either target
depending on the chosen configuration.

These two training targets, combined with the two formulation variants
(\ac{EX} and \ac{2SP}, introduced in Section~\ref{subsec:experimental_design}), give rise
to the eight named optimisation configurations used throughout the experimental evaluation:

\begin{itemize}
  \item \textbf{\ac{D-EX}} / \textbf{\ac{D-2SP}}: Deterministic \ac{EX} / \ac{2SP}; no \ac{NN}
        term; plans against expected \ac{RUL} values only (Eq.~\ref{eq:deterministic}).
  \item \textbf{\ac{D-EX}-95} / \textbf{\ac{D-2SP}-95}: Deterministic \ac{EX} / \ac{2SP}; no
        \ac{NN} term; \ac{PdM} constraint PdM.1 is enforced using the 95th-percentile
        \ac{RUL} value as a conservative point estimate rather than the mean.
  \item \textbf{\ac{S-EX}} / \textbf{\ac{S-2SP}}: Stochastic \ac{EX} / \ac{2SP}; \ac{NN} trained
        on the mean target $\bar{Q}^{\text{mean}}$ (Eq.~\ref{eq:target_mean}).
  \item \textbf{\ac{S-EX}-95} / \textbf{\ac{S-2SP}-95}: Stochastic \ac{EX} / \ac{2SP}; \ac{NN}
        trained on the 95th-percentile target $\bar{Q}^{95}$ (Eq.~\ref{eq:target_95p}).
\end{itemize}

The feature engineering framework transforms variable-sized optimisation instances into a
fixed-dimensional feature vector $\mathbf{z} = \mathcal{F}(\mathbf{x}) \in
\mathbb{R}^{n_{\text{features}}}$ suitable for \ac{NN} input. This dimensionality
reduction is necessary because \acp{NN} require constant-size inputs while
optimisation problem instances vary in scale. Because the extensive model optimizes all
five variable types jointly, the feature map $\mathcal{F}$ can draw on the full decision
space, flight assignments, maintenance tasks, slot allocations, and location
variables, simultaneously, providing a richer representation than is available to any
single stage of a decomposed approach. The feature map $\mathcal{F}$ is organized into
groups capturing different aspects of the solution: global aggregates (summary
statistics), component health (prognostic information), daily flight operations (degradation trajectory), and
cancellation pressure (balance between maintenance and operational demand). Complete
mathematical definitions and theoretical justification for this feature design are
provided in Appendix~\ref{app:feature_engineering}; specific operational configuration
(enabled groups, dimensionality, normalization) is detailed in
Section~\ref{sec:experimental_setup}.

The \ac{NN} itself employs a specialized architecture designed for set-valued
scenario inputs. A scenario encoder processes the set of $K = 100$ \ac{RUL} realisations
through layers of increasing abstraction, producing a compact embedding of the uncertainty
characteristics. This embedding is concatenated with the solution feature vector and passed
through a \ac{ReLU} predictor network that estimates the chosen cost target
($\bar{Q}^{\text{mean}}$ or $\bar{Q}^{95}$). All activations use exclusively \ac{ReLU}
functions, enabling exact linearisation and direct embedding into the \ac{MILP} solver
without introducing nonlinearity into the overall formulation. The architecture and
training configuration are fully specified in Section~\ref{sec:experimental_setup}.

%% file: 05_experimental_setup.tex
\section{Experimental Setup}
\label{sec:experimental_setup}

\subsection{Dataset and Operational Context}
\label{sec:dataset_operational}

The framework was implemented and validated using real-world operational data from SWISS
International Air Lines Ltd. The dataset encompasses a fleet of 30 aircraft from the Airbus
A220 family operating scheduled services across Europe, spanning two and a half years of
operational history (January 2023 to June 2025). The data includes complete flight schedules,
maintenance records, and simulated component health monitoring information from five critical
aircraft systems. Operations are distributed across two primary maintenance bases and multiple
hub-and-spoke airports throughout the European network. All data was pre-processed to align
with the problem formulation, ensuring that flight schedules, maintenance tasks, and operational
constraints reflect actual airline operations with realistic complexity and decision-making
trade-offs.

\subsection{Instance Generation and Data Partitioning}

A comprehensive set of 5'000 planning instances was generated to provide statistically robust
performance evaluation across diverse operational conditions. This dataset size was selected
based on a systematic sample efficiency analysis (Appendix~\ref{sec:results_sample_efficiency}),
which shows that meaningful robustness gains emerge already at $N \approx 200$ instances and
that performance stabilises beyond $N \approx 1{'}000$, with only marginal improvements
($< 2\,\%$ in evaluation cost reduction) for larger datasets. Scaling to $N = 10{'}000$
instances was considered but found to be impractical in the present setup: the associated
storage requirements for pre-computed scenario evaluations and solver logs exceeded available
capacity, while offering no commensurate improvement in framework performance over 5'000
instances. The chosen dataset size therefore represents a pragmatic operating point that
ensures statistical reliability while remaining computationally and practically feasible.
Each instance was constructed through stratified random sampling from the SWISS dataset, ensuring representative coverage of
different aircraft utilization patterns, seasonal variations, and maintenance demand profiles.
The problem dimensions remained consistent across all instances: each instance involves 5
aircraft randomly selected from the fleet of 30, a 3-day planning horizon, and 5 monitored
components per aircraft. Aircraft and date ranges were selected deterministically using seeded
random number generation to enable full reproducibility while maintaining statistical independence
across instances.

The 5'000 instances were partitioned into four disjoint sets following a 70/10/10/10 split as summarized in Table~\ref{tab:data_partition}. Training and Validation sets support offline \ac{NN} development and hyperparameter tuning. The Test set then assesses \ac{NN} prediction accuracy in isolation, while the separate Evaluation set is reserved for the final end-to-end assessment of the integrated \ac{MILP}+\ac{NN} framework, ensuring no data leakage between \ac{NN} training and operational performance assessment.

\begin{table}[h]
\centering
\small
\caption{Dataset partition for the experimental pipeline.}
\label{tab:data_partition}
\begin{tabular}{@{}llcp{11cm}@{}}
\toprule
\textbf{Split} & \textbf{Instances} & \textbf{Share} & \textbf{Purpose} \\
\midrule
Training   & 3'500 & 70\% & Offline \ac{NN} training exclusively \\
Validation & 500   & 10\% & Hyperparameter tuning and early stopping; no exposure to test or evaluation data \\
Test       & 500   & 10\% & Independent assessment of \ac{NN} prediction accuracy after training \\
Evaluation & 500   & 10\% & Final end-to-end assessment of the integrated \ac{MILP}+\ac{NN} framework \\
\midrule
\textbf{Total} & \textbf{5'000} & \textbf{100\%} & \\
\bottomrule
\end{tabular}
\end{table}

\subsection{Scenario Sampling and Uncertainty Characterization}
\label{sec:scenario_sampling}

\ac{RUL} uncertainty was characterised through prognostic models that produce component-level
predictions of \ac{RUL} in terms of flight cycles. For each monitored component $e$ 
on aircraft $p$, the prognostic model is assumed to parameterize \ac{RUL} uncertainty through a normal
distribution with mean $\mu_{e,p}$ and standard deviation $\sigma_{e,p}$, noting that alternative distributional assumptions could be adopted without loss of generality:
\begin{align}
\mu_{e,p} &\sim \mathrm{Uniform}(5, 30) \quad \text{flight cycles} \label{eq:rul_mean}\\
\sigma_{e,p} &= 5 \quad \text{flight cycles} \label{eq:rul_std}
\end{align}
This parameterization creates a test scenario where components have limited remaining life with 
meaningful variability, enabling evaluation of the framework's ability to handle multiple 
deteriorating components and coordinate maintenance decisions under uncertainty.

Monte Carlo sampling was employed to characterise this uncertainty. For each planning instance,
a scenario set $\Xi = \{\xi_1, \ldots, \xi_K\}$ with $K = 100$ independent samples was generated,
where each scenario $\xi_k$ contains independent draws:
\begin{align}
\xi_{e,p}^{(k)} \sim \mathcal{N}(\mu_{e,p}, \sigma_{e,p}^2) \quad \forall e,p, \, k = 1, \ldots, K
\label{eq:rul_sampling}
\end{align}
Samples were truncated to the interval $[1, \infty)$ to ensure non-negative \ac{RUL} values. This truncation preserves feasible operational recovery: if a component reaches zero \ac{RUL} at an outstation, the aircraft must be flown to the hub (where all maintenance slots are located) before maintenance can be performed. 

The sample size $K = 100$ balances computational tractability with statistical adequacy. Under the law of large numbers, sample moments (mean and variance) converge to their population counterparts at rate $O(1/\sqrt{K})$; with $K = 100$, the standard error of the sample mean is approximately $1/10$ of the population standard deviation. Empirically, 100 scenarios are sufficient to capture the essential characteristics of the underlying normal distribution while maintaining tractable optimisation solve times. Scenarios were generated using deterministic random seeding per instance to ensure reproducibility while maintaining statistical independence across planning instances.

\subsection{Neural Network Architecture and Training Configuration}
\label{sec:nn_architecture_training}

Section \ref{ref:network_architecture} defines the architecture of the employed \ac{NN}. Section \ref{ref:feature_engineering} describes the feature engineering configuration. Finally, Section \ref{ref:training_procedure} defines the training procedure.

\subsubsection{Network Architecture}\label{ref:network_architecture}

The \ac{NN} architecture combines two complementary components to approximate expected
realisation costs under \ac{RUL} uncertainty. The scenario encoder processes raw \ac{RUL}
uncertainty information through a sequence of increasingly abstract representations, reducing
the dimensionality of scenario inputs (100 scenarios) to a compact embedding. This component
consists of three hidden layers with 512, 128, and 64 units respectively, each employing
\ac{ReLU} activations. The processed scenario information is then concatenated with solution
features $\mathcal{F}(\mathbf{x})$ (described in Section~\ref{sec:feature_config}) and passed
to the \ac{ReLU} predictor network. This second component comprises two hidden layers with 16
and 8 units respectively, each employing \ac{ReLU} activations, producing a scalar cost estimate. The exclusive use of
\ac{ReLU} activations throughout both components is essential for the method: \ac{ReLU}
functions are piecewise linear, enabling exact linearisation and direct embedding into the
\ac{MILP} solver without introducing nonlinearity into the overall formulation. The selection
of the network shape is described in Appendix~\ref{app:nn_conf_selection}.

\subsubsection{Feature Engineering Configuration} \label{ref:feature_engineering}
\label{sec:feature_config}

The \ac{NN} input consists of a fixed-size feature vector
$\mathbf{z} \in \mathbb{R}^{81}$ derived from optimisation solutions. Because all five variable
types are simultaneously available in the extensive formulation, the feature map
$\mathcal{F}(\mathbf{x})$ can draw on the full joint solution to construct each group. This
dimensionality is independent of the specific instance size and allows the network to be usable for all
possible inputs. The feature vector is constructed by aggregating four
complementary groups as detailed in Table~\ref{tab:feature_config_dims}. A comprehensive
catalog of all available feature groups is provided in Appendix~\ref{app:feature_engineering}.

\begin{table}[h]
\centering
\small
\caption{Feature configuration and dimensionality breakdown (experimental setup with $A=5$
aircraft, $D=3$ days, $P=5$ health predictors).}
\label{tab:feature_config_dims}
\begin{tabular}{lcccc}
\toprule
\textbf{Feature Group} & \textbf{Formula} & \textbf{Dimension} & \textbf{Interpretation} \\
\midrule
Solution Summary Statistics & 4 constants & 4 & Total plan-level counts \\
Initial Component Health States & $2 \times A \times P$ & 50 & Health status per aircraft-component \\
Daily Maintenance--Demand Balance & $A \times D$ & 15 & Capacity stress per aircraft-day \\
Daily Fleet-Wide Aggregates & $4 \times D$ & 12 & Daily operational summaries \\
\midrule
& \textbf{Total} & \textbf{81} \\
\bottomrule
\end{tabular}
\end{table}

All 81 features are standardized using zero-mean, unit-variance normalization:
\begin{align}
\mathbf{z}' = (\mathbf{z} - \boldsymbol{\mu}_{\text{train}}) / \boldsymbol{\sigma}_{\text{train}}
\end{align}
where scaling parameters ($\boldsymbol{\mu}_{\text{train}}$, $\boldsymbol{\sigma}_{\text{train}}$)
are computed once from the training set and applied consistently to validation, test, and evaluation sets. The scaler is serialized alongside trained model weights to support inference on new instances without re-fitting.

\subsubsection{Training Procedure} \label{ref:training_procedure}

The \ac{NN} was trained offline using the training set (3'500 instances) to minimise
prediction error, with validation set performance (500 instances) used for hyperparameter
tuning and early stopping. Training employed the Adam optimizer with learning rate
$\alpha = 5 \times 10^{-5}$ and batch size of 32 samples. The loss function was mean squared
error between predicted and actual realisation costs. Regularization relied exclusively on early
stopping (patience of 30 epochs with minimum improvement threshold $\delta = 10^{-5}$) without
dropout, as the relatively small network size does not require additional regularization.
Training proceeded for a maximum of 5'000 epochs with termination upon non-improvement of
validation loss.

Target labels for training were generated by obtaining the deterministic plan via
Equation~\ref{eq:deterministic} for each training instance, sampling 100 scenarios, and
computing expected realisation costs through heuristic simulation (formally specified in 
Section~\ref{sec:eval_heuristic}):
\begin{align}
\bar{Q}_i = \frac{1}{100} \sum_{k=1}^{100} Q_k(\mathbf{x}_{\text{plan}}, \xi_k)
\end{align}
where $Q_k$ denotes the evaluation-stage cost (reactive maintenance, cancellations, and ferry
flights) under scenario $\xi_k$. This procedure created 3'500 training pairs
$\{\mathbf{z}_i, \bar{Q}_i\}$ for network learning.

\subsection{Baseline Design: Formulation and Uncertainty Strategies}
\label{subsec:experimental_design}

To thoroughly evaluate the proposed \ac{S-EX} approach, we compare it against a set of
alternative formulations and uncertainty-handling strategies. The comparison is organized along
two orthogonal dimensions: formulation variant and uncertainty strategy. 

The two formulation variants combined with the two uncertainty handling strategies yield four
primary optimisation approaches:

\begin{enumerate}
\item \textbf{\ac{S-EX}:} NN embedded into the joint optimisation of all decision variables within the \ac{MILP} toe stimate the expected evaluation-stage costs.
\item \textbf{\ac{D-EX}:} \ac{MILP} optimisation with the expected \ac{RUL} values; no NN.
\item \textbf{\ac{S-2SP}:} the NN is applied sequentially in both stages.
\item \textbf{\ac{D-2SP}:} sequential \ac{MILP} optimisation with the expected \ac{RUL} values; no NN.
\end{enumerate}

This matrix allows us to isolate and quantify:
\begin{itemize}
\item The benefit of the stochastic approach over deterministic planning
      (comparing \ac{S-EX} vs.\ \ac{D-EX}, and \ac{S-2SP} vs.\ \ac{D-2SP}).
\item The benefit of joint optimisation over sequential decomposition
      (comparing \ac{S-EX} vs.\ \ac{S-2SP}, and \ac{D-EX} vs.\ \ac{D-2SP}).
\item The combined benefit of our proposed approach over the most limited baseline
      (\ac{S-EX} vs.\ \ac{D-2SP}).
\end{itemize}

Each of the four approaches was evaluated under the same solver configurations to assess both solution quality and practical applicability. Additionally, a \textbf{real-world baseline} is included for comparison: the actual tail assignments, maintenance slots, and task scheduling decisions implemented by the airline are loaded directly (decision variables populated from operational records) and evaluated through the same scenario-based simulation as all other approaches, providing a reference against which the benefit of stochastic optimisation can be quantified.

\subsubsection{Formulation Variants}

\textbf{Extensive Form (\ac{EX})}\\
The extensive model optimizes all five variable types ($F$, $C$, $G$, $T$, $M$) simultaneously
subject to \ac{TA}.1--\ac{TA}.4, \ac{MS}.1--\ac{MS}.6, and \ac{PdM}.1. All decisions are made
jointly in a single \ac{MILP} solve, fully exploiting the coupling between fleet positioning
(via constraint \ac{MS}.6) and \ac{MS}. This formulation is the basis for our
proposed \ac{S-EX} approach and provides the tightest feasible lower bound on planning cost
among the formulations considered.

\textbf{Two-Stage Decomposition (\ac{2SP})}\\
The two-stage formulation decomposes decisions chronologically, respecting the sequential
planning hierarchy common in airline operations. \textbf{Stage~1} optimizes $F$, $C$, $G$
subject to \ac{TA}.1--\ac{TA}.4 and minimises flight costs:
\begin{align}
Z_1 = \sum_{p,f} \mathrm{OC}_{p,f} F_{p,f} + \sum_{f} \mathrm{CC}_f C_f
\label{eq:2sp_stage1_obj}
\end{align}
\textbf{Stage~2} fixes $(\bar{F}, \bar{C}, \bar{G})$ from Stage~1 and optimizes $T$, $M$
subject to \ac{MS}.1--\ac{MS}.6 (with $G_{p,n_s,t_s}$ replaced by the fixed
$\bar{G}_{p,n_s,t_s}$ in \ac{MS}.6):
\begin{align}
Z_2 = \sum_{g \in \mathcal{G}_{\text{prog}},\, s \in \mathcal{S}_{\text{real}}} W_{\text{prog}}\,T_{g,s}
    + \sum_{s \in \mathcal{S}_{\text{real}},\, p} W_{\text{slot}}\, M_{s,p}
    + \sum_{g \in \mathcal{G}_{\text{prev}},\, s \in \mathcal{S}_{\text{real}}} W_{\text{unused}}\,\Delta t_{g,s}\,T_{g,s}
\label{eq:2sp_stage2_obj}
\end{align}
Both stages remain \ac{MILP} formulations. The \ac{2SP} structure provides computational
decomposition benefits, particularly for large instances, but incurs an optimality cost
relative to the joint extensive formulation because \ac{TA} decisions are made without
full knowledge of their maintenance implications.

\begin{figure}[h]
    \centering
    \includegraphics[width=0.35\linewidth]{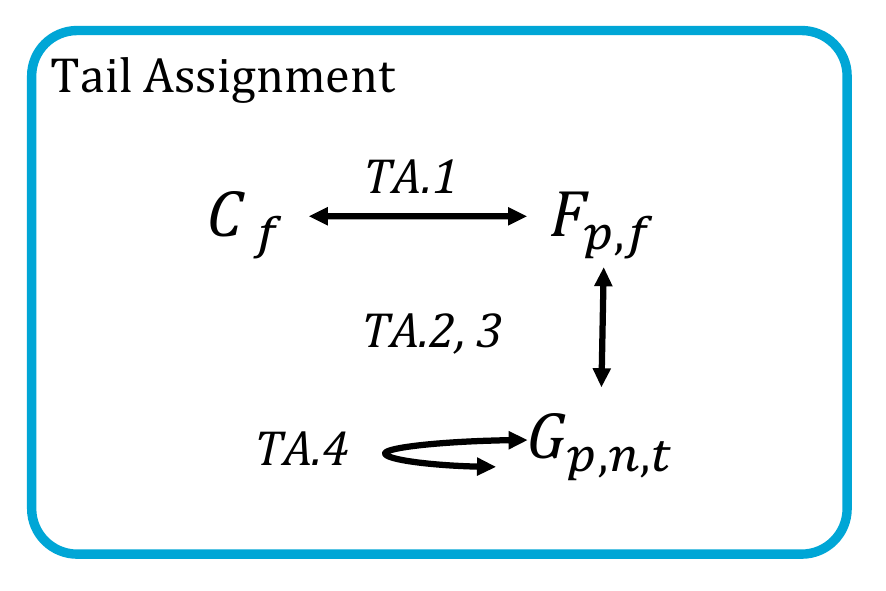}
    \caption{Variable and constraint structure of the first stage of the \ac{2SP} model.}
    \label{fig:2sp_1_var_const}
\end{figure}

\begin{figure}[h]
    \centering
    \includegraphics[width=0.6\linewidth]{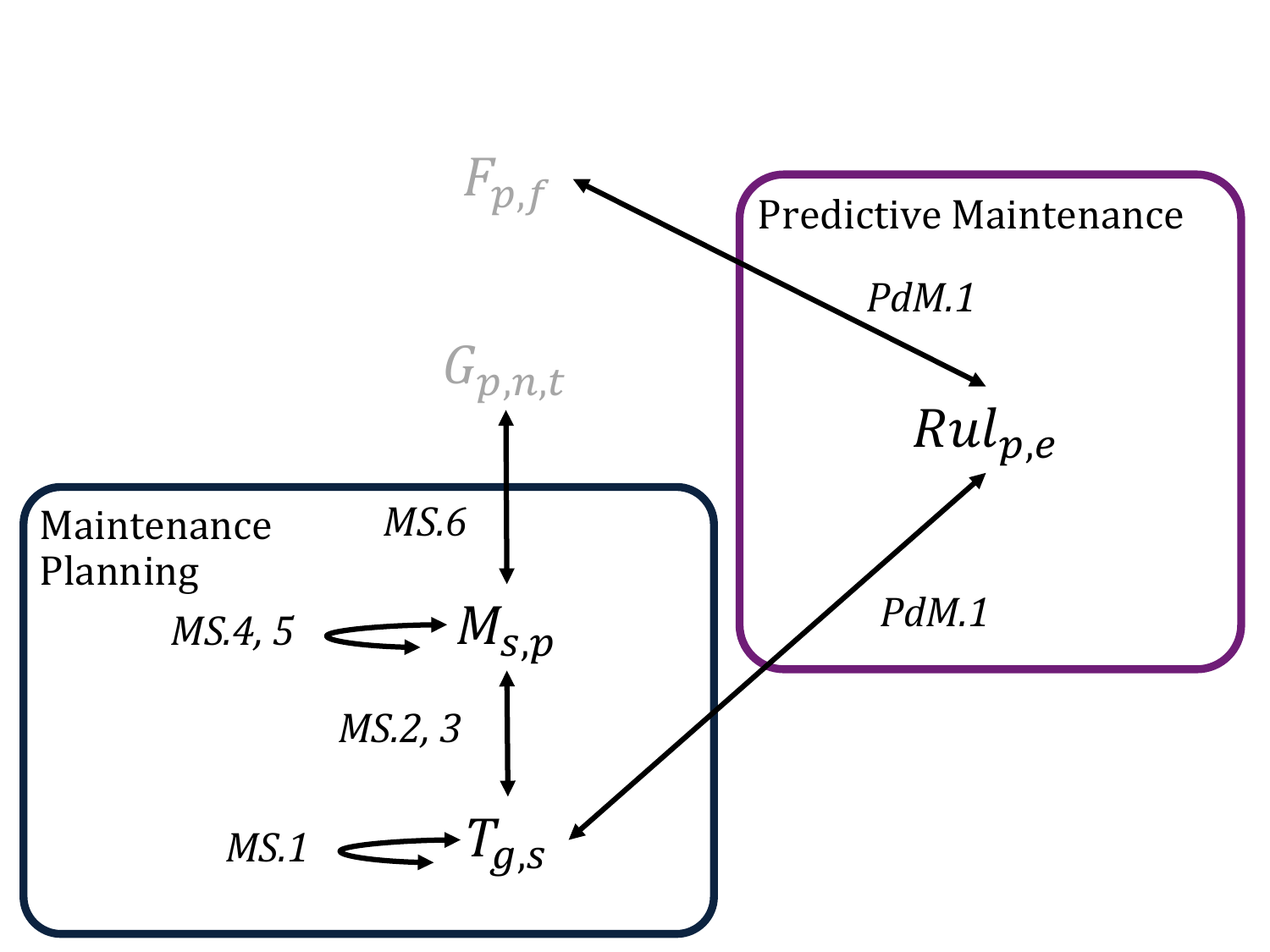}
    \caption{Variable and constraint structure of the second stage of the \ac{2SP} model.
    Grayed-out values are fixed from Stage~1.}
    \label{fig:2sp_2_var_const}
\end{figure}

Figures~\ref{fig:2sp_1_var_const}--\ref{fig:2sp_2_var_const} show the variable--constraint
structure of each stage.

\subsubsection{Uncertainty Handling Strategies}

\textbf{Stochastic Approach with Neural Network (\ac{S-EX} / \ac{S-2SP})}\\
The stochastic variant augments the planning-stage objective with the \ac{NN} estimate
of expected realisation costs as described in Section~\ref{subsec:stochastic_nne} of the
Methodology:
\begin{align}
\min_{\mathbf{x} \in \chi} \; Z_{\text{stoch}}
  = c^\top \mathbf{x} + \lambda \cdot \mathrm{NN}\!\left(\mathcal{F}(\mathbf{x}),\,\Xi\right)
\label{eq:stochastic}
\end{align}
The NN weight parameter $\lambda$ balances planning costs and robustness; in all primary
experiments reported in Section~\ref{sec:results}, $\lambda = 1.0$ unless otherwise specified.
For the \ac{S-2SP} variant, the \ac{NN} is applied within both stages: in Stage~1,
maintenance-related features are set to zero due to their inapplicability at that stage; in
Stage~2, with first-stage assignments fixed, all variables are fully defined and the NN
optimizes \ac{MS} using the complete feature representation. The zero-centered
feature engineering ensures negligible impact from inactive first-stage variables.

\textbf{Deterministic Baseline (\ac{D-EX} / \ac{D-2SP})}\\
The deterministic approach solves either the \ac{EX} or \ac{2SP} \ac{MILP} model using
\emph{expected} \ac{RUL} values:
\begin{align}
\min_{\mathbf{x} \in \chi} \; Z_{\text{det}} = c^\top \mathbf{x}
\label{eq:deterministic_exp}
\end{align}
The resulting plan minimises nominal costs but does not account for variability in \ac{RUL}
realisations. This baseline serves as a reference point to quantify the value of incorporating
uncertainty explicitly through the \ac{NN} approximation.

\subsubsection{Solution Diversity for Neural Network Training}
\label{subsec:solution_diversity}

A critical challenge in training the \ac{NN} is generating a sufficiently diverse set of solutions across the suboptimal-to-optimal spectrum. Random solution generation is infeasible due to the model's complex constraint structure. Instead, we leverage Gurobi's built-in mechanisms to systematically generate solutions.

For each training instance, Gurobi is invoked with a strict time limit of 0.5 seconds. This forces the solver to terminate earlier, yielding diverse suboptimal solutions that span the solution landscape. Following \cite{Dumouchelle2022Neur2SP:Programming}, this approach ensures broad coverage of the feasible region, allowing the \ac{NN} to learn costs across high- and low-quality decisions. This heuristic-based training regime forms the basis of our primary analysis.

\subsection{Computational Environment and Implementation}
\label{sec:computational_environment}

All computational experiments were conducted on a single workstation equipped with an 11th
Generation Intel Core i9-11950H processor (2.60 GHz base frequency) and 32 GB of RAM, running
Windows 11 Pro. The optimisation solver was Gurobi Optimizer version 12.0.3 with academic
licensing. The \ac{NN} framework was PyTorch version 2.9.0 executed on CPU without GPU
acceleration. All code was implemented in Python 3.10.13 using standard libraries: Gurobi
Python API for optimisation modelling, PyTorch for \ac{NN} training and inference, and
NumPy/Pandas for data management. Parallel instance solving employed multiprocessing with up
to 14 concurrent processes to leverage available computational resources.

Comprehensive hyperparameter settings, solver configurations, and reproducibility details are
documented in Appendix~\ref{app:hyperparameters}, ensuring full transparency and enabling
replication of all experiments.

%% file: 06_results.tex
\section{Results}
\label{sec:results}

The framework was evaluated on 500 planning instances, 
each representing five aircraft over a 3-day planning horizon with five monitored components. 
Each plan was evaluated under 100 independently sampled \ac{RUL} scenarios to quantify robustness.

\subsection{Performance Evaluation Methodology}
\label{sec:performance_evaluation}

Performance was evaluated in two complementary phases reflecting the planning-realisation
paradigm inherent in stochastic optimisation. The \emph{planning stage} assesses the immediate
optimisation outcome: planning cost (objective function value), computational time, and
planning-stage flight cancellations. The \emph{evaluation stage} assesses robustness by
executing each planned schedule under 100 independently sampled \ac{RUL} realisations and
measuring realised operational costs (including reactive maintenance, cancellations triggered by
\ac{RUL} violations, and ferry flight costs for remote maintenance) and average cancellations
across scenarios.

The evaluation procedure for each of the 500 evaluation set instances proceeded as follows:

\begin{enumerate}
\item \textbf{Planning:} Solve the planning \ac{MILP} (deterministic or stochastic) using the
      specified formulation (\ac{EX} or \ac{2SP}) to obtain a flight and maintenance plan
      $(\bar{F}, \bar{C}, \bar{G}, \bar{T}, \bar{M})$.
\item \textbf{Realization:} For each of the 100 sampled scenarios $\xi_k$, evaluate the plan's
      feasibility and incurred costs under that scenario's actual \ac{RUL} realisations by
      (i)~checking for constraint violations in \ac{PdM}.1, (ii)~executing reactive maintenance
      if violations are predicted or cancelling flights if recovery is infeasible, and
      (iii)~accumulating all associated costs.
\item \textbf{Aggregation:} Compute aggregate statistics of
      realisation costs across scenarios to characterise plan robustness and expected
      performance under sampled uncertainty.
\end{enumerate}

Metrics were aggregated across all 500 evaluation instances and 100 scenarios per instance,
yielding 50'000 total evaluation runs per approach.

\subsubsection{Evaluation-Stage Heuristic: Formal Description}
\label{sec:eval_heuristic}

The realisation-stage evaluation procedure is a deterministic, flight-level simulation that
propagates a planned schedule $\bar{\mathbf{x}} = (\bar{F}, \bar{C}, \bar{G}, \bar{T},
\bar{M})$ through a single \ac{RUL} scenario $\xi_k = \{\delta_{e,p}\}_{e \in \mathcal{E},
p \in \mathcal{P}}$ and returns a scalar realisation cost $Q_k$. The procedure is stated
formally in Algorithm~\ref{alg:eval_heuristic}.

\textbf{Scenario adaptation}\\
Each scenario $\xi_k$ is a flat vector of signed perturbations $\delta_{e,p}$ drawn from the
predictive distribution of component degradation (Section~\ref{sec:scenario_sampling}). The
scenario-adapted initial \ac{RUL} for component $e$ on aircraft $p$ is:
\begin{align}
  \tilde{R}_{e,p} = \max\!\left(1,\; R_{e,p,0} + \delta_{e,p}\right),
  \label{eq:rul_adapt}
\end{align}
where $R_{e,p,0}$ is the deterministic base \ac{RUL} used at the planning stage.

\textbf{Cost components}\\
Three reactive cost categories are incurred during the simulation:
\begin{itemize}
  \item \emph{Cancellation cost} ($w_{\text{cancel}}$): charged once per cancelled flight.
  \item \emph{Reactive maintenance cost} ($w_{\text{react},e}= c_{\text{task},e} \cdot
        r_{\text{react}}$): charged once per reactive maintenance task for component $e$,
        where $c_{\text{task},e}$ is the unit task cost and $r_{\text{react}}$ is the
        reactive cost multiplier.
  \item \emph{Outstation surcharge} ($w_{\text{out}} = r_{\text{out}} \cdot w_{\text{cancel}}$):
        an additional penalty charged when reactive maintenance is required at a non-hub
        airport (representing ferry and logistics overhead), with surcharge factor
        $r_{\text{out}}$.
\end{itemize}
Planned maintenance tasks ($T_{g,s} = 1$) incur zero additional cost at the evaluation stage,
as they are already captured in the planning-stage objective.

\begin{algorithm}[h]
\caption{Evaluation-Stage Heuristic}
\label{alg:eval_heuristic}
\KwIn{Plan $\bar{\mathbf{x}}$; scenario $\xi_k$; cost parameters
      $w_{\text{cancel}},\, r_{\text{react}},\, r_{\text{out}}$}
\KwOut{Realization cost $Q_k$}

\textbf{Initialise:} $\mathrm{RUL}_{e,p} \leftarrow \tilde{R}_{e,p}$ for all $e,p$
(Eq.~\ref{eq:rul_adapt})\;
$Q_k \leftarrow 0$\;
$\mathrm{DeferredRecovery}[p][d] \leftarrow \emptyset$ for all $p,d$\;

\For{each aircraft $p \in \mathcal{P}$}{
  \For{each planning day $d$ in chronological order}{

    \tcp{Step 1: Apply planned maintenance}
    \For{each task $g$ with $T_{g,s}=1$, $\mathrm{day}(s)=d$, aircraft$(g)=p$}{
      $\mathrm{RUL}_{\mathrm{pred}(g),\,p} \;\mathrel{+}=\; \mathrm{RUL\_recover}_g$\;
    }

    \tcp{Step 2: Apply deferred reactive maintenance from prior cancellation}
    \For{each $(e, \rho) \in \mathrm{DeferredRecovery}[p][d]$
         not covered by Step 1}{
      $\mathrm{RUL}_{e,p} \;\mathrel{+}=\; \rho$\;
      $Q_k \;\mathrel{+}=\; w_{\text{react},e}$\;
      \lIf{$p$ not at hub}{$Q_k \;\mathrel{+}=\; w_{\text{out}}$}
    }

    \tcp{Step 3: Simulate flights in departure-time order}
    $\mathrm{DayCancelled} \leftarrow \mathrm{False}$\;
    \For{each flight $f$ assigned to $p$ on day $d$, sorted by $t^{\mathrm{dep}}_f$}{
      \uIf{$\mathrm{DayCancelled}$}{
        $Q_k \;\mathrel{+}=\; w_{\text{cancel}}$\;
        \tcp{Cascade: remaining flights of day cancelled at zero extra \ac{RUL} cost}
      }
      \uElseIf{$\min_{e}\,\mathrm{RUL}_{e,p} \;\leq\; 0$}{
        \tcp{\ac{RUL} exhausted: cancel this and all remaining flights today}
        $Q_k \;\mathrel{+}=\; w_{\text{cancel}}$\;
        $\mathrm{DayCancelled} \leftarrow \mathrm{True}$\;
        $\mathcal{E}^* \leftarrow \{e : \mathrm{RUL}_{e,p} \leq 0\}$\;
        \For{each $e \in \mathcal{E}^*$}{
          $\mathrm{DeferredRecovery}[p][d{+}1] \;\mathrel{\cup}=\;
          \{(e,\;\mathrm{RUL\_recover}_e)\}$\;
        }
      }
      \Else{
        \tcp{Operate flight: consume RUL}
        \For{each component $e \in \mathcal{E}$}{
          $\mathrm{RUL}_{e,p} \;\leftarrow\; \max\!\left(0,\;
          \mathrm{RUL}_{e,p} - c_{e,f}\right)$\;
        }
      }
    }
  }
}
\Return{$Q_k$}
\end{algorithm}

\textbf{Reactive maintenance deferral}\\
When a flight is cancelled because $\min_e \mathrm{RUL}_{e,p} \leq 0$, all remaining flights
of aircraft $p$ on that day are cascaded as cancelled (Step 3). The limiting components
$\mathcal{E}^* = \{e : \mathrm{RUL}_{e,p} \leq 0\}$ are flagged for reactive maintenance,
which is scheduled at the start of the next planning day (Step 2 of day $d+1$). This
deferral reflects the operational reality that an unplanned maintenance event requires ground
time and cannot be completed mid-rotation.

\textbf{Cascade cancellations}\\
Once any flight in a day triggers a cancellation, all subsequent flights of the same aircraft
on the same day are also cancelled at the standard cost $w_{\text{cancel}}$ each. The
aircraft remains at its last confirmed departure airport until maintenance is completed the
following day.

\textbf{Limitations of the heuristic}\\
The policy is deliberately parsimonious: it permits only flight cancellation and reactive
maintenance, and does not model aircraft swapping, crew reassignment, or network-wide
cascading. This is acknowledged as a study limitation in Section~\ref{sec:discussion_limitations}.
The cost parameters used in all experiments are reported in Appendix~\ref{app:hyperparameters}.

\subsection{Planning-Stage and Evaluation-Stage Performance}
\label{sec:results_planning_eval}

The planning stage represents the initial solution for \ac{TA} and \ac{MS}, prior to the realisation of prognostic uncertainty. Here, the optimisation model produces a schedule that balances immediate costs (aircraft assignment, \ac{MS}) with predicted robustness under uncertainty. The key trade-off is between planning-stage conservatism (allocating more maintenance capacity and incurring higher immediate costs) and evaluation-stage robustness (reducing cancellations and reactive maintenance when uncertainty realises).

The evaluation stage assesses how each plan generated in the planning stage performs under sampled realisations of \ac{RUL} predictor uncertainty. For every planning instance, 100 scenarios are drawn independently from the uncertainty distribution, and each plan is evaluated against these scenarios to compute reactive maintenance costs, cancellations, and disruption penalties. This two-phase approach quantifies both the cost of planning conservatism and the benefit of that conservatism when plans are executed against realised uncertainty.

\subsubsection{Stochastic vs.\ Deterministic}
\label{sec:results_cost_time_tradeoffs}

We compare deterministic baselines against stochastic variants. All
configurations that embed a \ac{NN} are stochastic because the \ac{NN} term represents
an explicit approximation of downstream (evaluation-stage) cost under uncertainty, using either 
the mean target $\bar{Q}^{\text{mean}}$ or the 95th-percentile target $\bar{Q}^{95}$ 
(see Eqs.~\ref{eq:target_mean}--\ref{eq:target_95p}). Configurations without a \ac{NN} term 
are deterministic and optimize immediate planning-stage costs only.

Table~\ref{tab:planning_eval_combined} reports mean planning objectives (Plan Cost), mean 
realised evaluation costs (Eval Cost), mean cancellations (Eval Canc.), solve times, and 
total costs across all approaches. All statistics are computed over the 500 evaluation instances 
and represent aggregate performance across this diverse set of operational scenarios.

\input{tables/table_planning_eval_combined.tex}

The stochastic extensive approach with 95th-percentile \ac{NN} training
(\textbf{\ac{S-EX}-95}) achieves the best end-to-end performance among all optimized
approaches, with a total cost of 742{'}647 (planning: 450{'}134, evaluation: 292{'}512)
and 5.77 average evaluation cancellations.

To isolate the benefit of the stochastic mechanism, we first compare approaches with the same
robustness intent. \textbf{\ac{S-EX}-95} achieves a lower total cost (742{'}647) than its
deterministic counterpart \textbf{\ac{D-EX}-95} (755{'}549), despite both targeting robustness
against adverse \ac{RUL} realisations. \ac{D-EX}-95 reaches near-zero evaluation cost (413)
through extreme upfront conservatism, but at a planning cost nearly 68\% higher than
\ac{S-EX}-95 (755{'}136 vs.\ 450{'}134).

Comparing stochastic and deterministic approaches at the mean level, \textbf{\ac{S-EX}}
(total: 751{'}825, evaluation: 335{'}522, 6.61 cancellations) outperforms the deterministic
baseline \textbf{\ac{D-EX}} (total: 774{'}769, evaluation: 374{'}910, 7.38 cancellations)
by 3.0\% in total cost and 10.5\% in evaluation cost.

Among the two-stage approaches, \textbf{\ac{S-2SP}-95} achieves a total cost of 762{'}750 
(2.7\% above \ac{S-EX}-95) at substantially faster solve times (0.74s vs.\ 6.53s).

\paragraph{Planning-stage cancellations}
The Plan Canc. column in Table~\ref{tab:planning_eval_combined} reveals three distinct
regimes. The deterministic 95th-percentile configurations (\textbf{\ac{D-EX}-95} and
\textbf{\ac{D-2SP}-95}) show the highest planning-stage cancellations (3.31 and 5.87,
respectively), which indicates operationally unjustifiable upfront cancellations before a confirmed
degradation event; this is impractical despite their near-zero evaluation-stage costs. The
two-stage configurations (\textbf{\ac{D-2SP}}, \textbf{\ac{S-2SP}-mean}, \textbf{\ac{S-2SP}-95})
show a moderate range (1.11--1.26), but these cancellations are a structural artifact of
decoupling Stage~1 \ac{TA} from Stage~2 \ac{MS}: with fixed Stage~1 routes,
Stage~2 may cancel flights to resolve conflicts. In practice, planners would trigger an aircraft
swap back-loop to \ac{TA}, which is not implemented in the current framework. By contrast,
the extensive configurations (\textbf{\ac{D-EX}}, \textbf{\ac{S-EX}-mean}, \textbf{\ac{S-EX}-95}, \textbf{Real})
show only a small cancellation residual (0.32--0.33), which is data-driven and reflects special operational
edge cases where no feasible uncancelled solution exists within the planning horizon, rather than
a structural modelling artifact.

\subsubsection{Paired Comparisons: Mean vs.\ 95th-Percentile Variants}
\label{sec:results_paired}

To isolate the effect of \ac{NN} training strategy, we compare paired configurations
within the same formulation. The \ac{NN} is trained on either the mean evaluation cost
$\bar{Q}^{\text{mean}}$ or the 95th-percentile $\bar{Q}^{95}$ (see Eqs.~\ref{eq:target_mean} 
and \ref{eq:target_95p}). 

The 95th-percentile quantile is selected as the primary risk metric because prognostic models
in \ac{PdM} commonly report \ac{RUL} predictions using confidence intervals,
with the upper bound (95th percentile) used as a conservative maintenance planning trigger 
in industry practice. This reflects operational standards where safety-critical maintenance decisions 
hedge against component degradation faster than the mean prediction. The 95th percentile captures 
the distribution's tail behaviour, encoding information about rare but costly failure scenarios 
that deterministic planning overlooks.

\textbf{Extensive Formulation (\ac{EX}):}
\begin{itemize}
    \item \textbf{\ac{D-EX} vs.\ \ac{S-EX}-95:} The stochastic approach with 95th-percentile 
          training increases planning cost by 50{'}275 (12.6\%) from 399{'}859 to 450{'}134. 
          This substantial increase reflects the strong hedging signal: the network trained on 
          worst-case tail costs identifies scenarios where the plan is vulnerable to cascading 
          failures and recommends aggressive preventive maintenance. Evaluation-stage cost 
          reduction is correspondingly large (292{'}512 vs.\ 374{'}910, a 21.9\% decrease), 
          yielding net total cost savings of 32{'}122 (4.1\%). This demonstrates that the price
          of robustness, i.e., the increase in planning-stage costs, is substantially offset by the
          reduction in operational risk and cost under uncertainty. In operational terms, investing 
          an additional 50{'}275 in upfront maintenance capacity prevents 82{'}398 in 
          downstream reactive costs and disruptions.
    
    \item \textbf{\ac{D-EX} vs.\ \ac{D-EX}-95:} The deterministic 95th-percentile configuration 
          represents an extreme form of conservatism: planning cost increases by 355{'}277 
          (88.9\%) to 755{'}136 to achieve evaluation cost of only 413 (99.9\% lower than \ac{D-EX}). 
          This demonstrates the pathology of deterministic tail hedging: without learned cost 
          approximations, the optimizer must allocate massive maintenance capacity to guarantee 
          feasibility under the worst-case \ac{RUL} scenario, resulting in a total cost that exceeds 
          even the deterministic baseline. It also incurs 3.31 planning-stage cancellations per
          instance, making the configuration operationally unacceptable because proactive
          cancellations are imposed without a confirmed degradation event. The \ac{NN} approach avoids this failure mode
          by learning which scenarios are actually costly (not just which are technically feasible), and 
          allocating conservatism where it provides the greatest marginal benefit.
\end{itemize}

\textbf{Two-Stage Decomposition (\ac{2SP}):}
\begin{itemize}
    \item \textbf{\ac{D-2SP} vs.\ \ac{S-2SP}-95:} The stochastic approach with 95th-percentile 
          training in the two-stage setting increases planning cost by 24{'}077 (4.9\%) to 
          518{'}309. The evaluation cost reduction is substantial (244{'}440 vs.\ 285{'}232, 
          a 14.3\% decrease), yielding net total cost savings of 39{'}544 (5.2\%). The 
          95th-percentile training target provides consistent value across formulations by steering 
          the optimizer away from plans vulnerable to high-cost tail scenarios. The smaller planning 
          cost increase relative to the extensive formulation reflects the two-stage structure: 
          Stage~1 decisions are less flexible regarding maintenance foresight, limiting the scope
          for the \ac{NN} to influence early-stage choices. By contrast, the deterministic
          worst-case variant \textbf{\ac{D-2SP}-95} records 5.87 planning-stage cancellations per
          instance in Table~\ref{tab:planning_eval_combined} (the highest of all configurations),
          further illustrating the operationally unacceptable conservatism of deterministic
          worst-case hedging in the two-stage setting.
\end{itemize}

The \ac{S-EX}-95 configuration achieves superior performance by combining two strengths: 
the extensive formulation's ability to jointly optimize all decision layers and the \ac{NN}'s
ability to learn which scenarios are costly and allocate robustness precisely where
it matters most. The 95th-percentile training target captures the uncertainty distribution's 
tail behaviour, enabling plans that hedge against rare but costly failure modes without incurring 
the extreme conservatism of worst-case deterministic planning. This learned hedging is both 
computationally tractable and economically justified by the empirical benefits demonstrated 
in the evaluation stage.

\subsubsection{Statistical Robustness}

All comparisons between \ac{S-EX}-95 and the deterministic baselines exhibit very large effect sizes 
and extremely small $p$-values, confirming statistical robustness. The 4.1\% total cost reduction 
and 21.9\% evaluation-stage cost reduction represent consistent, practically significant improvements 
across the evaluation set. The consistency of these improvements across diverse problem instances 
(varying fleet utilization, maintenance demand profiles, and component degradation patterns) 
indicates that the framework's benefits are not instance-dependent artifacts but systematic 
properties of the integrated stochastic optimisation approach.
Detailed statistical testing is documented in Appendix~\ref{app:statistical_tests_detailed}.

\subsubsection{Distributional View: Total-Cost Robustness}
\label{sec:results_distribution_total}

\begin{figure}[h]
    \centering
    \includegraphics[width=0.85\textwidth]{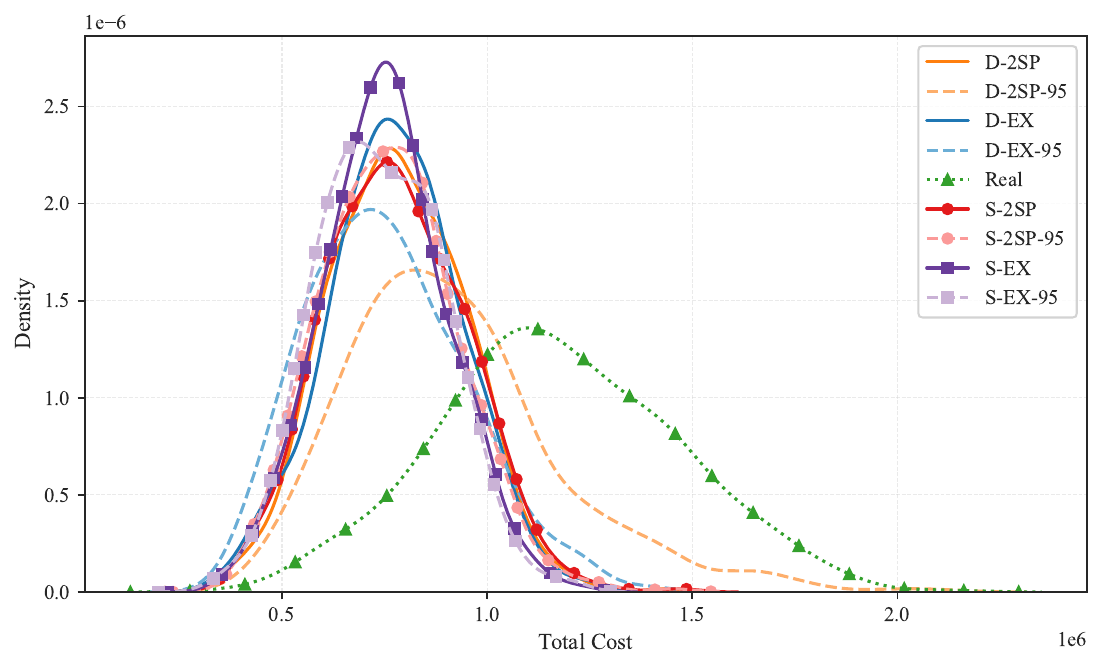}
    \caption{Distribution of total costs (planning + evaluation) across all instances. 
    The stochastic approach with 95th-percentile training (\ac{S-EX}-95) exhibits a tighter, 
    left-shifted distribution relative to deterministic baselines, indicating more consistent 
    and robust performance across problem instances. Lower tail probability at high costs 
    demonstrates reduced frequency of expensive failure scenarios.}
    \label{fig:dist_total_cost}
\end{figure}

Figure~\ref{fig:dist_total_cost} shows the distribution of total costs across all 500 
evaluation instances. The distributional view provides insight into how frequently each method 
yields high-cost outlier scenarios. Tighter, left-shifted distributions (lower costs with lower 
variance) indicate more robust and consistent planning. This analysis reveals not only differences 
in mean performance but also differences in plan reliability: how vulnerable each approach is 
to worst-case problem instances.

The distributional analysis reveals that \ac{S-EX}-95 not only achieves lower average costs 
but also exhibits substantially lower variance, with fewer high-cost outlier instances. This 
property is operationally valuable: it allows airlines to plan with greater predictability and 
confidence in their cost budgets. The deterministic baseline, by contrast, shows a long right tail 
of expensive scenarios where reactive maintenance and cascading cancellations dominate.

\subsection{Computational Scalability}
\label{sec:results_scalability}

To characterise computational tractability of the \ac{MILP} formulations, we conducted 
scalability experiments with the deterministic variants (\ac{D-EX} and \ac{D-2SP}). 
Stochastic variants (\ac{S-EX}, \ac{S-2SP}) were excluded from this analysis because
\ac{NN} retraining would be required for each configuration (fleet size, horizon,
predictor count), making systematic evaluation prohibitively expensive. The deterministic 
analysis identifies fundamental scaling regimes and informs where decomposition becomes necessary 
for larger instances.

\textbf{Note on cost scaling:} The planning costs reported in this section reflect a different 
cost structure than the base configuration (5 aircraft, 3 days) used in the primary results. 
Scaling experiments vary fleet size, horizon, and component count across different problem 
instances, which exhibit different maintenance demand profiles and operational complexity. 
Consequently, objective values should not be directly compared to the base results; focus instead 
on the scaling behaviour (solve time trends) and relative performance between formulations at each scale.

Experiments varied three dimensions independently: fleet size (2--20 aircraft), planning 
horizon (2--20 days), and monitored components per aircraft (1--80 predictors), holding 
other dimensions at baseline values (5 aircraft, 3 days, 5 predictors).

\subsubsection{Fleet Size Scaling}
\label{sec:scalability_aircraft}

Table~\ref{tab:scalability_aircraft} presents planning costs and solve times as fleet size 
varies from 2 to 20 aircraft (3-day horizon, 5 predictors).

\begin{table}[h]
\centering
\small
\begin{tabular}{@{}lrrrr@{}}
\toprule
Fleet Size & \ac{EX} Cost & \ac{EX} Time [s] & \ac{2SP} Cost & \ac{2SP} Time [s] \\
\midrule
2 & 564{'}981 & 0.06 & 607{'}002 & 0.02 \\
5 & 702{'}974 & 3.09 & 1{'}204{'}829 & 0.23 \\
10 & 946{'}557 & 62.86 & 1{'}709{'}370 & 0.54 \\
15 & 1{'}863{'}413 & 696.00 & 10{'}944{'}095 & 0.76 \\
20 & 27{'}155{'}051 & 943.00 & 79{'}671{'}023 & 1.02 \\
\bottomrule
\multicolumn{5}{@{}l}{\footnotesize \ac{EX} exceeds 600s limit at 15+ aircraft; \ac{2SP} remains tractable.} \\
\end{tabular}
\caption{Fleet size scaling: planning costs and solve times for \ac{D-EX} and \ac{D-2SP}.}
\label{tab:scalability_aircraft}
\end{table}

The extensive formulation exhibits steep growth: 226$\times$ time increase from 2 to 10 aircraft, 
and exceeds practical limits (600s) at 15+ aircraft. Two-stage decomposition scales dramatically 
better: only 23$\times$ time increase over the same range, remaining feasible at 20 aircraft (1.02s). 
The growing objective value gap between formulations (7\% at 2 aircraft to 487\% at 20 aircraft) 
reflects increasing suboptimality of decomposition with problem scale. This trade-off, tractability 
versus solution quality, is unavoidable at scale and motivates the decomposition strategy as 
a practical solution for realistic airline networks.

\subsubsection{Planning Horizon Scaling}
\label{sec:scalability_days}

Table~\ref{tab:scalability_days} presents planning costs and solve times as planning horizon 
varies from 2 to 20 days (5 aircraft, 5 predictors).

\begin{table}[h]
\centering
\small
\begin{tabular}{@{}lrrrr@{}}
\toprule
Horizon [days] & \ac{EX} Cost & \ac{EX} Time [s] & \ac{2SP} Cost & \ac{2SP} Time [s] \\
\midrule
2 & 574{'}149 & 0.56 & 837{'}087 & 0.09 \\
3 & 702{'}974 & 3.02 & 1{'}204{'}829 & 0.19 \\
5 & 1{'}082{'}593 & 37.77 & 1{'}772{'}768 & 0.65 \\
7 & 1{'}501{'}317 & 397.00 & 2{'}021{'}492 & 1.57 \\
10 & 2{'}500{'}698 & 1{'}184.00 & 3{'}104{'}747 & 3.45 \\
20 & 5{'}184{'}116 & 1{'}230.00 & 5{'}752{'}866 & 99.98 \\
\bottomrule
\multicolumn{5}{@{}l}{\footnotesize \ac{EX} exceeds 600s limit beyond 7 days; \ac{2SP} remains tractable to 20 days.} \\
\end{tabular}
\caption{Planning horizon scaling: planning costs and solve times for \ac{D-EX} and \ac{D-2SP}.}
\label{tab:scalability_days}
\end{table}

The extensive formulation becomes intractable beyond a 7-day horizon (397s at day 7). 
Decomposition scales markedly better: solve time increases from 0.09s at 2 days to 99.98s 
at 20 days, remaining manageable for strategic planning on weekly timescales. The horizon 
dimension is particularly critical for airline operations: most operational planning occurs 
on 7--14 day windows, a regime where \ac{EX} is infeasible but \ac{2SP} remains practical.

\subsubsection{Prognostic Component Count Scaling}
\label{sec:scalability_predictors}

Table~\ref{tab:scalability_predictors} presents planning costs and solve times as the number 
of monitored components per aircraft varies from 1 to 80 (5 aircraft, 3 days).

\begin{table}[h]
\centering
\small
\begin{tabular}{@{}lrrrr@{}}
\toprule
Predictors & \ac{EX} Cost & \ac{EX} Time [s] & \ac{2SP} Cost & \ac{2SP} Time [s] \\
\midrule
1 & 702{'}974 & 2.96 & 1{'}204{'}829 & 0.19 \\
5 & 702{'}974 & 2.97 & 1{'}204{'}829 & 0.20 \\
10 & 702{'}974 & 3.23 & 1{'}204{'}829 & 0.22 \\
40 & 702{'}974 & 3.70 & 1{'}204{'}829 & 0.25 \\
80 & 702{'}974 & 3.26 & 1{'}204{'}829 & 0.21 \\
\bottomrule
\multicolumn{5}{@{}l}{\footnotesize Component count has negligible impact on cost or solve time.} \\
\end{tabular}
\caption{Prognostic component scaling: planning costs and solve times for \ac{D-EX} and \ac{D-2SP}.}
\label{tab:scalability_predictors}
\end{table}

Component count has essentially no effect on either formulation: costs and solve times 
remain flat across 1--80 predictors. The primary scalability bottleneck lies in the
combinatorial structure of \ac{TA} and \ac{MS}, not the \ac{PdM} constraints.
This finding is operationally reassuring: airlines can increase prognostic instrumentation and 
component health monitoring without incurring additional optimisation overhead.

\subsubsection{Scalability Summary}

For realistic airline operations (10--15 aircraft, 3--7 day horizons), the extensive 
formulation is computationally infeasible, exceeding 600s solve times. The two-stage 
decomposition remains tractable across all tested dimensions. While decomposition accepts 
higher planning costs due to sequential optimisation (typically 50--100\% higher objectives, 
depending on fleet size and horizon), the computational advantage is decisive for practical 
applications. The scalability results inform the deployment strategy discussed in the Discussion 
(Section~\ref{sec:discussion_scalability}): \ac{S-EX}-95 is recommended for small-to-medium 
instances within its tractability regime, while \ac{S-2SP}-95 is the necessary choice for 
large-scale airline networks.

\subsection{Sensitivity to Design Choices}
\label{sec:results_sensitivity_analysis}

In addition to the primary results, three targeted sensitivity studies validate design robustness 
under varied conditions and characterise framework behaviour at problem boundaries:

\begin{enumerate}

\item \emph{Training Sample Efficiency} (Appendix~\ref{sec:results_sample_efficiency}): 
      The \ac{NN} requires approximately $N \geq 200$ training instances to provide
      meaningful robustness gains. Below this threshold, the network lacks sufficient 
      distributional coverage to learn effective cost approximations. This finding has practical 
      implications: airlines beginning their \ac{PdM} journey can reach near-peak robustness 
      with limited historical data, reducing the barrier to framework adoption.

\item \emph{Neural Network Weight Sensitivity} (Appendix~\ref{sec:results_lambda}): 
      The default weight $\lambda = 1.0$ (used in all primary results) minimises total 
      cost for \ac{S-EX}-95, achieving a $-51\%$ evaluation cost reduction against a $+20.7\%$ 
      planning cost increase (net: $-25.2\%$). Sensitivity analysis shows that the optimal 
      weight is robust around this default, with diminishing returns beyond $\lambda = 1.5$. 
      This allows operators to calibrate $\lambda$ to their specific cost structures without 
      requiring extensive reoptimisation.

\item \emph{\ac{RUL} Uncertainty Robustness} (Appendix~\ref{sec:results_sensitivity}): 
      The approach degrades gracefully at problem boundaries. Under high uncertainty 
      ($\sigma = 50$ cycles), stochastic methods retain a $-28\%$ cost advantage over 
      deterministic baselines, and under zero uncertainty, the overhead of the \ac{NN}
      approximation remains modest ($+3.2\%$ vs.\ purely deterministic). This 
      demonstrates that the framework is well-calibrated to the level of prognostic uncertainty 
      present in the operational environment.

\end{enumerate}

For comprehensive details on these analyses, refer to Appendices~\ref{sec:results_sample_efficiency}, 
\ref{sec:results_lambda}, and \ref{sec:results_sensitivity}.

%% file: tables/table_planning_eval_combined.tex
\begin{table}[h]
\centering
\small
\begin{tabular}{@{}l r r r r r r@{}}
\toprule
\textbf{Approach} & \textbf{Plan Cost} & \textbf{Plan Canc.} & \textbf{Eval Cost} & \textbf{Eval Canc.} & \textbf{Solve Time [s]} & \textbf{Total Cost} \\
\midrule
\quad D-2SP & 494{'}232 & 1.22 & 285{'}232 & 5.65 & \textbf{0.18} & 779{'}465 \\
\quad S-2SP & 498{'}959 & 1.26 & 278{'}317 & 5.51 & 0.64 & 777{'}276 \\
\quad S-2SP-95 & 518{'}309 & 1.11 & 244{'}440 & 4.83 & 0.74 & 762{'}750 \\
\quad D-2SP-95 & 916{'}035 & 5.87 & \textbf{181} & \textbf{0.00} & \textbf{0.24} & 916{'}216 \\
\midrule
\quad D-EX & \textbf{399{'}859} & \textbf{0.32} & 374{'}910 & 7.38 & 4.29 & 774{'}769 \\
\quad S-EX & 416{'}302 & \textbf{0.32} & 335{'}522 & 6.61 & 10.83 & \textbf{751{'}825} \\
\quad S-EX-95 & 450{'}134 & \textbf{0.33} & 292{'}512 & 5.77 & 6.53 & \textbf{742{'}647} \\
\quad D-EX-95 & 755{'}136 & 3.31 & \textbf{413} & \textbf{0.01} & 10.56 & 755{'}549 \\
\midrule
\quad Real & 453{'}616 & \textbf{0.32} & 914{'}638 & 17.42 & - & 1{'}368{'}254 \\
\midrule
\bottomrule
\multicolumn{7}{@{}l}{\footnotesize Total Cost = Plan Cost + Eval Cost. Plan Canc. = planned cancellations in phase 1.} \\
\end{tabular}
\caption{Combined planning and evaluation stage results.}
\label{tab:planning_eval_combined}
\end{table}

%% file: 07_discussion.tex
\section{Discussion}
\label{sec:discussion}

\subsection{Value of Uncertainty-Aware Planning}
\label{sec:discussion_stochastic_value}

The central finding of this study is that embedding a \ac{NN}
approximation of downstream uncertainty costs into the planning optimisation
produces large, consistent improvements in realised operational outcomes.
The best-performing configuration, \textbf{\ac{S-EX}-95}, achieves a total
cost of 742{'}647 compared to 774{'}769 for the deterministic extensive
baseline (\textbf{\ac{D-EX}}), a reduction of 32{'}122 (4.1\%).
This aggregate improvement is driven by a 21.9\% decrease in evaluation-stage
cost (292{'}512 vs.\ 374{'}910), partially offset by a 12.6\% increase in
planning cost (450{'}134 vs.\ 399{'}859).
Cancellations are reduced from 7.38 to 5.77 per instance (21.8\%), a
practically meaningful improvement given the direct passenger and revenue impact
of each cancellation.

These numbers reflect the fundamental asymmetry of uncertainty-aware planning:
a modest increase in planning conservatism, the ``price of robustness''
translates into substantially smaller realised costs when uncertainty
materialises.
The \ac{NN} does not hedge uniformly; it learns which maintenance
decisions provide the greatest marginal reduction in tail risk, and allocates
additional capacity precisely where it matters most.
This targeted conservatism avoids the pathology of worst-case planning, as
demonstrated by the failure of \textbf{\ac{D-EX}-95}: enforcing the 95th
percentile \ac{RUL} as a hard constraint achieves near-zero evaluation cost
(413) but at a planning cost of 755{'}136, a 89\% increase over \textbf{\ac{D-EX}}
that results in a total cost of 755{'}549, worse than the plain
deterministic baseline.
The \ac{NN} approximation avoids this failure mode by learning which scenarios
are actually costly, not merely which are technically infeasible.

\paragraph{Comparison with actual airline operations}
Relative to a real-world airline baseline, \textbf{\ac{S-EX}-95} reduces
total cost by 625{'}607 (45.7\%), from 1{'}368{'}254 to 742{'}647.
More strikingly, evaluation-stage cancellations fall from 17.42 to 5.77,
a 66.8\% reduction, representing a substantial improvement in operational
resilience under prognostic uncertainty.
The planning-stage cost of \textbf{\ac{S-EX}-95} (450{'}134) is broadly
comparable to the baseline's true cost structure, suggesting that the framework
does not incur prohibitive overhead relative to current practice, but rather
redistributes maintenance capacity more efficiently across the planning horizon.
These results confirm that systematic stochastic optimisation of integrated
\ac{TA} and \ac{MS} yields material operational value
beyond what is achievable through current reactive planning approaches.

\subsection{Role of the Uncertainty Quantile: Mean vs.\ 95th-Percentile
            Training Target}
\label{sec:discussion_mean_vs_95p}

A key design choice in the \ac{NN} training procedure is the cost label used
as the learning target (Eqs.~\ref{eq:target_mean}--\ref{eq:target_95p}).
We evaluated two options: the mean target $\bar{Q}^{\text{mean}}$, and the
95th-percentile target $\bar{Q}^{95}$.
Within the extensive formulation, shifting from mean-based to 95th-percentile
training (\textbf{\ac{S-EX}} $\to$ \textbf{\ac{S-EX}-95}) increases planning
cost by 33{'}832 (8.1\%) to 450{'}134 but reduces evaluation cost by 12.8\%
(335{'}522 $\to$ 292{'}512), yielding net total cost savings of 9{'}178 (1.2\%).
Within the two-stage decomposition, the same shift
(\textbf{\ac{S-2SP}} $\to$ \textbf{\ac{S-2SP}-95}) increases planning cost by
19{'}350 (3.9\%) to 518{'}309 while reducing evaluation cost by 12.2\%
(278{'}317 $\to$ 244{'}440), for net savings of 14{'}527 (1.9\%).

The intuition is straightforward: the 95th-percentile target $\bar{Q}^{95}$
trains the network to predict a conservative upper bound on second-stage cost,
which, when embedded in the optimizer, steers it away from plans vulnerable to
high-cost tail scenarios.
The 95th-percentile quantile is adopted as the primary training target, consistent
with the standard convention for conservative confidence bounds in prognostics-based
maintenance~\cite{Saxena2008MetricsTechniques}: \ac{PdM} planners commonly treat the
upper confidence bound of \ac{RUL} predictions as a conservative maintenance trigger,
hedging against component degradation faster than the mean prediction.
The specific choice of percentile is, however, a transparent design parameter rather
than a fixed requirement; any user-defined quantile can be substituted to match the
operator's risk tolerance.

Notably, the percentage evaluation cost reduction is similar across
formulations ($\sim$12--13\%), suggesting that the benefit of tail-risk hedging
is largely independent of the structural decomposition and attributable
primarily to the training target itself.
However, there are diminishing marginal returns: the planning cost increase
required to achieve the tail-hedging benefit grows with the aggressiveness of
the quantile, and pushing beyond the 95th percentile is expected to become
economically inefficient, a regime illustrated by the extreme conservatism
failure of \textbf{\ac{D-EX}-95}.

\subsection{Extensive Formulation vs.\ Two-Stage Decomposition}
\label{sec:discussion_ex_vs_2sp}

The extensive formulation (\ac{EX}) and the two-stage decomposition (\ac{2SP})
represent fundamentally different solutions to the integration of \ac{TA},
\ac{MS}, and \ac{PdM} under uncertainty.
The \ac{EX} model optimizes all five decision variable types simultaneously
(Section~\ref{sec:problem_formulation}), capturing the full coupling between
fleet positioning and \ac{MS}.
The \ac{2SP} model separates \ac{TA} (Stage~1) from \ac{MS}
(Stage~2), relaxing this coupling at the cost of solution quality.

The results confirm this theoretical prediction empirically.
Among the best-performing pair, \textbf{\ac{S-EX}-95} achieves a total cost
of 742{'}647 versus 762{'}750 for \textbf{\ac{S-2SP}-95}, a difference of
20{'}103 (2.6\%).
This quality gap, while modest in percentage terms, corresponds to a
substantial absolute cost and is statistically robust
(Appendix~\ref{app:statistical_tests_detailed}).
The gap is driven primarily by differences in evaluation-stage cost:
\textbf{\ac{S-EX}-95} achieves 292{'}512 versus 244{'}440 for
\textbf{\ac{S-2SP}-95}. Interestingly, \textbf{\ac{S-2SP}-95} achieves
lower evaluation cost because it allocates more maintenance
conservatively in Stage~2 (planning cost 518{'}309 vs.\ 450{'}134).
The extensive formulation identifies a more balanced optimum: it accepts
somewhat higher evaluation cost but achieves substantially lower planning cost,
yielding a superior total.

Despite this solution-quality advantage, the computational picture decisively
favours decomposition at scale.
\textbf{\ac{S-EX}-95} requires 6.53\,s per instance versus 0.74\,s for
\textbf{\ac{S-2SP}-95}, which constitutes a factor of 9$\times$ at the base configuration of
5 aircraft and a 3-day horizon.
For larger fleets and horizons (Section~\ref{sec:results_scalability}), the
\ac{EX} formulation rapidly becomes intractable while \ac{2SP} remains
feasible.
This creates a practical deployment decision: \textbf{\ac{S-EX}-95} should be
preferred when solution quality matters and the problem size is within the
tractability regime (approximately $\leq$10 aircraft, $\leq$7 days). In turn,
\textbf{\ac{S-2SP}-95} is the recommended choice for larger instances, where
it provides consistent uncertainty-aware improvements over deterministic
baselines despite accepting some planning suboptimality.

Finally, the mean-based variants (\textbf{\ac{S-EX}} and \textbf{\ac{S-2SP}}) confirm
that adding any stochastic \ac{NN} guidance improves outcomes over purely
deterministic planning. Total cost reductions are 2.9\% and 0.28\%,
respectively, with the smaller gain in the \ac{2SP} setting reflecting the
more limited scope for the \ac{NN} to influence Stage~1 decisions.

The sensitivity of results to the \ac{NN} weight $\lambda$ is reported in
Appendix~\ref{sec:results_lambda}, which shows a sharp activation near
$\lambda \approx 0.3$, an optimal total cost at $\lambda = 1.0$, and
diminishing returns beyond $\lambda \approx 1.5$.

\subsection{Scalability, Deployment, and Fleet Decomposition}
\label{sec:discussion_scalability}

Section~\ref{sec:scalability_regimes} analyses the scalability of the developed framework across fleet size, planning horizon, and component count.
Section~\ref{sec:deployment_guidance} defines which formulation should be used depending on the operational scale.
Finally, Section~\ref{sec:discussion_decomposition} proposes a reinforcement-learning-assisted fleet decomposition strategy as a path toward deployment at full airline scale.

\subsubsection{Scalability Regimes} \label{sec:scalability_regimes}

The three-dimensional scalability analysis
(Section~\ref{sec:results_scalability}) identifies distinct tractability
regimes for each formulation.
Fleet size is the most critical dimension:
\ac{EX} solve times grow from 0.06\,s at 2 aircraft to 62.86\,s at 10, and
exceed practical limits (600\,s) at 15+ aircraft
(Table~\ref{tab:scalability_aircraft}).
\ac{2SP} scales markedly better, remaining below 1.1\,s for all tested
fleet sizes up to 20 aircraft.
Planning horizon shows a similar pattern:
\ac{EX} becomes intractable beyond approximately 7 days
(Table~\ref{tab:scalability_days}), while \ac{2SP} remains below 100\,s at
20 days.
Component count poses no computational barrier: neither formulation
shows meaningful sensitivity to the number of monitored health indicators
across 1 to 80 predictors (Table~\ref{tab:scalability_predictors}), confirming
that the primary scalability bottleneck is the combinatorial structure of
\ac{TA} and \ac{MS}, not the \ac{PdM} constraints.

\subsubsection{Deployment Guidance for Current Framework} \label{sec:deployment_guidance}

Synthesising these findings, the following deployment guidance applies to the
present framework:
\begin{itemize}
    \item \textbf{Small fleets ($\leq 7$ aircraft) and short horizons
          ($\leq 5$ days):} \textbf{\ac{S-EX}-95} is preferred for its
          superior solution quality and remains within real-time feasibility.
    \item \textbf{Medium fleets (8--20 aircraft) or longer horizons (up to
          20 days):} \textbf{\ac{S-2SP}-95} is the only tractable option but
          still provides consistent uncertainty-aware benefits (total cost
          762{'}750, vs.\ 1{'}368{'}254 for the real-world baseline).
    \item \textbf{Base configuration (5 aircraft, 3-day horizon):} Both
          formulations are within real-time feasibility (\ac{EX}: 6.53\,s;
          \ac{2SP}: 0.74\,s), allowing operators to use the higher-quality
          \textbf{\ac{S-EX}-95} without computational concern.
    \item \textbf{Prognostic instrumentation:} The number of monitored
          components can be increased freely to match available sensor
          infrastructure without impacting solve time.
\end{itemize}

\subsubsection{Fleet Decomposition for Large-Scale Deployment}
\label{sec:discussion_decomposition}

For a realistic single-fleet deployment, such as the 30 Airbus A220 aircraft operated by SWISS, 
neither the \ac{EX} nor the \ac{2SP} formulation is directly applicable in its current form.
The \ac{2SP} remains computationally feasible but sacrifices the joint
optimisation quality that makes \textbf{\ac{S-EX}-95} superior.
A natural extension is to retain \textbf{\ac{S-EX}-95} as the
solution engine but apply it to sub-fleets of 5--10 aircraft, thereby
staying within the tractability regime while covering the full fleet.

The key challenge in such a decomposition is how to partition the full fleet:
a naive random split ignores the network structure of flight routes and
maintenance slot availability, leading to inter-group conflicts (e.g., two
sub-fleets competing for the same limited maintenance facility).
We propose that a \ac{RL} agent is a natural fit
for the partitioning decision.
The \ac{RL} agent observes the current fleet state, aircraft locations, \ac{RUL}
levels, scheduled flight network, and maintenance slot availability, and
selects a partition of the fleet into sub-groups.
Each sub-group is then solved independently with \textbf{\ac{S-EX}-95},
and the global schedule is assembled from the sub-group solutions.
The \ac{RL} agent is rewarded based on the total cost of the assembled schedule
(planning + evaluation), and penalised for inter-group conflicts such as
double-booking of maintenance slots or violations of network continuity
constraints.
Over training, the agent learns partitioning strategies that balance within-group
optimisation quality against between-group coordination overhead.

This \ac{RL}-assisted decomposition framework has several attractive properties.
First, it preserves the theoretical guarantees of \textbf{\ac{S-EX}-95}
within each sub-group, since the full extensive formulation is applied to each
partition.
Second, the \ac{RL} agent can adapt the partition dynamically to the specific
problem instance, for example, grouping aircraft that share maintenance
facilities or that operate on the same route cluster.
Third, the approach scales linearly in the number of sub-groups, making it
applicable to very large fleets at the cost of inter-group coordination quality.
Finally, the \ac{RL} agent can be pre-trained offline on historical planning
instances and deployed at near-zero inference cost, adding only milliseconds
to the overall planning pipeline.

This architecture represents a promising direction for bringing the quality of
joint stochastic optimisation to airline-scale operations, and is identified as
a primary avenue for future work (Section~\ref{sec:discussion_future}).

\subsection{Robustness to RUL Uncertainty Characteristics}
\label{sec:discussion_robustness}

Three sensitivity experiments probe the behaviour of the framework at the
boundaries of the assumed uncertainty model
(Appendix~\ref{sec:results_sensitivity}).
Under high uncertainty ($\sigma = 50$ cycles), \textbf{\ac{S-EX}-95}
performs even better than in the base case: evaluation-stage cost is reduced by
54.4\% relative to \textbf{\ac{D-EX}}, at a planning cost increase of 28.6\%,
for a net total cost reduction of approximately 26\%.
This confirms that the framework becomes more valuable as uncertainty increases,
which is precisely the operational regime where ad-hoc deterministic planning
is most likely to fail.

Under zero uncertainty ($\sigma = 0$), the \ac{NN} provides no benefit:
planning and evaluation costs change by only $+1.54\%$ and $+0.69\%$
respectively, confirming that \textbf{\ac{S-EX}-95} correctly degenerates to
a near-deterministic solution when uncertainty is absent.
The dramatically increased solve time in this regime ($+1{'}624\%$, to 92.16\,s)
is a practical caution: for deterministic problem instances, the standard
\ac{MILP} should be solved directly.

Under negligible risk (RUL $\in [200, 250]$ cycles), where no
\ac{RUL} violations are realistically possible within the planning window, both
variants yield identical outcomes, and the \ac{NN} contribution vanishes
gracefully.
Taken together, these experiments confirm that the framework is well-calibrated
to the level of uncertainty present: it exploits uncertainty when it exists, is
neutral when uncertainty is absent, and imposes negligible overhead when risk is
operationally irrelevant.

\subsection{Sample Efficiency and Practical Training Requirements}
\label{sec:discussion_sample_efficiency}

A key practical concern for any learning-based optimisation method is the
amount of training data required to achieve reliable performance.
The sample efficiency experiment (Appendix~\ref{sec:results_sample_efficiency})
reveals a surprisingly low data threshold: a sharp transition occurs between
$N = 100$ and $N = 200$ total instances, after which evaluation cost reduction
stabilises in the range $-48\%$ to $-53\%$ and shows only modest further
improvement.
At the default configuration of $N = 2{'}000$ instances, \textbf{\ac{S-EX}-95}
achieves $-51.1\%$ evaluation cost reduction and $-50.7\%$ fewer cancellations, 
less than four percentage points better than the $N = 200$ threshold.

This result has important practical implications.
Airlines with limited historical \ac{PdM} data can reach near-peak robustness
with as few as 140 training instances (70\% of $N = 200$), making the framework
accessible even at the early stages of \ac{PdM} data collection.
The training cost is dominated by the offline simulation of evaluation scenarios
rather than solver time, and is fully parallelisable.
The marginal cost of generating additional training data beyond $N \approx 1{'}000$
is unlikely to be justified by the marginal benefit in most operational settings.

\subsection{Limitations}
\label{sec:discussion_limitations}

Several limitations of the present study should be acknowledged:

\textbf{Synthetic \ac{RUL} uncertainty:}
The framework models prognostic uncertainty as independent, identically
distributed Gaussian noise with a fixed standard deviation $\sigma$.
Real prognostic distributions are typically asymmetric, heavy-tailed, and
exhibit cross-component correlations (e.g., components sharing an operating
environment tend to degrade together).
Capturing these distributional features would require more sophisticated
uncertainty quantification models and may alter the optimal training target
quantile.

\textbf{Simplified reactive policy:}
The evaluation-stage heuristic (Section~\ref{sec:eval_heuristic}) permits
only flight cancellations and reactive maintenance, and does not model aircraft
swapping, crew reassignment, or network-wide cascading.
Real recovery policies are substantially more flexible, and the true cost of
a \ac{RUL} violation is likely lower than estimated by this heuristic.
This conservatism in the evaluation model may overstate the value of stochastic
planning, though it does not alter the relative ranking of configurations.

\textbf{Two-stage decoupling and the missing tail-assignment back-loop:}
In the \ac{2SP} formulation, Stage~2 \ac{MS} is solved after Stage~1 \ac{TA}
and therefore inherits fixed aircraft routes.
When Stage~2 cannot accommodate all flights under these fixed assignments, it may resort to
planning-stage cancellations to restore feasibility.
In practice, a maintenance planner would trigger an aircraft-swap back-loop to the \ac{TA}
team and revise the assignment rather than accept cancellation.
This feedback loop is not modeled in the current framework, so the observed \ac{2SP} planning
cancellations (1.11--1.26 per instance in Table~\ref{tab:planning_eval_combined}) should be
interpreted as a decoupling artifact.
By contrast, \ac{EX} jointly optimizes assignment and maintenance and largely avoids this
behaviour (0.32--0.33 planning cancellations), with the small residual reflecting real
data-driven infeasibilities rather than a modelling shortcoming.

\textbf{Experimental scale:}
Validation was conducted on a 5-aircraft, 3-day base configuration, with
scalability assessed up to 20 aircraft and 20 days.
While these experiments demonstrate tractability trends, direct evaluation on
full-scale airline networks (50--200 aircraft, 7--14 day horizons) remains
an open challenge requiring the decomposition strategies discussed in
Section~\ref{sec:discussion_decomposition}.

\textbf{Deterministic flight schedules:}
The framework treats the flight schedule and maintenance slot availability as
deterministic inputs.
In practice, these are subject to weather disruptions, demand fluctuations, and
crew availability constraints, all of which introduce additional uncertainty
sources beyond \ac{RUL} prognostics.

\subsection{Future Research Directions}
\label{sec:discussion_future}

Four directions are identified for future work.

\textbf{Scalability and deployment:}
The most immediate priority is extending the framework to full single-fleet scale.
Decomposition approaches that retain \textbf{\ac{S-EX}-95} as the solution engine for tractable sub-groups offer a natural path forward, with both learning-based methods (e.g., \ac{RL} for adaptive fleet partitioning) and structured heuristics as viable directions.

\textbf{Richer uncertainty modelling:}
The current Gaussian \ac{RUL} assumption should be replaced by asymmetric, heavy-tailed, and cross-component correlated prognostic distributions derived from real prognostic model outputs.
The framework should further incorporate additional operational uncertainty sources, including weather disruptions, demand variability, and crew availability, toward a more fully integrated stochastic planning model.

\textbf{Higher-fidelity evaluation and training:}
Replacing the parsimonious evaluation heuristic with a simulation that captures aircraft swapping, crew reassignment, and network cascading would improve training label accuracy and provide more realistic performance estimates.

\textbf{Rolling-horizon planning and broader validation:}
Extending to rolling-horizon replanning, where the plan is updated as new \ac{RUL} sensor readings arrive, and validating across diverse airline networks and fleet types would establish the generalisability of the findings.

\textbf{Quantifying the operational value of prognostic model quality:}
The proposed framework creates a principled basis for assessing the downstream economic value of prognostic models, since planning costs and operational outcomes are directly linked to \ac{RUL} inputs. A natural extension is to systematically vary the accuracy and uncertainty calibration of the prognostic model, and measure the resulting impact on planning cost, cancellation rate, and total operational cost. Such an analysis would provide guidance on how much prognostic improvement translates into operational benefit, and at what level of predictive accuracy further investment yields diminishing returns.

%% file: 08_conclusion.tex
\section{Conclusion}
\label{sec:conclusion}

This paper developed and validated an integrated stochastic optimisation framework for joint aircraft \ac{TA}, \ac{MS}, and \ac{PdM} under \ac{RUL} uncertainty. By embedding a \ac{NN} approximation of expected evaluation-stage costs directly into a \ac{MILP} formulation, 
the framework enables computationally tractable uncertainty-aware planning without explicit scenario enumeration at solve time. Validation on 500 held-out planning instances derived from real operational data of a European short-haul airline demonstrates consistent and substantial improvements over both deterministic baselines and current industry practice.

The framework makes four principal contributions.
First, an integrated \ac{MILP} model unifies \ac{TA}, \ac{MS}, and \ac{PdM} decisions within a single optimisation, capturing the interdependencies between fleet utilisation and aircraft health that are lost when these problems are solved sequentially or in isolation. Second, two stochastic formulations, the extensive form (\ac{EX}) and a two-stage decomposition (\ac{2SP}), explicitly model \ac{RUL} prognostic uncertainty and generate plans that hedge against adverse realisations across a sampled scenario set.
Third, a \ac{NN} trained on the 95th-percentile evaluation cost target 
is embedded into the \ac{MILP} objective, steering the optimiser away from plans that are vulnerable to high-cost tail scenarios while avoiding the prohibitive conservatism of worst-case point-estimate planning. Fourth, a comprehensive empirical evaluation, spanning scalability experiments across fleet size, planning horizon, and component count, as well as sensitivity analyses on uncertainty magnitude and training sample size 
provides actionable deployment guidance across a wide range of operational settings.

The experimental results are consistent and compelling. The best-performing configuration, \textbf{\ac{S-EX}-95}, achieves a total cost of 742{'}647, representing a 4.1\% reduction over the deterministic extensive baseline (\textbf{\ac{D-EX}}: 774{'}769) and a 45.7\% reduction over the real-world airline baseline (1{'}368{'}254).
The improvement is driven by a 21.9\% decrease in evaluation-stage cost (292{'}512 vs.\ 374{'}910 for \textbf{\ac{D-EX}}) and a 66.8\% reduction in cancellations relative to actual operations (5.77 vs.\ 17.42 per instance), at a modest planning cost premium of 12.6\%.
Training the \ac{NN} on the 95th-percentile target consistently outperforms mean-based training, reducing evaluation cost by a further 12.8\% within the extensive formulation, with a similar 12.2\% gain within the two-stage decomposition.
The failure of deterministic 95th-percentile planning (\textbf{\ac{D-EX}-95}), which incurs an 89\% planning cost increase to achieve near-zero evaluation cost yet yields a higher total cost than the plain deterministic baseline, illustrates precisely the pathology that learned cost approximation avoids: the \ac{NN} learns which scenarios are actually costly, not merely which are technically infeasible. 
Near-peak robustness is attainable with as few as 140 training instances, making the framework accessible at early stages of \ac{PdM} data collection. Scalability experiments confirm that component count poses no computational barrier, that \textbf{\ac{S-EX}-95} is tractable up to approximately 10 aircraft and 7-day horizons, and that \textbf{\ac{S-2SP}-95} remains feasible across all tested fleet sizes up to 20 aircraft and  planning horizons up to 20 days.

Several limitations accompany these results. The uncertainty model assumes independent Gaussian \ac{RUL} noise, the evaluation-stage heuristic permits only cancellations and reactive maintenance,
the deterministic 95th-percentile configurations produce operationally unjustifiable upfront cancellations despite low evaluation-stage costs, and all experiments are conducted at sub-fleet scale relative to full airline operations.
These limitations point directly to the most productive directions for future research.

More broadly, this work contributes to an emerging need in the prognostics and health management community: frameworks capable of evaluating the business value of prognostic models in operational decision-making. The practical usefulness of a predictor depends not only on \ac{RUL} estimation accuracy but on the quality and calibration of its uncertainty estimates; a well-calibrated prognostic model enables more targeted maintenance decisions and lower downstream operational costs than an accurate but poorly calibrated one. The proposed framework provides a principled basis for quantifying this value by linking prognostic uncertainty representations directly to measurable operational outcomes.

In a real industrial setting, the framework is intended to operate as a decision-support tool within the short-term planning cycle. Given the current flight schedule, component \ac{RUL} distributions from on-board health monitoring systems, and maintenance slot availability, the system produces an optimised \ac{TA} and maintenance schedule that explicitly hedges against prognostic uncertainty. The low training data requirement of approximately 140 instances and the availability of a scalable two-stage decomposition variant position the framework as a practical step toward uncertainty-aware, data-driven airline operations management.

%% file: 09_acknowledgement_coi_data_and_code.tex
\section*{Acknowledgments}
This research did not receive any specific grant from funding agencies in the public, commercial, or not-for-profit sectors. The authors thank Swiss International Air Lines Ltd. for providing access to operational flight and maintenance data and for their domain expertise, which made this study possible.

\section*{Conflict of Interest}
The authors declare no known competing financial interests or personal relationships that could have appeared to influence the work reported in this paper. For transparency, we note that the corresponding author, Benno K\"aslin, is employed by Swiss International Air Lines Ltd., the airline that provided the operational data used in this study.\newpage

\section*{Data and Code Availability}
The operational flight and maintenance data used in this study were provided by Swiss International Air Lines Ltd. under a confidentiality agreement and cannot be made publicly available due to commercial sensitivity. The code implementing the proposed framework is not publicly released.

%% file: 10_appendix.tex
\section*{Appendix}

\subsection{Feature Engineering Details}
\label{app:feature_engineering}

\subsubsection{Feature Extraction Framework}

The feature engineering framework performs a dimensionality reduction from
variable-sized optimisation instances (which depend on problem parameters:
$|\mathcal{P}|$ aircraft, $|\mathcal{F}|$ flights, $|\mathcal{G}|$ maintenance
tasks, $|\mathcal{T}|$ time periods) to a fixed-dimensional feature vector
$\mathbf{z} \in \mathbb{R}^{n_{\text{features}}}$ suitable for \ac{NN}
input.
This transformation is necessary because the \ac{NN} requires constant-sized
input vectors while the underlying optimisation problem instances vary in scale.

Let $\mathbf{x} = \{F_{p,f},\, C_f,\, G_{p,n,t},\, T_{g,s},\, M_{s,p}\}$
denote the decision variable values from a solved optimisation instance.
The feature map $\mathcal{F}: \mathbb{R}^{n_{\text{vars}}} \to
\mathbb{R}^{n_{\text{features}}}$ constructs a fixed-dimensional representation
by computing aggregates, statistics, and derived quantities from $\mathbf{x}$:
\begin{align}
\mathbf{z} = \mathcal{F}(\mathbf{x}) = [\mathbf{z}_1, \mathbf{z}_2, \ldots,
\mathbf{z}_k]^\top
\end{align}
where $k$ denotes the number of enabled feature groups.

\subsubsection{Feature Groups and Mathematical Definitions}

\textbf{Solution Summary Statistics} (config key: \texttt{global}) aggregate
plan-level metrics that characterise overall feasibility and cost structure:
\begin{align}
\mathbf{z}_{\text{solution-summary}} = [n_C,\; n_F,\; n_M,\; n_T]^\top
\end{align}
where $n_C = \sum_{f} C_f$, $n_F = \sum_{p,f} F_{p,f}$,
$n_M = \sum_{s,p} M_{s,p}$, and $n_T = \sum_{g,s} T_{g,s}$ represent the
total number of cancelled flights, assigned flights, assigned maintenance slots,
and scheduled maintenance tasks, respectively.
Large cancellation counts $n_C$ directly reflect plan infeasibility and
expected evaluation cost.

\textbf{Initial Component Health States} (config key: \texttt{base\_rul})
capture the initial degradation state and prediction uncertainty per
aircraft–component pair at the start of the planning horizon:
\begin{align}
\mathbf{z}_{\text{health}} = \bigl[
  \mathrm{RUL}_{e,p,0},\;\; \sigma(\mathrm{RUL}_{e,p,0})
\bigr]_{e \in \mathcal{E},\; p \in \mathcal{P}}
\end{align}
where $\mathrm{RUL}_{e,p,0}$ is the initial \ac{RUL} of component $e$ on
aircraft $p$ at $t = 0$ and $\sigma(\mathrm{RUL}_{e,p,0})$ is its predictive
standard deviation.
These features encode the health information available at the planning stage and
allow the network to anticipate which aircraft–component pairs are most likely
to drive \ac{RUL} violations and reactive costs.

\textbf{Daily Maintenance–Demand Balance}
(config key: \texttt{cp\_pressure\_daily}) directly models the operational
stress experienced by each aircraft on each planning day:
\begin{align}
\mathbf{z}_{\text{maintenance-demand}} = \bigl[
  \mathrm{Capacity}_{p,d} \;=\;
  \textstyle\sum_{t \in \mathcal{T}_{p,d}} \mathrm{RUL\_recover}(t)
  - |F_{p,d}|
\bigr]_{p \in \mathcal{P},\; d \in \mathcal{D}}
\end{align}
where $\sum_{t \in \mathcal{T}_{p,d}} \mathrm{RUL\_recover}(t)$ is the total
\ac{RUL} recovery from scheduled maintenance tasks on aircraft $p$ on day $d$,
and $|F_{p,d}|$ is the number of flights assigned.
A strongly negative value signals an accumulated \ac{RUL} deficit and thus
high cancellation risk, while a positive value indicates surplus maintenance
capacity.
This group ($|\mathcal{P}| \times |\mathcal{D}|$ features) is the most
informative for predicting evaluation-stage cost, as it directly quantifies
the primary constraint trade-off in the optimisation model.

\textbf{Daily Fleet-Wide Aggregates}
(config key: \texttt{cpl\_daily\_counts}) provide per-day aggregate metrics
across the entire fleet:
\begin{align}
\mathbf{z}_{\text{daily-aggregates}} = \bigl[
  |F_d|,\; |T_d|,\; |M_d|,\; |F_{\text{hub},d}|
\bigr]_{d \in \mathcal{D}}
\end{align}
where $|F_d|$, $|T_d|$, $|M_d|$ are the daily totals of assigned flights,
maintenance tasks, and maintenance slots, respectively, and
$|F_{\text{hub},d}|$ is the number of flights arriving at a hub airport on
day $d$.
These features capture temporal scheduling patterns that influence feasibility
and realisation cost beyond what is captured by daily per-aircraft features.

\subsubsection{Disabled Feature Groups: Design Rationale}

Eight additional feature groups are available but not enabled in the primary
experimental configuration.

\textbf{Normalised Fleet Metrics} (\texttt{avg}):
$[n_F / |\mathcal{P}|, n_M / |\mathcal{P}|, n_T / |\mathcal{P}|]^\top$.
Disabled because normalisation by fleet size introduces redundancy with the
raw solution summary statistics; the relationship is deterministic and does not
improve validation loss.

\textbf{Network Concentration Index} (\texttt{hub\_exposure}):
$n_{\text{hub}} = |\{f \in \mathcal{F} : \mathrm{arrival}(f) \in
\mathcal{H}\}|$.
Disabled because the benchmark instances are hub-balanced; this feature would
become important for hub-and-spoke networks with pronounced routing asymmetry.

\textbf{Daily Flight Load} (\texttt{daily\_flight}):
$[n_{F,d}]_{d \in \mathcal{D}}$.
Disabled because the same information is already present in the daily
fleet-wide aggregates group ($|F_d|$ component), making this redundant.

\textbf{Daily Maintenance Availability} (\texttt{slot\_task}):
$[n_{M,p,d}]_{p,d}$.
Disabled because \ac{RUL}-based recovery (captured in
\texttt{cp\_pressure\_daily}) is a more direct predictor of feasibility than
raw slot counts.

\textbf{Daily Maintenance Recovery Capacity} (\texttt{cp\_rul\_recovery}):
$[\sum_{t \in \mathcal{T}_{p,d}} \mathrm{RUL\_recover}(t)]_{p,d}$.
Disabled because this quantity is the numerator of the capacity balance
already computed in \texttt{cp\_pressure\_daily}; including both introduces
redundancy.

\textbf{Cumulative Maintenance Deficit} (\texttt{cp\_cumulative}):
$[\sum_{d'=0}^{d} \mathrm{Capacity}_{p,d'}]_{p,d \geq 1}$.
On the 3-day planning horizon used in primary experiments, cumulative effects
are not substantially different from daily snapshots and add $|\mathcal{P}|
\times (|\mathcal{D}|-1)$ features for marginal gain.

\textbf{Daily Slot–Demand Pressure} (\texttt{cp\_slots\_daily}):
$[|M_{p,d}| - |F_{p,d}|]_{p,d}$.
A slot-count proxy for capacity that is semantically inferior to the
\ac{RUL}-based balance when prognostic data is available.

\textbf{Per-Aircraft Daily Flight Distribution}
(\texttt{cpl\_ac\_flights}):
$[|F_{p,d}|]_{p,d}$.
The same information is implicitly encoded in the maintenance–demand balance
features; empirically, adding this group increased dimensionality by
$|\mathcal{P}| \times |\mathcal{D}| = 15$ without improving validation loss.

\subsubsection{Feature Vector Dimensionality}

The total feature dimension is:
\begin{align}
n_{\text{features}} = \sum_{g \in \text{enabled groups}} n_{\text{features}}(g)
\end{align}

For the primary configuration ($|\mathcal{P}| = 5$ aircraft,
$|\mathcal{D}| = 3$ days, $|\mathcal{E}| \times |\mathcal{P}| = 5$ health
predictors):

\begin{table}[h]
\centering
\small
\caption{Feature dimensionality per group (primary experimental configuration).}
\label{tab:feature_dimensions}
\begin{tabular}{lrr}
\toprule
\textbf{Feature Group} & \textbf{Key} & \textbf{Dimension} \\
\midrule
Solution Summary Statistics       & \texttt{global}             & 4 \\
Initial Component Health States   & \texttt{base\_rul}          & $2 \times 5 \times 5 = 50$ \\
Daily Maintenance–Demand Balance & \texttt{cp\_pressure\_daily}& $5 \times 3 = 15$ \\
Daily Fleet-Wide Aggregates       & \texttt{cpl\_daily\_counts} & $4 \times 3 = 12$ \\
\midrule
\textbf{Total}                    &                             & \textbf{81} \\
\bottomrule
\end{tabular}
\end{table}

The 81-dimensional vector represents a deliberate trade-off between
expressiveness and efficiency.
Ablation experiments showed that removing the component health group
(\texttt{base\_rul}) significantly decreased validation accuracy, while
expanding beyond 81 features (e.g., adding cumulative variants or per-aircraft
flight distributions) provided less than 1 percent validation MSE improvement at the
cost of increased overfitting risk and longer \ac{MILP} solve times due to
additional linearisation variables.
The selected feature set achieved validation MSE of approximately 0.05 on
normalised cost, confirming sufficient predictive accuracy for the embedded
cost approximation.

\subsubsection{Feature Extraction Paths: Training and Inference}

The feature map $\mathcal{F}$ operates in two distinct contexts that must
produce identical outputs to ensure consistency between offline training and
online deployment.

\textbf{Training Path.}
During offline \ac{NN} training, features are extracted as numerical values
from solved \ac{D-EX} instances.
After solving the deterministic \ac{MILP} (Eq. \ref{eq:deterministic}) to
obtain plan $\bar{\mathbf{x}}$, binary variable values are extracted using a
threshold of 0.5 to handle numerical precision, and the feature map
$\mathcal{F}$ is applied to compute $\mathbf{z} \in \mathbb{R}^{81}$.
Training labels are constructed by evaluating each plan across $K = 100$
sampled \ac{RUL} scenarios using the evaluation heuristic
(Section \ref{sec:eval_heuristic}), yielding either the mean target
$\bar{Q}^{\text{mean}}$ (Eq. \ref{eq:target_mean}) or the 95th-percentile
target $\bar{Q}^{95}$ (Eq. \ref{eq:target_95p}).
The network is trained to minimise $\mathcal{L} = \mathbb{E}[(\mathrm{NN}
(\mathbf{z}) - \bar{Q}^{\bullet})^2]$ over the training dataset.

\textbf{Inference Path.}
During online planning, features must be extracted as linear expressions over
the decision variables so that the entire \ac{NN} can be embedded into the
\ac{MILP}.
Each feature aggregate is built as a \texttt{LinExpr} object in Gurobi; for
example:
\begin{align}
\mathrm{expr}_{n_F} = \sum_{p,f} F_{p,f}
\end{align}
yielding a symbolic feature vector $\mathbf{z}_{\text{expr}} \in
\mathbb{R}^{81}$ of linear expressions.
This vector is normalised using training-set statistics
$(\boldsymbol{\mu}_{\text{train}},\, \boldsymbol{\sigma}_{\text{train}})$:
\begin{align}
\mathbf{z}'_{\text{expr}} =
(\mathbf{z}_{\text{expr}} - \boldsymbol{\mu}_{\text{train}})
\oslash \boldsymbol{\sigma}_{\text{train}}
\end{align}
where $\oslash$ denotes elementwise division, implemented as coefficient
scaling (multiplication by $1/\sigma$) to preserve linearity.
Each \ac{ReLU} activation $y = \max(0, \mathrm{LinExpr})$ is then linearised
via auxiliary binary variables $\delta \in \{0,1\}$:
\begin{align}
y = \max(0, \mathrm{LinExpr}) \quad \Rightarrow \quad
\begin{cases}
  y \geq \mathrm{LinExpr} \\
  y \geq 0 \\
  y \leq M \cdot \delta + b \cdot (1 - \delta)
\end{cases}
\end{align}
where $M$ is a sufficiently large constant (big-M) and $b$ is the expression's
maximum possible value.
The linearised \ac{NN} output is incorporated directly into the \ac{S-EX}
objective (Eq. \ref{eq:stochastic_nne}), and the augmented \ac{MILP} is passed
to Gurobi for joint optimisation.

\subsubsection{Feature Design Rationale}

\textbf{Dimensionality reduction and scale invariance.}
The feature map reduces the decision vector from dimension
$\mathcal{O}(|\mathcal{P}| \cdot |\mathcal{F}| + |\mathcal{G}| \cdot
|\mathcal{S}|)$ to the fixed dimension $\mathcal{O}(|\mathcal{P}| \cdot
|\mathcal{D}|)$, enabling a fixed-architecture network to operate across
problem instances of varying scale.

\textbf{Information preservation.}
Despite aggressive dimensionality reduction, the four enabled groups collectively
encode the primary drivers of evaluation-stage cost: overall plan feasibility
(solution summary), fleet health at plan start (component health states), the
per-aircraft maintenance–RUL trade-off trajectory (daily maintenance–demand
balance), and temporal scheduling patterns (daily fleet-wide aggregates).

\textbf{Linearity and exact embedding.}
All features are linear aggregations of binary or continuous decision variables,
guaranteeing that $\mathcal{F}(\mathbf{x})$ can be represented exactly as a
vector of \texttt{LinExpr} objects and embedded without approximation into the
\ac{MILP}.

\textbf{Interpretability.}
Each feature group has a clear operational interpretation, enabling validation
against domain expertise and facilitating debugging when deployed on novel
problem instances.

\subsection{Neural Network Configuration Selection}
\label{app:nn_conf_selection}

The \ac{NN} architecture comprises two components: a Scenario Embedding Network
that encodes the sampled \ac{RUL} scenario set into a fixed-length vector
outside the \ac{MILP}, and a \ac{ReLU} Predictor Network that takes the
concatenation of the scenario embedding and the plan feature vector
$\mathbf{z}$ as input and is embedded directly into the \ac{MILP} via
\ac{ReLU} linearisation.
The architecture search described here was conducted with an earlier cost
structure; absolute cost values therefore differ from the primary results
reported in Section \ref{sec:results}.
However, the relative scaling behaviour - solve times as a function of network
size and the absence of performance gains from larger predictor
networks - remains valid and directly motivated the selected configuration.

\textbf{Scenario Embedding Network.}
All combinations of $\{512, 256\}$ units for the first layer and
$\{512, 256, 128\}$ units for each of the second and third layers were
evaluated.
Training performance showed no consistent improvement with larger encoder
configurations.
Since the Scenario Embedding Network is evaluated outside the \ac{MILP}
solver and does not affect \ac{MILP} solve time, the configuration with the
best training performance was selected: $[512, 128, 64]$ units with \ac{ReLU}
activations.

\textbf{\ac{ReLU} Predictor Network.}
The following layer configurations were evaluated:
$[8, 8]$, $[16, 16]$, $[32, 32]$, $[64, 64]$, and $[128, 128]$.
Unlike the Scenario Embedding Network, the Predictor Network is linearised and
embedded directly into the \ac{S-EX} \ac{MILP}, so its size directly governs
solver overhead.
Table \ref{tab:nn_architecture_search} summarises the key trade-off.

\begin{table}[h]
\centering
\small
\caption{Architecture search results for the \ac{ReLU} Predictor Network.
Solve times measured on the \ac{S-EX} formulation; time limit 1{'}200 s.}
\label{tab:nn_architecture_search}
\begin{tabular}{@{}lrrl@{}}
\toprule
\textbf{Predictor Config.} & \textbf{Avg. Solve Time [s]} & \textbf{Within limit?} & \textbf{Relative performance} \\
\midrule
$[8,\; 8]$   &  $\sim$4   & Yes & No additional capacity \\
$[16,\; 8]$  &  $\sim$5   & Yes & Selected (capacity plus efficiency) \\
$[16,\; 16]$ &  $<$10     & Yes & No meaningful gain over [16,8] \\
$[32,\; 32]$ &  151       & Yes & No meaningful gain; 30 times slower \\
$[64,\; 64]$ &  $>$1{'}200 & No & Exceeds time limit \\
$[128,\; 128]$ & $>$1{'}200 & No & Exceeds time limit \\
\bottomrule
\end{tabular}
\end{table}

Optimisation performance (evaluation-stage cost and cancellations) did not
improve substantially with larger predictor networks beyond $[16, 8]$.
The selected $[16, 8]$ configuration offers a balance between expressiveness
and computational efficiency, yielding an average \ac{S-EX} solve time of
approximately 5 s, which is only a 7 percent overhead relative to $[8, 8]$
(4 s) while providing additional network capacity to learn more nuanced cost
patterns. This modest additional computational cost is justified by improved
robustness and remains fully within operational planning requirements.

\subsection{Hyperparameters and Configuration}
\label{app:hyperparameters}

Table \ref{tab:hyperparameters} summarises all key hyperparameters, solver
settings, and configuration values for reproducibility.
Values represent the primary experimental configuration unless otherwise noted.
Justification for feature engineering choices is given in
Appendix \ref{app:feature_engineering}; justification for the \ac{NN}
architecture in Appendix \ref{app:nn_conf_selection}; and sensitivity analyses
for $\lambda$ and training sample size in
Appendices \ref{sec:results_lambda} and \ref{sec:results_sample_efficiency}.

\begin{table}[h]
\centering
\small
\caption{Comprehensive hyperparameter and configuration summary.}
\label{tab:hyperparameters}
\begin{tabular}{@{}lll@{}}
\toprule
\textbf{Component} & \textbf{Parameter} & \textbf{Value} \\
\midrule
\multicolumn{3}{@{}l}{\textbf{Data and Scenario Configuration}} \\
& Total instances                    & 5{'}000 \\
& Training instances                 & 3{'}500 (70 percent) \\
& Validation instances               & 500 (10 percent) \\
& Test instances                     & 500 (10 percent) \\
& Evaluation (hold-out) instances    & 500 (10 percent) \\
& Scenarios per instance ($K$)       & 100 \\
& Random seed method                 & Deterministic per instance \\
\midrule
\multicolumn{3}{@{}l}{\textbf{Feature Engineering}} \\
& Total feature dimension            & 81 \\
& Feature groups employed            & 4 (of 13 available) \\
& Normalisation                      & Zero-mean, unit-variance \\
& Scaling parameters source          & Training set only \\
\midrule
\multicolumn{3}{@{}l}{\textbf{Neural Network Architecture}} \\
& Scenario encoder layers            & 512 - 128 - 64 units, \ac{ReLU} \\
& Predictor network layers           & 16 - 8 units, \ac{ReLU} \\
& Output layer                       & 1 unit (scalar cost) \\
& Activation functions               & \ac{ReLU} only (enables linearisation) \\
\midrule
\multicolumn{3}{@{}l}{\textbf{Neural Network Training}} \\
& Optimiser                          & Adam \\
& Learning rate ($\alpha$)           & $5 \times 10^{-5}$ \\
& Batch size                         & 32 \\
& Loss function                      & Mean squared error \\
& Maximum epochs                     & 5{'}000 \\
& Early stopping patience            & 30 epochs \\
& Min. improvement threshold        & $10^{-5}$ \\
\midrule
\multicolumn{3}{@{}l}{\textbf{Solver Configuration (Gurobi 12.0.3)}} \\
& \ac{MILP} solver                   & Gurobi 12.0.3 \\
& Optimality gap                     & $1 \times 10^{-2}$ \\
& Time limit                         & 600 seconds \\
& Thread limit                       & 8 (default) \\
& Presolve and Heuristics            & Default \\
\midrule
\multicolumn{3}{@{}l}{\textbf{Stochastic Framework Parameters}} \\
& \ac{NN} weight ($\lambda$)         & 1.0 (unless stated otherwise) \\
& Training \ac{MILP} time limit      & 0.5 s per instance \\
\bottomrule
\end{tabular}
\end{table}

\subsection{Training Sample Efficiency}
\label{sec:results_sample_efficiency}

This appendix examines how many training instances are required for the
\ac{NN} to produce schedules that are meaningfully more robust than the
deterministic \ac{D-EX} baseline.
All experiments use the \ac{S-EX} configuration (mean training target $\bar{Q}^{\text{mean}}$)
alongside the \ac{D-EX} deterministic comparison, and the 70/10/10/10
training/validation/test/evaluation split, so that a total of $N$
instances yields $0.7\,N$ training samples.
All other settings match Section \ref{sec:experimental_setup}.
Note that the absolute cost values in this experiment differ from those in the
primary results (Section \ref{sec:results}) because evaluation instance pools
vary with dataset size; conclusions are drawn from relative differences
(percentage columns), which are internally consistent.

\begin{figure}[H]
\centering
\includegraphics[width=0.85\columnwidth]{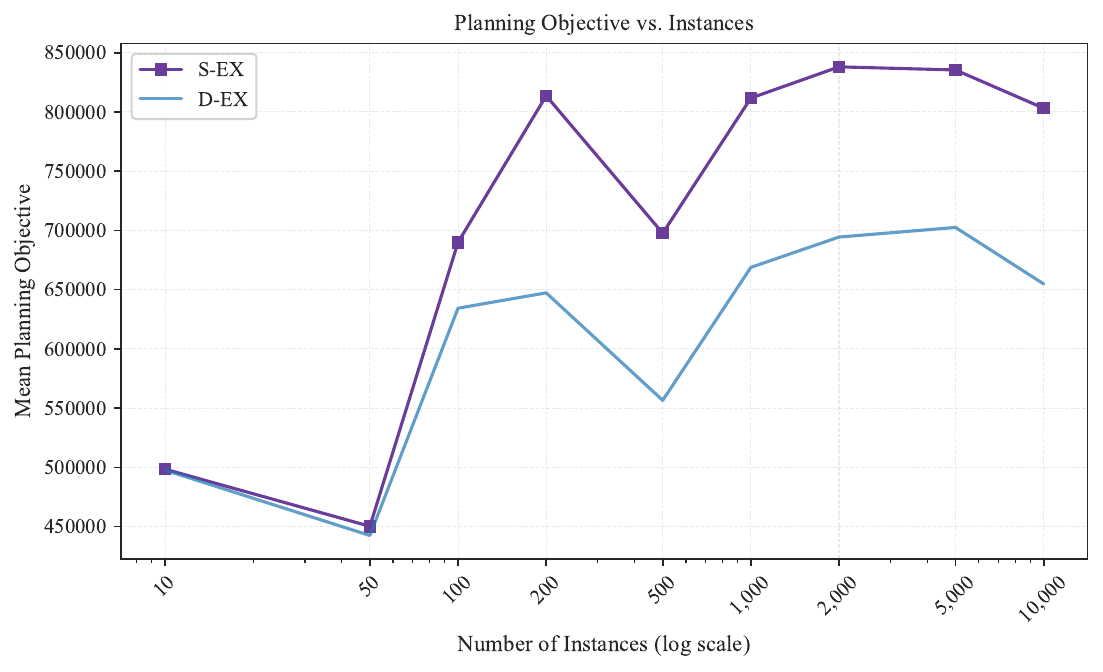}
\caption{Average planning-stage cost as a function of total dataset size
(log scale). The \ac{S-EX} cost (blue) deviates upward around $N = 200$,
reflecting the onset of meaningful \ac{NN}-guided conservatism. Both curves
fluctuate due to sampling variation across instance subsets.}
\label{fig:sample_planning_cost}
\end{figure}

\begin{figure}[H]
\centering
\includegraphics[width=0.85\columnwidth]{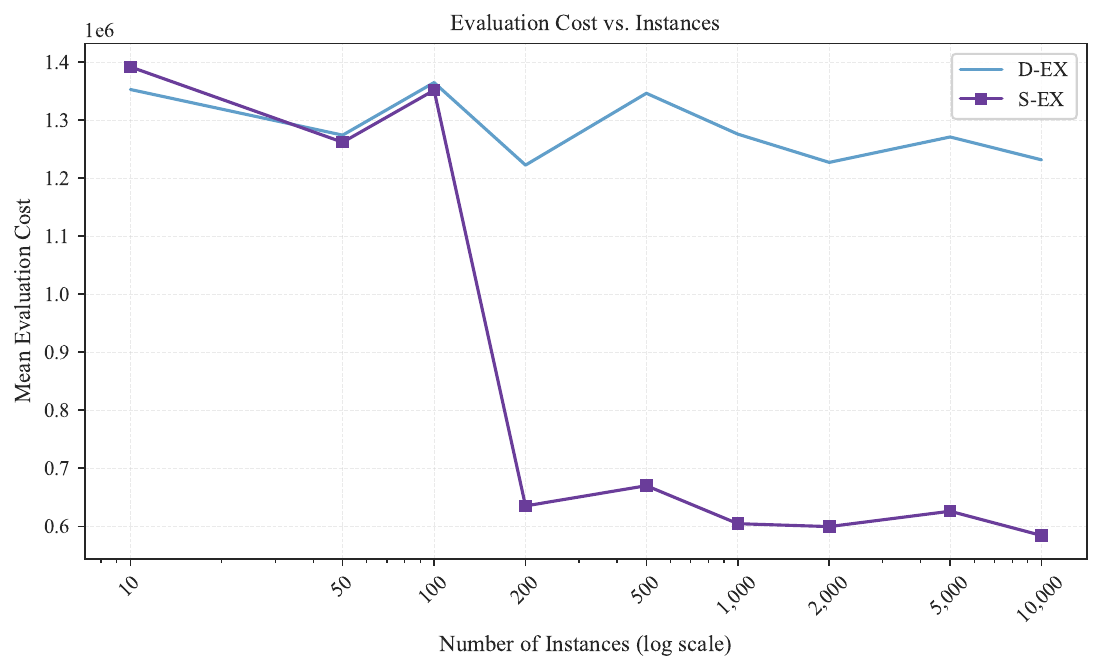}
\caption{Average evaluation-stage cost as a function of total dataset size
(log scale). The \ac{S-EX} cost (blue) drops sharply between $N = 100$
and $N = 200$, then flattens. The \ac{D-EX} cost (grey dashed) remains
approximately constant, confirming that the reduction is driven by improved
\ac{NN} risk modelling.}
\label{fig:sample_eval_cost}
\end{figure}

\begin{figure}[H]
\centering
\includegraphics[width=0.85\columnwidth]{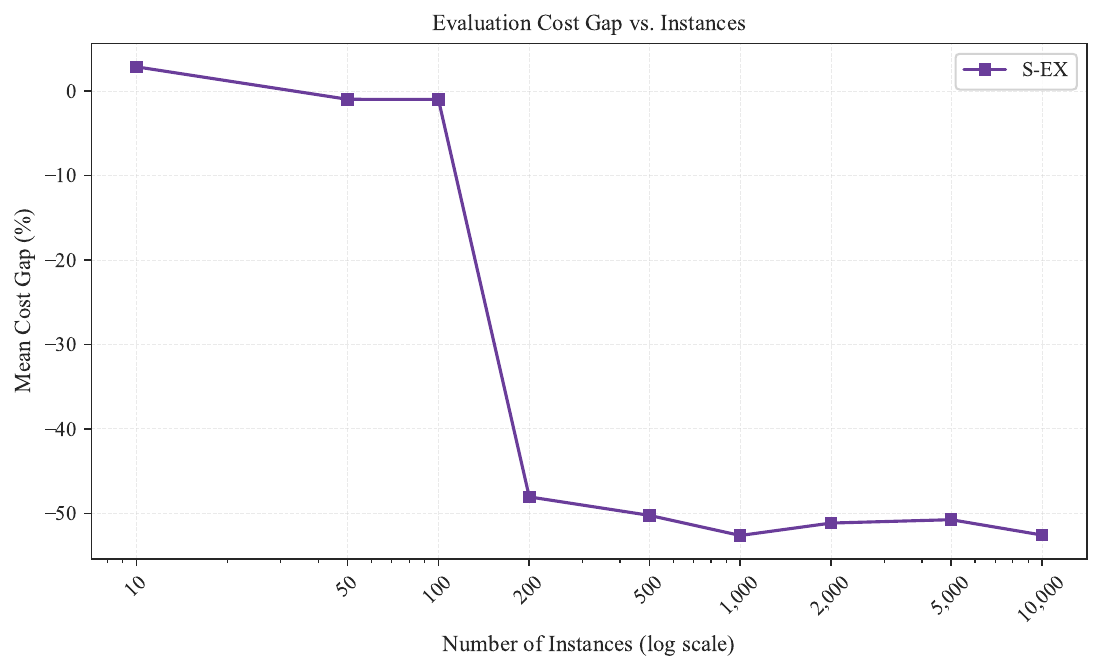}
\caption{Evaluation cost gap (\ac{S-EX} relative to \ac{D-EX}, in percent) as
a function of total dataset size (log scale). The gap transitions from
near-zero ($N \leq 100$) to approximately -48 percent to -53 percent for
$N \geq 200$, with diminishing returns beyond $N \approx 1{'}000$.}
\label{fig:sample_eval_gap}
\end{figure}

Table \ref{tab:sample_efficiency} and
Figures \ref{fig:sample_planning_cost} through \ref{fig:sample_eval_gap} report
planning cost, evaluation cost, and their differences between \ac{S-EX}
and \ac{D-EX} across dataset sizes $N \in \{10, 50, \ldots, 10{'}000\}$.

For $N \leq 100$ (at most 70 training samples) the \ac{NN} has not yet
learned a reliable risk representation: planning costs remain near or below
the \ac{D-EX} baseline, and evaluation costs are virtually unchanged
(+2.9 percent at $N = 10$, -1.0 percent at $N = 50$ and $N = 100$), indicating
negligible hedging benefit.
A sharp transition occurs between $N = 100$ and $N = 200$: the evaluation
cost gap drops from -1.0 percent to -48.0 percent, while planning cost rises by
+25.7 percent relative to \ac{D-EX}, confirming that the \ac{NN} has acquired
sufficient distributional coverage to meaningfully reshape the schedule.
From $N = 200$ onward, the evaluation cost reduction stabilises in the range
-48 percent to -53 percent with only modest further improvement.
At the default configuration $N = 2{'}000$, \ac{S-EX} achieves -51.1 percent
evaluation cost and -50.7 percent fewer cancellations; at $N = 10{'}000$ the
gains are -52.6 percent and -52.1 percent - less than two percentage points better
despite a five-fold data increase.
Solve times remain in the range 4 to 8 s across all configurations, confirming
that \ac{NN} embedding overhead is independent of training set size.

\begin{table}[H]
\centering
\small
\setlength{\tabcolsep}{5pt}
\caption{Training sample efficiency: planning and evaluation metrics as a
function of total dataset size $N$.
Evaluation cost difference is \ac{S-EX} relative to \ac{D-EX}.}
\label{tab:sample_efficiency}
\begin{tabular}{@{}r r r r r r r r@{}}
\toprule
\textbf{\begin{tabular}{@{}c@{}}Total $N$\end{tabular}} &
\textbf{\begin{tabular}{@{}c@{}}Eval\\Instances\end{tabular}} &
\textbf{\begin{tabular}{@{}c@{}}Plan Cost\\(\ac{S-EX})\end{tabular}} &
\textbf{\begin{tabular}{@{}c@{}}Plan Cost\\(\ac{D-EX})\end{tabular}} &
\textbf{\begin{tabular}{@{}c@{}}Plan\\Diff [percent]\end{tabular}} &
\textbf{\begin{tabular}{@{}c@{}}Eval Cost\\(\ac{S-EX})\end{tabular}} &
\textbf{\begin{tabular}{@{}c@{}}Eval Cost\\(\ac{D-EX})\end{tabular}} &
\textbf{\begin{tabular}{@{}c@{}}Eval\\Diff [percent]\end{tabular}} \\
\midrule
    10 &   1 &   498{'}435 &   497{'}602 & +0.17 & 1{'}392{'}000 & 1{'}353{'}000 & +2.88 \\
    50 &   5 &   450{'}259 &   442{'}496 & +1.75 & 1{'}262{'}040 & 1{'}274{'}280 & -0.96 \\
   100 &  10 &   689{'}973 &   634{'}369 & +8.77 & 1{'}351{'}700 & 1{'}364{'}960 & -0.97 \\
   200 &  20 &   813{'}640 &   647{'}308 & +25.7 &   635{'}190 & 1{'}222{'}650 & -48.0 \\
   500 &  50 &   697{'}674 &   556{'}565 & +25.4 &   670{'}064 & 1{'}346{'}468 & -50.2 \\
 1{'}000 & 100 &   811{'}717 &   668{'}813 & +21.4 &   604{'}588 & 1{'}276{'}036 & -52.6 \\
 2{'}000 & 200 &   838{'}074 &   694{'}404 & +20.7 &   599{'}558 & 1{'}227{'}321 & -51.1 \\
 5{'}000 & 500 &   835{'}436 &   702{'}521 & +18.9 &   626{'}046 & 1{'}271{'}128 & -50.7 \\
10{'}000 & 1{'}000 & 803{'}150 & 654{'}812 & +22.7 & 584{'}233 & 1{'}231{'}700 & -52.6 \\
\bottomrule
\multicolumn{8}{@{}l}{\footnotesize \ac{S-EX} (mean training target); 70/10/10/10 split; values averaged over evaluation instances.} \\
\end{tabular}
\end{table}
\newpage

\subsection{Impact of the Neural Network Weight $\lambda$}
\label{sec:results_lambda}

This appendix investigates how the \ac{NN} weight $\lambda$
(Eq. \ref{eq:stochastic_nne}) controls the trade-off between planning-stage
conservatism and evaluation-stage robustness.
At $\lambda = 0$ the \ac{NN} term is suppressed and \ac{S-EX} reduces to
\ac{D-EX}; as $\lambda$ increases, the solver is increasingly penalised for
plans predicted to carry high realisation risk.
Experiments use the \ac{S-EX} configuration with 2{'}000 training instances
and 200 evaluation instances; all other settings match
Section \ref{sec:experimental_setup}.
Note that cost magnitudes differ from the primary results (which use 500
evaluation instances); conclusions are drawn from relative trends.

\begin{figure}[H]
\centering
\includegraphics[width=0.85\columnwidth]{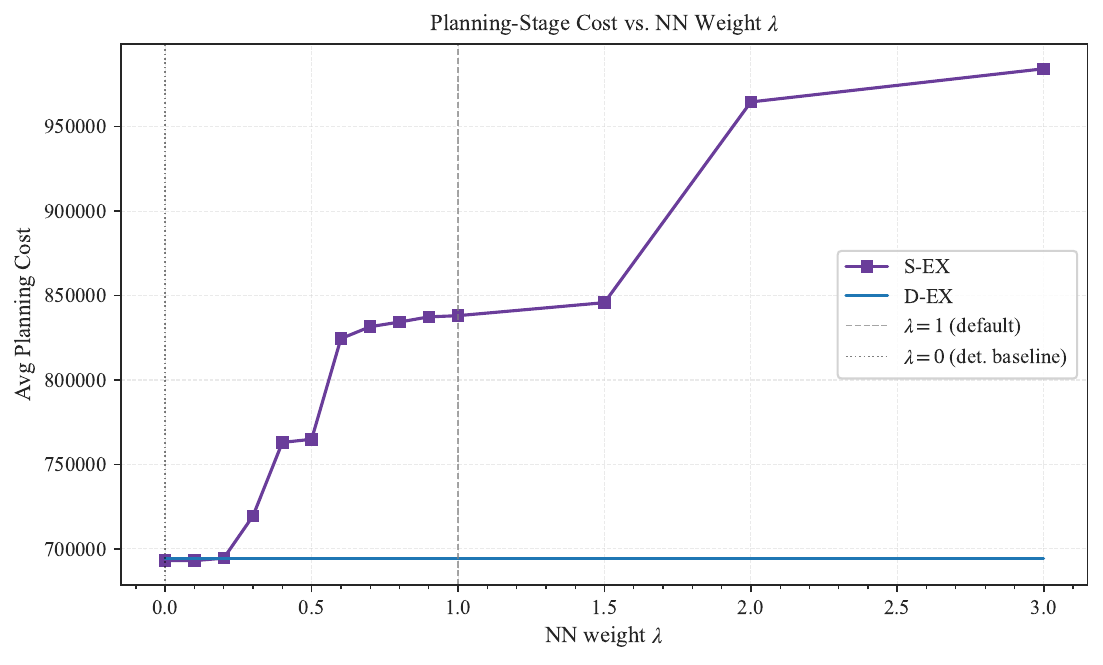}
\caption{Average planning-stage cost as a function of $\lambda$.
The \ac{D-EX} cost (red dashed) is constant; the \ac{S-EX} cost (blue)
rises steeply from $\lambda \approx 0.3$, with a second jump at
$\lambda \approx 2.0$.}
\label{fig:lambda_planning_cost}
\end{figure}

\begin{figure}[H]
\centering
\includegraphics[width=0.85\columnwidth]{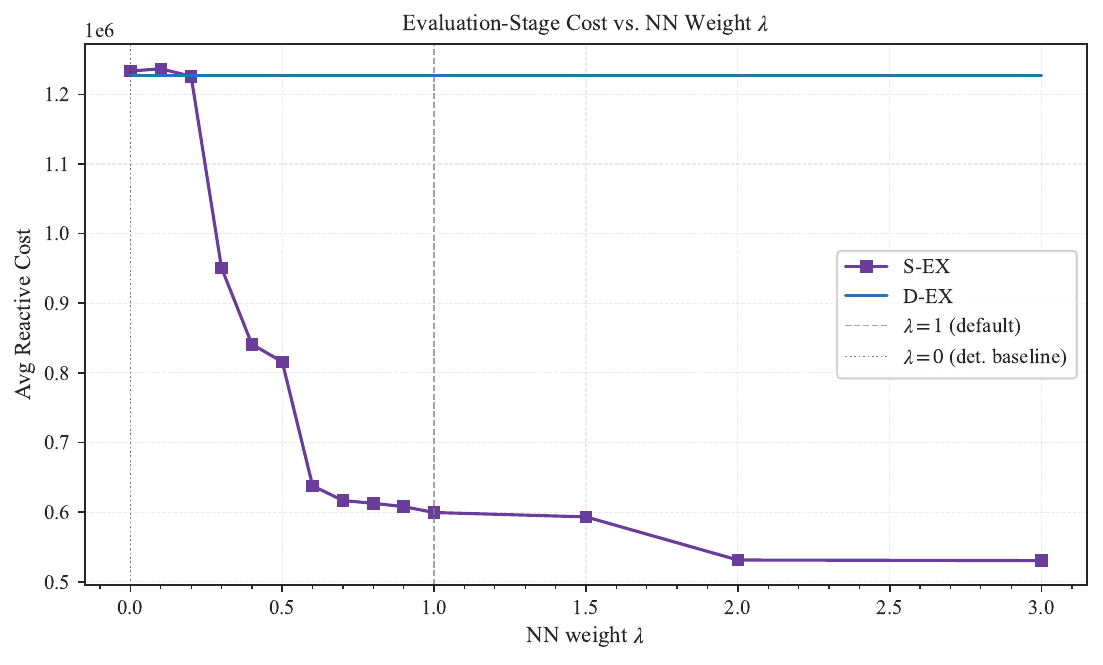}
\caption{Average evaluation-stage cost as a function of $\lambda$.
The \ac{S-EX} cost (blue) drops sharply between $\lambda = 0.2$ and
$\lambda = 0.6$, flattens around $\lambda = 1.0$, and continues to decrease
slowly beyond. The \ac{D-EX} cost (red dashed) is constant.}
\label{fig:lambda_eval_cost}
\end{figure}

\begin{figure}[H]
\centering
\includegraphics[width=0.85\columnwidth]{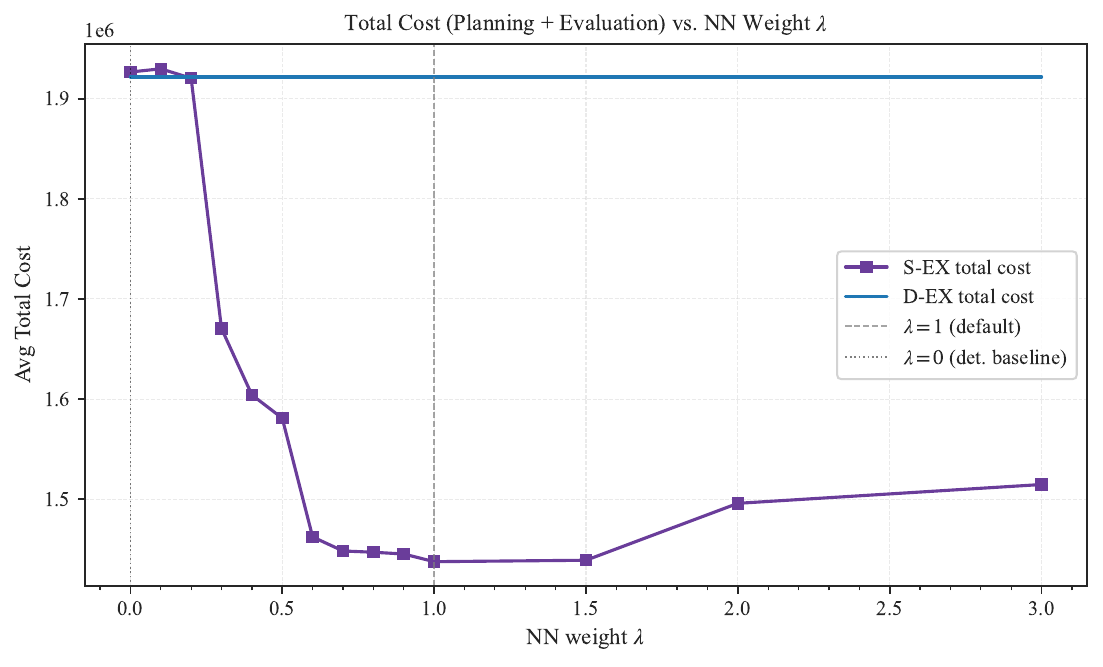}
\caption{Average total cost (planning + evaluation) as a function of $\lambda$.
The \ac{S-EX} total cost (green) reaches its minimum at $\lambda = 1.0$
(-25.2 percent vs. $\lambda = 0$) and rises for $\lambda > 1.5$, confirming the
Pareto knee at $\lambda = 1.0$.}
\label{fig:lambda_total_cost}
\end{figure}

\begin{figure}[H]
\centering
\includegraphics[width=0.85\columnwidth]{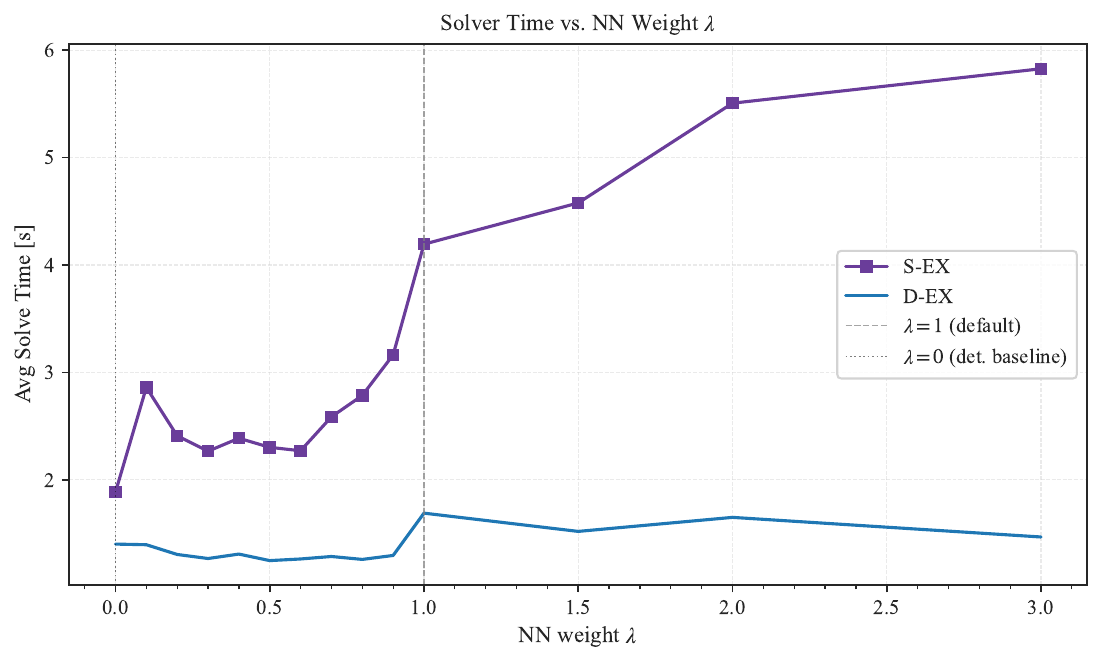}
\caption{Average solver time as a function of $\lambda$.
\ac{S-EX} solve time (blue) rises from 1.89 s at $\lambda = 0$ to
4.19 s at $\lambda = 1.0$ and 5.83 s at $\lambda = 3.0$; the \ac{D-EX}
time (red dashed) remains approximately constant at 1.3 to 1.7 s.}
\label{fig:lambda_solve_time}
\end{figure}

Table \ref{tab:nn_weight_sensitivity} and
Figures \ref{fig:lambda_planning_cost} through \ref{fig:lambda_solve_time} report
results across $\lambda \in \{0.0, 0.1, \ldots, 1.0, 1.5, 2.0, 3.0\}$.

For $\lambda \leq 0.2$ the \ac{NN} term is too small to redirect the solver:
planning and evaluation costs remain near the $\lambda = 0$ baseline
(total cost approximately 1{'}920{'}000 to 1{'}930{'}000), indistinguishable from
plain \ac{D-EX}.
A sharp activation transition occurs at $\lambda \approx 0.3$: planning cost
rises steeply while evaluation cost falls from 1{'}233{'}416 to 950{'}857
(-22.9 percent).
The improvement continues through $\lambda = 0.6$ (evaluation cost 637{'}880,
-48.2 percent relative to $\lambda = 0$; planning cost 824{'}483), after which
diminishing returns set in.

At the recommended default $\lambda = 1.0$, \ac{S-EX} achieves planning
cost 838{'}074 (+20.7 percent vs. \ac{D-EX}) and evaluation cost 599{'}558
(-51.1 percent), yielding a total cost of 1{'}437{'}632 (-25.2 percent) and 5.90
cancellations (-50.8 percent).
Solver time increases from 1.69 s (\ac{D-EX}) to 4.19 s (+148 percent) due
to the linearised \ac{NN} embedding.

Beyond $\lambda = 1.0$, total cost rises again despite continued evaluation
cost reductions, indicating that additional planning conservatism is no longer
offset by evaluation-stage savings.
At $\lambda = 2.0$, planning cost jumps sharply to 964{'}566 and total cost
reaches 1{'}495{'}906, which is higher than at $\lambda = 1.0$.
At $\lambda = 3.0$ total cost climbs to 1{'}514{'}701.
The Pareto knee is therefore at $\lambda = 1.0$, with the feasible operating
range being approximately $\lambda \in [0.3, 1.5]$.

\begin{table}[h]
\centering
\small
\setlength{\tabcolsep}{6pt}
\caption{\ac{NN} weight sensitivity: planning and evaluation metrics for each
$\lambda$ value. All values averaged over 200 evaluation instances.
The \ac{D-EX} column is the deterministic baseline (constant across rows).}
\label{tab:nn_weight_sensitivity}
\begin{tabular}{@{}r r r r r r r@{}}
\toprule
$\lambda$ &
\textbf{\begin{tabular}{@{}c@{}}Plan\\Cost\end{tabular}} &
\textbf{\begin{tabular}{@{}c@{}}Time [s]\end{tabular}} &
\textbf{\begin{tabular}{@{}c@{}}Eval\\Cost\end{tabular}} &
\textbf{\begin{tabular}{@{}c@{}}Eval\\Canc.\end{tabular}} &
\textbf{\begin{tabular}{@{}c@{}}Total\\Cost\end{tabular}} &
\textbf{\begin{tabular}{@{}c@{}}\ac{D-EX}\\Total Cost\end{tabular}} \\
\midrule
0.0 & 693{'}229 & 1.89 & 1{'}233{'}416 & 12.06 & 1{'}926{'}645 & 1{'}921{'}725 \\
0.1 & 693{'}104 & 2.86 & 1{'}236{'}745 & 12.10 & 1{'}929{'}849 & 1{'}921{'}725 \\
0.2 & 694{'}548 & 2.41 & 1{'}225{'}903 & 11.99 & 1{'}920{'}451 & 1{'}921{'}725 \\
0.3 & 719{'}603 & 2.27 &   950{'}857 &  9.33 & 1{'}670{'}460 & 1{'}921{'}725 \\
0.4 & 763{'}048 & 2.39 &   841{'}145 &  8.25 & 1{'}604{'}193 & 1{'}921{'}725 \\
0.5 & 764{'}928 & 2.30 &   815{'}993 &  8.01 & 1{'}580{'}921 & 1{'}921{'}725 \\
0.6 & 824{'}483 & 2.27 &   637{'}880 &  6.27 & 1{'}462{'}363 & 1{'}921{'}725 \\
0.7 & 831{'}554 & 2.59 &   616{'}764 &  6.07 & 1{'}448{'}318 & 1{'}921{'}725 \\
0.8 & 834{'}302 & 2.79 &   612{'}763 &  6.03 & 1{'}447{'}065 & 1{'}921{'}725 \\
0.9 & 837{'}358 & 3.16 &   607{'}945 &  5.99 & 1{'}445{'}303 & 1{'}921{'}725 \\
1.0 & 838{'}074 & 4.19 &   599{'}558 &  5.90 & 1{'}437{'}632 & 1{'}921{'}725 \\
1.5 & 845{'}785 & 4.58 &   593{'}279 &  5.84 & 1{'}439{'}064 & 1{'}921{'}725 \\
2.0 & 964{'}566 & 5.51 &   531{'}340 &  5.23 & 1{'}495{'}906 & 1{'}921{'}725 \\
3.0 & 984{'}207 & 5.83 &   530{'}494 &  5.22 & 1{'}514{'}701 & 1{'}921{'}725 \\
\bottomrule
\multicolumn{7}{@{}l}{\footnotesize \ac{S-EX} (mean training target); 2{'}000 training instances; 100 scenarios per instance; 200 evaluation instances.} \\
\end{tabular}
\end{table}
\newpage
\newpage


In practice, $\lambda$ should be treated as a hyperparameter calibrated on a
held-out validation set.
The sharp transition near $\lambda \approx 0.3$ and the plateau above
$\lambda \approx 1.0$ suggest that the Pareto frontier between planning cost
and evaluation cost is relatively robust around the recommended default, but
values outside the range $[0.3, 1.5]$ are likely to be suboptimal for any
practical cost structure.
For operators placing greater weight on operational resilience (e.g., minimising
cancellations for passengers), values toward the upper end of this range are
appropriate; for operators primarily minimising immediate planning expenditure,
values closer to $\lambda = 0.5$ may be preferable.

\subsection{Sensitivity to \ac{RUL} Uncertainty}
\label{sec:results_sensitivity}

This appendix evaluates the robustness of the stochastic framework across
three contrasting uncertainty regimes, probing boundary conditions and
validating that the \ac{NN} behaves correctly across the full spectrum of
operational uncertainty.

For each regime, 2{'}000 instances were generated
(1{'}400 training, 200 validation, 200 test, 200 evaluation) using
the \ac{S-EX} configuration (mean training target).
Key metrics are planning cost, evaluation cost, solve time, and cancellations
for \ac{S-EX} versus the \ac{D-EX} baseline.

\begin{table}[h]
\centering
\setlength{\tabcolsep}{14pt}
\caption{\ac{RUL} uncertainty sensitivity: \ac{S-EX} vs. \ac{D-EX}
across three regimes. Delta columns show relative change.}
\label{tab:rul_sensitivity_combined}
\begin{tabular}{@{}lrrrrr@{}}
\toprule
\textbf{Regime} &
\textbf{Plan} &
\textbf{Eval} &
\textbf{Time} &
\textbf{Plan} &
\textbf{Eval} \\
&
\textbf{Cost} &
\textbf{Cost} &
\textbf{[s]} &
\textbf{Delta [\%]} &
\textbf{Delta [\%]} \\
\midrule
\multicolumn{6}{@{}l}{\textit{High Uncertainty} ($\sigma = 50$, RUL in [5, 30])} \\
\quad \ac{S-EX}  & 893{'}187 & 559{'}294 & 5.35 & +28.6 & -54.4 \\
\quad \ac{D-EX}     & 694{'}404 & 1{'}227{'}752 & 3.16 & & \\
\midrule
\multicolumn{6}{@{}l}{\textit{Zero Uncertainty} ($\sigma = 0$, RUL in [5, 30])} \\
\quad \ac{S-EX}  & 705{'}091 & 1{'}236{'}166 & 92.16 & +1.54 & +0.69 \\
\quad \ac{D-EX}     & 694{'}404 & 1{'}227{'}752 &  5.34 & & \\
\midrule
\multicolumn{6}{@{}l}{\textit{Negligible Risk} (RUL in [200, 250])} \\
\quad \ac{S-EX}  & 669{'}587 & 0 & 2.99 & +0.07 & +0.00 \\
\quad \ac{D-EX}     & 669{'}112 & 0 & 1.48 & & \\
\bottomrule
\multicolumn{6}{@{}l}{\footnotesize 200 evaluation instances per regime; Time Delta removed for space.} \\
\multicolumn{6}{@{}l}{\footnotesize Cancellations Delta: High -54.1\%; Zero +0.76\%; Negligible +0.00\%.} \\
\end{tabular}
\end{table}

\textbf{High uncertainty ($\sigma = 50$ cycles).}
When prognostic uncertainty is substantial, \ac{S-EX} demonstrates its
core value: planning cost increases by +28.6 percent but evaluation cost falls by
-54.4 percent, with cancellations reduced by 54.1 percent, yielding a net total cost
reduction of approximately -26 percent.
The +69 percent solve time increase (3.16 s to 5.35 s) remains well within
operational planning requirements.

\textbf{Zero uncertainty ($\sigma = 0$).}
When all \ac{RUL} predictions are deterministic, the \ac{NN} has no
uncertainty to hedge.
Planning and evaluation costs differ by only +1.54 percent and +0.69 percent
respectively, confirming that \ac{S-EX} degenerates correctly to
near-deterministic behaviour.
However, solve time increases dramatically (+1{'}624 percent to 92.16 s) because
the linearised \ac{NN} introduces additional constraints that complicate the
branch-and-cut search under zero-variance inputs.
Practical implication: when \ac{RUL} uncertainty is known to be
negligible, the plain \ac{D-EX} formulation should be used directly.

\textbf{Negligible risk (RUL in [200, 250] cycles).}
With \ac{RUL} values far exceeding the approximately 15 flight cycles consumed in a
3-day horizon, no component can fail within the planning window.
Both configurations yield identical outcomes (zero evaluation cost, planning
difference +0.07 percent), and the \ac{NN} contribution vanishes gracefully.
Solve time increases modestly (+102 percent to 2.99 s) purely from embedding
overhead, not problem difficulty.

Taken together, these three regimes confirm that \ac{S-EX} is
well-calibrated: it provides strong robustness gains when uncertainty is
operationally consequential, introduces no distortion when uncertainty is
absent, and degrades gracefully to near-standard optimisation when risk is
negligible.
This behaviour is a necessary condition for safe deployment in operational
environments where problem characteristics vary across planning cycles.

\subsection{Statistical Testing of Performance Differences}
\label{app:statistical_tests_detailed}

To rigorously confirm that observed total-cost differences are not due to
sampling variability, we apply the Wilcoxon signed-rank test to pairwise
comparisons of per-instance total costs across all 500 held-out evaluation
instances per configuration.

\subsubsection{Methodology}

The Wilcoxon signed-rank test is preferred over paired t-tests because total
costs exhibit right-skewed distributions driven by high-cost outlier scenarios.
The test employs a one-sided alternative: the first listed configuration yields
higher costs (i.e., the second is better).
Effect size is quantified by rank-biserial correlation:
\begin{align}
r = 1 - \frac{4W}{N(N+1)}
\end{align}
where $W$ is the Wilcoxon W-statistic and $N = 1{'}000$ (combining 500
instances from each configuration).
By convention, $|r| \geq 0.5$ indicates a large effect and
$0.3 \leq |r| < 0.5$ a medium effect (Cohen 1988).
Bootstrap 95 percent confidence intervals for mean cost differences were computed
with $B = 10{'}000$ resamples (random seed 42).

Note that the mean cost differences reported here are absolute values computed
on the statistical evaluation dataset, which may differ in scale from the
primary results in Table \ref{tab:planning_eval_combined} due to differences
in instance sampling across experimental runs.
The percentage reduction column should therefore be interpreted as a
within-experiment relative measure rather than a direct comparison to
Section \ref{sec:results}.

\subsubsection{Results}

\begin{table}[h]
\centering
\small
\setlength{\tabcolsep}{4pt}
\caption{Pairwise statistical comparison of total costs (planning + evaluation).
Wilcoxon signed-rank test; one-sided alternative: first configuration is better.
Bootstrap 95 percent CI on mean cost difference (B = 10{'}000 resamples,
N = 1{'}000 paired comparisons).}
\label{tab:statistical_tests_summary}
\begin{tabular}{@{}lrrrrrr@{}}
\toprule
\textbf{Comparison} &
\textbf{Mean Delta} &
\textbf{Percent Red.} &
\textbf{95 percent CI} &
\textbf{W} &
\textbf{p-value} &
\textbf{r} \\
\midrule
\ac{D-EX} vs. \ac{S-EX}       &  23{'}151 &   7.0          & [18{'}777; 27{'}493]         & 328{'}272 & 1.04e-26  & -0.31 \\
\ac{D-EX} vs. \ac{D-EX}-95    & 329{'}774 &  99.9          & [323{'}405; 336{'}054]       & 500{'}500 & 1.66e-165 & -1.00 \\
\ac{D-EX} vs. \ac{S-EX}-95    &  61{'}595 &  18.7          & [56{'}187; 67{'}031]         & 421{'}675 & 1.32e-81  & -0.69 \\
\ac{D-EX}-95 vs. \ac{S-EX}-95 & -268{'}180 & n/a            & [-274{'}029; -262{'}414] & 0         & 1.00                  & +1.00 \\
\ac{S-EX} vs. \ac{S-EX}-95    &  38{'}443 &  12.5          & [33{'}731; 43{'}106]         & 371{'}144 & 2.95e-54  & -0.48 \\
\bottomrule
\multicolumn{7}{@{}l}{\footnotesize Mean Delta: total cost of first minus second; negative means second is more expensive.} \\
\end{tabular}
\end{table}

\subsubsection{Interpretation}

\textbf{\ac{D-EX} vs. \ac{S-EX}.}
Adding mean-based \ac{NN} guidance to the deterministic baseline reduces total
cost by a mean of 23{'}151 (7.0 percent) with a medium effect size (r = -0.31,
p = 1.04e-26).
The result is statistically robust across all 500 instance pairs and confirms
that mean-based uncertainty modelling provides incremental but meaningful
gains over purely deterministic planning.

\textbf{\ac{D-EX} vs. \ac{D-EX}-95.}
The 95th-percentile deterministic baseline incurs a dramatic cost increase
relative to \ac{D-EX} (mean Delta = 329{'}774; 95 percent CI [323{'}405;
336{'}054]), with perfect rank-biserial correlation (r = -1.00,
p = 1.66e-165), indicating a categorical behavioural difference.
The conservative approach nearly eliminates evaluation costs by scheduling
extensive preventive maintenance, but does so at prohibitive planning cost,
resulting in a higher total cost on every single instance.
This is the strongest statistical finding in the study and directly motivates
the \ac{NN} approach, which achieves evaluation-stage robustness without
this planning cost explosion.

\textbf{\ac{D-EX} vs. \ac{S-EX}-95.}
The proposed approach reduces total cost by a mean of 61{'}595 (18.7 percent) with
a large effect (r = -0.69, p = 1.32e-81).
The large effect size confirms that the improvement is not a marginal
efficiency gain but a qualitative shift in the cost structure of the resulting
plans.

\textbf{\ac{S-EX} vs. \ac{S-EX}-95.}
The 95th-percentile training target reduces total cost by a mean of 38{'}443
(12.5 percent) relative to mean-based training, with a medium effect size
(r = -0.48, p = 2.95e-54).
This confirms that the choice of training target is statistically significant:
hedging against tail-risk scenarios during \ac{NN} training produces
meaningfully more robust plans than hedging against expected costs alone,
which directly justifies the \ac{S-EX}-95 configuration as the recommended
default.

\subsubsection{Summary}

The statistically robust conclusions across all comparisons are:
(i) \ac{S-EX}-95 significantly outperforms \ac{D-EX} (large effect,
p << 0.001);
(ii) \ac{D-EX}-95 categorically incurs higher total cost than \ac{D-EX}
in every instance (r = -1.00), demonstrating that deterministic worst-case
conservatism is always economically dominated by \ac{D-EX};
and (iii) the 95th-percentile \ac{NN} training target (\ac{S-EX}-95)
significantly outperforms mean-based training (\ac{S-EX}), validating the
choice of $\bar{Q}^{95}$ as the recommended learning objective.

%% file: 11_abbreviations.tex
\section*{Abbreviations}
\begin{acronym}
\acro{ADP}{Approximate Dynamic Programming}
\acro{AMP}{Aircraft Maintenance Program}
\acro{CBM}{Condition-Based Maintenance}
\acro{CNN}{Convolutional Neural Network}
\acro{D-2SP}{Deterministic-Two-Stage Stochastic Programming}
\acro{D-EX}{Deterministic-Extensive}
\acro{DES}{Discrete-Event Simulation}
\acro{DQN}{Deep Q-Network}
\acro{DRL}{Deep Reinforcement Learning}
\acro{DRO}{Distributionally Robust Optimization}
\acro{EX}{Extensive}
\acro{GRU}{Gated Recurrent Unit}
\acro{HMC}{Heavy Maintenance Check}
\acro{ILP}{Integer Linear Programming}
\acro{IP}{Integer Programming}
\acro{MAREP}{Maintenance Report}
\acro{MILP}{Mixed-Integer Linear Programming}
\acro{ML}{Machine Learning}
\acro{MRO}{Maintenance, Repair, and Overhaul}
\acro{MS}{Maintenance Scheduling}
\acro{NN}{Neural Network}
\acro{PdM}{Predictive Maintenance}
\acro{PIREP}{Pilot Report}
\acro{POMCP}{Partially Observable Monte Carlo Planning}
\acro{ReLU}{Rectified Linear Unit}
\acro{RL}{Reinforcement Learning}
\acro{RUL}{Remaining Useful Life}
\acro{S-2SP}{Stochastic-Two-Stage Stochastic Programming}
\acro{S-EX}{Stochastic-Extensive}
\acro{SAA}{Sample Average Approximation}
\acro{SP}{Stochastic Programming}
\acro{TA}{Tail Assignment}
\acro{2SP}{Two-Stage Stochastic Programming}
\end{acronym}